\documentclass[twocolumn, aps,prd,preprintnumbers,superscriptaddress,showpacs,nofootinbib]{revtex4-1}
\usepackage{amsmath,amsthm,amssymb,mathptmx}
\usepackage[dvipdfmx]{graphicx}
\usepackage{color}
\usepackage{float}
\usepackage{booktabs}
\usepackage{bm}

\newcommand{\sigmav}{\sigma_{\rm v}}
\allowdisplaybreaks[1]

\begin{document}
\title{Precision cosmology with redshift-space bispectrum: a perturbation theory based model at one-loop order} 

\author{Ichihiko~Hashimoto}
\affiliation{Yukawa Institute for Theoretical Physics, Kyoto University, Kyoto 606-8502, Japan}

\author{Yann~Rasera}
\affiliation{LUTH, Observatoire de Paris, PSL Research University, CNRS, Universit\'e Paris Diderot, Sorbonne Paris Cit\'e, 5 place Jules Janssen, 92195 Meudon, France}
\affiliation{Center for Gravitational Physics, Yukawa Institute for Theoretical Physics, Kyoto University, Kyoto 606-8502, Japan}

\author{Atsushi~Taruya}
\affiliation{Center for Gravitational Physics, Yukawa Institute for Theoretical Physics, Kyoto University, Kyoto 606-8502, Japan}
\affiliation{Kavli Institute for the Physics and Mathematics of the Universe (WPI), Todai institute
for Advanced Study, University of Tokyo, Kashiwa, Chiba 277-8568, Japan}

\date{\today}

\begin{abstract}
The large-scale matter distribution in the late-time Universe exhibits gravity-induced non-Gaussianity, and the bispectrum, three-point cumulant is expected to contain significant cosmological information. In particular, the measurement of the bispectrum helps to tighten the constraints on dark energy and modified gravity through the redshift-space distortions (RSD). In this paper, extending the work by Taruya, Nishimichi \& Saito (2010, Phys.Rev.D {\bf 82}, 063522), we present a perturbation theory (PT) based model of redshift-space matter bispectrum that can keep the non-perturbative damping effect under control. Characterizing this non-perturbative damping by a univariate function with single free parameter, the PT model of the redshift-space bispectrum is tested against a large set of cosmological $N$-body simulations, finding that the predicted monopole and quadrupole moments are in a good agreement with simulations at the scales of baryon acoustic oscillations (well beyond the range of agreement of standard PT). The validity of the univariate ansatz of the damping effect is also examined, and with the PT calculation at next-to-leading order, the fitted values of the free parameter is shown to consistently match those obtained from the PT model of power spectrum by Taruya, Nishimichi \& Saito (2010). 
\end{abstract}

\preprint{YITP-17-46}
\pacs{98.80.-k,\,\,98.65.Dx}
\keywords{cosmology, large-scale structure}

\maketitle


\section{Introduction}
\label{sec:intro}
The galaxy clustering observed via the spectroscopic measurements appears distorted due to the peculiar velocity of galaxies. This systematic effect is known as the {\it redshift-space distortions} (RSD) (e.g., Refs.~\cite{1983ApJ...267..465D,Kaiser:1987qv,Hamilton:1992zz,Scoccimarro:2004tg}), which manifestly break the statistical isotropy. While the RSD complicates the interpretation of the galaxy clustering data, it now attracts much attention as a probe of gravity on cosmological scales. This is because, on large scales, the effect is accounted by the linear theory, and the size of the anisotropies is related to the growth of structure (e.g., Refs.~\cite{Kaiser:1987qv,Hamilton:1992zz}).  To be precise, the measurement of RSD offers an opportunity to determine the parameter $f\,\sigma_8$ (e.g., Ref.~\cite{Percival:2008sh,Song:2008qt,Yamamoto:2008gr}). Here, $f$ is the linear growth rate defined by $f\equiv\,d\ln D_+/d\ln a$ with $D_+$ and $a$ being respectively the linear growth factor and scale factor of the Universe, and $\sigma_8$ is the normalization amplitude of the linear power spectrum smoothed over $8\,h^{-1}$\,Mpc. Combining the RSD with the measurement of baryon acoustic oscillations (BAO) imprinted on galaxy clustering (e.g., \cite{EisensteinHu1998,Meiksin:1998ra,Eisenstein:2005su}), we can simultaneously estimate both the geometric distances and growth of structure (e.g., \cite{Taruya:2010mx,2012MNRAS.426.2719R,Oka:2013cba,Beutler:2013yhm}), from which we are able to scrutinize the theory of gravity that accounts for cosmic expansion and structure formation (e.g., Refs.~\cite{Linder:2007nu,Seo:2003pu,Blake:2003rh,Glazebrook:2005mb,2015PhRvD..92d3522S}).

The measurements of RSD taken from various observations now cover a wide range of redshifts out to $z\sim1.4$ (e.g., Refs.~\cite{Guzzo:2008ac,Okumura:2015lvp,2013A&A...557A..54D,Beutler:2016arn,Blake:2011rj}, see also Sec.~5.5 of Ref.~\cite{Planck2015_XIII}), and the results are broadly consistent with general relativity. But a closer look at the constrained values of $f\,\sigma_8$ suggests a mild tension with the Planck $\Lambda$CDM model \cite{2015arXiv150201589P}, indicating some systematics or potentially new physics. A further improvement on the RSD measurement is thus important, mitigating both the statistical errors and systematics. This is one of the main reasons why there are various projects aiming at precisely measuring RSD which will uncover a large cosmic volume.

With a gigantic survey volume, the next-generation galaxy redshift surveys (e.g. Euclid, LSST) will therefore offer an opportunity to precisely measure the power spectrum or correlation function at an unprecedented precision, and provided a high-precision template, a robust and tight constraint on $f\,\sigma_8$ will be expected. Furthermore, one promising point is that taking advantage of a large-volume data, a high-statistical measurement of the higher-order statistics such as bispectrum or three-point correlation function is made available, and combining it with power spectrum measurement further helps to tighten the constraint on $f\,\sigma_8$.

It is widely known that the nonlinearity of gravity generically incorporates the non-Gaussian statistical nature into the large-scale structures, and a part of the cosmological information initially encoded in the two-point statistics is leaked to the higher-order statistics. In this respect, the combination of the bispectrum data with power spectrum in the cosmological data analysis is a natural first step to efficiently extract the cosmological information from the large-scale structures. There have been various studies based on the Fisher matrix analysis to quantify the statistical impact of the bispectrum on the cosmological parameter estimation, showing that adding the bispectrum indeed plays an important role to break parameter degeneracy (e.g., Refs.~\cite{Sefusatti:2006pa,2013MNRAS.429..344K,Sato:2013mq,Greig:2012zw}). In particular, the impact of combining bispectrum measurement is demonstrated for the RSD measurement by Refs.~\cite{Song:2015gca,2017MNRAS.tmp..138G}, showing that even if we restrict the data to the large-scale modes, the constraint on $f\,\sigma_8$ will be improved by more than a factor of two.

However, most of these analysis has been demonstrated in a very simplified setup, assuming that the observed bispectrum is described by the perturbation theory (PT) at leading order \cite{Scoccimarro:1999ed}, on top of a simple prescription for galaxy bias. While such a simplified forecast study helps to understand the ability and/or potential of the planned galaxy surveys, there are a number of remarks and cautions in a practical data analysis. One important point to be noted is the theoretical template of redshift-space bispectrum. While the perturbative description is supposed to be adequate in real space at the weakly nonlinear stage of gravitational clustering, the situation becomes more subtle in redshift space, because the redshift space is related to the real space through nonlinear mapping. As a result, in terms of the real-space quantities, even the redshift-space power spectrum cannot be simply expressed as the large-scale two-point statistics of the underlying fields, and is significantly affected by the small-scale physics (e.g., Ref.~\cite{Scoccimarro:2004tg}). Hence, a sophisticated treatment is needed for a reliable theoretical template even in the weakly nonlinear regime \cite{Taruya:2010mx, Reid:2011ar,Vlah:2012ni,Matsubara:2007wj,Matsubara:2008wx,Carlson:2012bu,Matsubara:2013ofa,2013PhRvD..87h3509T}.

The aim of this paper is to address this issue in the case of bispectrum, and based on the next-to-leading order calculations, we present a perturbation-theory model of redshift-space bispectrum, which consistently incorporates the effect coming from the small-scale virial motion. While modeling the redshift-space bispectrum has been already investigated based on fitting formula \cite{2012JCAP...02..047G,2001MNRAS.325.1312S} and halo model (e.g., Refs.~\cite{2008PhRvD..78b3523S,Yamamoto:2016anp}), analytical treatment would be certainly powerful in characterizing the anisotropies of the redshift-space bispectrum (see e.g., \cite{2012JCAP...06..018R} for discussion on resummed PT treatment). In particular, with perturbation theory, one can give an accurate description for the large scales of our interest. Albeit its limitation, the PT-based modeling therefore deserves further consideration, and we present, for the first time, monopole and quadrupole moments of the redshift-space bispectrum at one-loop order, which agree well with $N$-body simulations at BAO scales.

This paper is organized as follows. In Sec.~\ref{sec:P_and_B}, we begin by briefly reviewing the relation between real and redshift spaces, and derive the exact formulas for power spectrum and bispectrum in redshift space. Based on the standard PT, the expressions for redshift-space bispectrum at one-loop order are also presented. Then, in Sec.~\ref{ourmodel}, extending the work by Ref.~\cite{Taruya:2010mx}, we give the PT-based model of redshift-space bispectrum, which consistently includes both the next-to-leading order PT corrections and non-perturbative damping effect coming from the small-scale virial motion. Rewriting the exact formulas with the cumulants, we derive the expressions for our PT model valid at one-loop order. In Sec.~\ref{sec:comparison}, the proposed PT model is quantitatively tested against a large suite of cosmological $N$-body simulations, and the validity of the assumption is checked in detail. Finally, Sec.~\ref{sec:sum} is devoted to discussion and conclusion on our important findings. Throughout the paper, we assume $\Lambda$CDM model and adopt the cosmological parameters determined from seven-year WMAP results \cite{2011ApJS..192...18K}: 
 $h=0.72$, $\Omega_{\rm m}=0.2573$, $\Omega_b=0.04356$, $\sigma_8=0.801$ and $n_s=0.963$.

\section{Power spectrum and Bispectrum in redshift space}
\label{sec:P_and_B}

In this section, we begin by writing down the exact expressions of power spectrum and bispectrum for matter density field in redshift space. We then consider the standard PT calculations, and based on the explicit expressions for bispectrum up to the next-to-leading order called one-loop, we discuss its limitation and applicability.

\subsection{Exact formula of power spectrum and bispectrum}

Throughout the paper, we consider the distant-observer case, in which observer's line-of-sight direction is described by a specific direction given by the unit vector $\hat{\bm{z}}$. The mapping between real-space position $\bm{r}$ and redshift-space position $\bm{s}$ is then given by
\begin{align}
\bm{s}=\bm{r}+\frac{(1+z)\,v_z(\bm{r})}{H(z)}\hat{\bm{z}},
\label{eq:red_real_spaces} 
\end{align}
where $z$ is the redshift, $v_z$ is the line-of-sight component of the velocity field, $\bm{v}$ (i.e., $v_z=\bm{v}\cdot \bm{z}$). The quantity $H$ is the Hubble constant. 
Denoting respectively the density fields in real and redshift spaces by $\delta(\bm{r})$ and $\delta^{({\rm s})}(\bm{s})$, the mass conservation implies
$\{1+\delta^{({\rm s})}(\bm{s})\}d^3\bm{s}=\{1+\delta(\bm{r})\}d^3\bm{r}$, which leads to 
\begin{align}
\delta^{({\rm s})}(\bm{s})=\left|\frac{\partial\bm{s}}{\partial\bm{r}}\right|^{-1}\{1+\delta(\bm{r})\}-1.
\label{eq:density_red_real_spaces} 
\end{align}
Since we are interested in the statistical quantities in Fourier-space,  we rewrite the above relation in terms of the Fourier counterpart: 
\begin{align}
\delta^{\rm (s)}(\bm{k})=\int d\bm{r}^3\left\{\delta(\bm{r})-\frac{\nabla_zv_z(\bm{r})}{aH(z)}\right\}e^{i(k\mu v_{\rm z}/H+\bm{k}\cdot\bm{r})},
\label{fredels}
\end{align}
where the variable $\mu$ is defined as the directional cosine of the angle between the line-of-sight $\hat{\bm{z}}$ and wave vector $\bm{k}$.

Using Eq.~(\ref{fredels}), the exact formulas for the power spectrum and bispectrum in redshift space, expressed in terms of the real-space quantities, are derived. Recalling that the power spectrum and bispectrum are respectively defined as
\begin{align}
&(2\pi)^3\delta_{\rm D}(\bm{k}+\bm{k}')P^{\rm (s)}(\bm{k})\equiv\langle\delta^{\rm (s)}(k)\delta^{\rm (s)}(k')\rangle,
\label{def_pow}
\\
&(2\pi)^3\delta_{\rm D}(\bm{k}_1+\bm{k}_2+\bm{k}_3)B^{\rm (s)}(\bm{k}_1,\bm{k}_2,\bm{k}_3)
\nonumber
\\
&\equiv\langle\delta^{\rm (s)}(k_1)\delta^{\rm (s)}(k_2)\delta^{\rm (s)}(k_3)\rangle,
\label{def_bis}
\end{align}
substituting Eq.~(\ref{fredels}) into the above leads to
\begin{align}
&P^{\rm (s)}(\bm{k})=\int d\bm{r}e^{i\bm{k}\cdot\bm{r}_{12}}\nonumber\\
&\times\biggl\langle e^{-ifk_{z}u_{12z}}\left\{\delta(\bm{r}_1)+f\nabla_z u_z(\bm{r}_1)\right\}\left\{\delta(\bm{r}_2)+f\nabla_z u_z(\bm{r}_2)\right\}\biggr\rangle,\label{power_first}
\end{align}
for the redshift-space power spectrum, and
\begin{align}
&B^{\rm (s)}(\bm{k}_1,\bm{k}_2,\bm{k}_3)=\int d\bm{r}_{13}\bm{r}_{23}e^{i(\bm{k}_1\cdot\bm{r}_{13}+\bm{k}_2\cdot\bm{r}_{23})}\nonumber\\
&\times\biggl\langle e^{-if(k_{1z}u_{13z}+k_{2z}u_{23z})}\left\{\delta(\bm{r}_1)+f\nabla_z u_z(\bm{r}_1)\right\}\nonumber\\
&\times\left\{\delta(\bm{r}_2)+f\nabla_z u_z(\bm{r}_2)\right\}\left\{\delta(\bm{r}_3)+f\nabla_z u_z(\bm{r}_3)\right\}\biggr\rangle,
\label{bis_first}
\end{align}
for the redshift-space bispectrum. Here, we define the normalized peculiar velocity $\bm{u}(\bm{r})=-\bm{v}(\bm{r})/\{faH\,(z)\}$ and denote the pairwise normalized velocity of the separation $\bm{r}_{ij}=\bm{r}_i-\bm{r}_j$ by $\bm{u}_{ij}\equiv\bm{u}_i-\bm{u}_j$. The function $f$ is the linear growth rate defined by $f\equiv d\ln{D(z)}/d\ln{a}$, with $D(z)$ being the linear growth factor.

The above expressions show that albeit the simple relation (\ref{eq:red_real_spaces}), the power spectrum and bispectrum in redshift space are rather intricate statistical relation. Qualitatively, the amplitude of the power spectrum is known to be boosted by the additional term of the velocity field at large scales (Kaiser effect), and to be suppressed at small scales by an exponential damping factor (Fingers-of-God effect). Since the structure of the expressions is of the same form in both the power spectrum and bispectrum, we expect that the redshift-space bispectrum possesses similar qualitative features. The additional complexity in bispectrum is, however, that it is no longer characterized simply by the shape of the triangle, i.e., length of three wave vectors $k_1$, $k_2$ and $k_3$, or length of vectors $k_1$ and $k_2$ and their angle $\theta_{12}\equiv\cos^{-1}(\bm{\hat{k}}_1\cdot\bm{\hat{k}}_2)$. On top of these three variables, one needs two more variables to describe the orientation of the triangular shape with respect to the line-of-sight direction.  In this respect, the identification and separation of the Kaiser and Finger-of-God effects are rather non-trivial, and a careful treatment is required for an accurate modeling of redshift-space bispectrum.

\subsection{Standard perturbation theory in redshift space}
\label{1-loop_sPT}

In this paper, we employ the perturbation theory technique to construct an analytic model of redshift-space bispectrum relevant at large scales. To do this, we here present the standard PT results at next-to-leading order called one-loop, and discuss its limitation and the issues to be improved. We will then present, in next section, a PT model of redshift-space bispectrum which properly incorporates the effect of FoG damping in a non-perturbative manner.

To derive the standard PT results of redshift-space bispectrum, let us first expand the redshift-space density field in Fourier space. Regarding the real-space density and velocity fields as perturbative quantities, a systematic expansion of Eq.~(\ref{fredels}) leads to 
\begin{align}
&\delta^{\rm (s)}(\bm{k})=\sum_{n=0}\int \frac{d^3\bm{q}_1}{(2\pi)^3}\cdots\int \frac{d^3\bm{q}_n}{(2\pi)^3} \delta_{\rm D}(\bm{k}-\bm{q}_{1\cdots n})
\nonumber
\\
&\times \bigl\{\delta(\bm{k})+f\mu^2\theta(\bm{k})\bigr\}
\,\frac{(f\mu k)^n}{n!}\frac{\mu_1}{q_1}\theta(\bm{q}_1)\cdots\frac{\mu_n}{q_n}\theta(\bm{q}_n)
\label{teylar}
\end{align}
with $\bm{q}_{1\cdots n}$ being $\bm{q}_{1}+\cdots+\bm{q}_{n}$.
In the above, we assumed the irrotational flow of the velocity field, and introduced the velocity-divergence field defined by $\theta(\bm{x})=-\nabla\cdot\bm{v}/(f\,aH)$. Then, we further expand the real-space quantities in terms of the standard PT kernels: 
\begin{align}
&\delta(\bm{k})=\sum_{n=1}\int\frac{d^3\bm{p}_1\cdots d^3\bm{p}_n}{(2\pi)^{3n}}\,\delta_{\rm D}(\bm{k}-\bm{p}_{1\cdots n})\,F_n(\bm{p}_1,\cdots,\bm{p}_n)
\nonumber\\
&\qquad\qquad\times
\,\delta_{\rm L}(\bm{p}_1)\cdots\delta_{\rm L}(\bm{p}_n),
\label{eq:delta_n_SPT}
\\
&\theta(\bm{k})=\sum_{n=1}\int\frac{d^3\bm{p}_1\cdots d^3\bm{p}_n}{(2\pi)^{3n}}\,\delta_{\rm D}(\bm{k}-\bm{p}_{1\cdots n})\,G_n(\bm{p}_1,\cdots,\bm{p}_n)
\nonumber\\
&\qquad\qquad\times
\,\delta_{\rm L}(\bm{p}_1)\cdots\delta_{\rm L}(\bm{p}_n),
\label{eq:theta_n_SPT}
\end{align}
where $\delta_{\rm L}$ is the linear density field, and $F_n$ and $G_n$ are the standard PT kernels (see for example \cite{Bernardeau:2002}). Plugging the above expressions into Eq.~(\ref{teylar}), reorganizing the perturbative expansion in powers of $\delta_{\rm L}$ leads to
\begin{align}
&\delta^{\rm (s)}(\bm{k})=\sum_{n=1}\int \frac{d^3\bm{q}_1}{(2\pi)^3}\cdots\int \frac{d^3\bm{q}_n}{(2\pi)^3}\delta_D(\bm{k}-\bm{q}_{1\cdots n})\nonumber\\
&\qquad
\times Z_n(\bm{q}_1,\cdots,\bm{q}_n)\delta_{\rm L}(\bm{q}_1)\cdots\delta_{\rm L}(\bm{q}_n).
\label{RSD_fluc}
\end{align}
Here, the $Z_n$ are the so-called redshift-space PT kernel, and these are expressed in terms of the real-space PT kernels, $F_n$ and $G_n$ in a rather intricate way. The explicit expression for $Z_n$ is presented in Appendix \ref{app:kernel}. With the PT expression of the density field in Eq.~(\ref{RSD_fluc}), the redshift-space power spectrum and bispectrum are expanded as
\begin{align}
P^{\rm (s)}(\bm{k})&= P^{\rm lin}+P^{\rm 1\mbox{-}loop}+\cdots,
\label{eq:pkred}
\\
 B^{\rm (s)}(\bm{k}_1,\bm{k}_2,\bm{k}_3)&= B^{\rm tree}+B_{222}^{\rm 1\mbox{-}loop}+
B_{\rm 321\mbox{-}I}^{\rm 1\mbox{-}loop}+B_{\rm 321\mbox{-}II}^{\rm 1\mbox{-}loop}
\nonumber
\\
&\quad+B_{411}^{\rm 1\mbox{-}loop}+\cdots.
\label{eq:bkred}
\end{align}
Each term at the right-hand-side is given by 
\begin{widetext}
\begin{align}
P^{\rm lin}(\bm{k})&=\{Z_1(\bm{k})\}^2P_{\rm L}(k),\label{Ptree}\\
P^{\rm 1\mbox{-}loop}(\bm{k})&=\int\frac{d^3p}{\left(2\pi\right)^3}\{Z_2\left(\bm{p},\bm{k}-\bm{p}\right)\}^2P_{\rm L}(p)P_{\rm L}(|\bm{k}-\bm{p}|) 
+ 2Z_1(\bm{k})P_{\rm L}(k)\int\frac{d^3p}{\left(2\pi\right)^3}\{Z_3\left(\bm{p},-\bm{p},\bm{k}\right)\}\,P_{\rm L}(p).
\label{P1loop}
\end{align}
for the power spectrum, and  
\begin{align}
B^{\rm tree}(\bm{k}_1,\bm{k}_2,\bm{k}_3)&=2Z_2(\bm{k}_1,\bm{k}_2)Z_1(\bm{k}_1)Z_1(\bm{k}_2)P_{\rm L}(k_1)P_{\rm L}(k_2)+2\,\,{\rm perms}\,\,(\bm{k}_1\leftrightarrow \bm{k}_2\leftrightarrow \bm{k}_3),
\label{tree}
\\
B^{\rm 1\mbox{-}loop}_{222}\left(\bm{k}_1,\bm{k}_2,\bm{k}_3\right)&=\int\frac{d^3p}{\left(2\pi\right)^3}Z_2\left(\bm{p},\bm{k}_1-\bm{p}\right)Z_2\left(-\bm{p},\bm{k}_2+\bm{p}\right)Z_2\left(-\bm{k}_1+\bm{p},-\bm{k}_2-\bm{p}\right)
P_{\rm L}\left(p\right)P_{\rm L}\left(|\bm{k}_1-\bm{p}|\right)P_{\rm L}\left(|\bm{k}_2+\bm{p}|\right)
\nonumber\\
&\quad 
+2\,\, {\rm perms}\,\, (\bm{k}_1\leftrightarrow \bm{k}_2\leftrightarrow \bm{k}_3),
\label{loop1-222}
\\
B^{\rm 1\mbox{-}loop}_{\rm 321\mbox{-}I}\left(\bm{k}_1,\bm{k}_2,\bm{k}_3\right)&= Z_1\left(\bm{k}_1\right)P_{\rm L}\left(k_1\right)\int\frac{d^3p}{\left(2\pi\right)^3}Z_2\left(\bm{p},\bm{k}_2-\bm{p}\right)Z_3\left(-\bm{k}_1,-\bm{p},-\bm{k}_2+\bm{p}\right)P_{\rm L}\left(p\right)P_{\rm L}\left(|\bm{k}_2-\bm{p}|\right)
\nonumber
\\
&\quad +5\,\,{\rm perms}\,\,(\bm{k}_1\leftrightarrow \bm{k}_2\leftrightarrow \bm{k}_3),
\label{loop1-321-I}\\
B^{\rm 1\mbox{-}loop}_{\rm 321\mbox{-}II}\left(\bm{k}_1,\bm{k}_2,\bm{k}_3\right)&= Z_1\left(\bm{k}_1\right)Z_2\left(\bm{k}_1,\bm{k}_2\right)P_{\rm L}\left(k_1\right)P_{\rm L}\left(k_2\right)\int\frac{d^3p}{\left(2\pi\right)^3}Z_3\left(\bm{k}_2,\bm{p},-\bm{p}\right)P_{\rm L}\left(p\right)
+5\,\,{\rm perms}\,\,(\bm{k}_1\leftrightarrow \bm{k}_2\leftrightarrow \bm{k}_3),
\label{loop1-321-II}
\\
B^{\rm 1\mbox{-}loop}_{411}\left(\bm{k}_1,\bm{k}_2,\bm{k}_3\right)&= Z_1\left(\bm{k}_1\right)Z_1\left(\bm{k}_2\right)P_{\rm L}\left(k_1\right)P_{\rm L}\left(k_2\right)\int\frac{d^3p}{\left(2\pi\right)^3}Z_4\left(-\bm{k}_1,-\bm{k}_2,\bm{p},-\bm{p}\right)P_{\rm L}\left(p\right)
+2\,\,{\rm perms}\,\,(\bm{k}_1\leftrightarrow \bm{k}_2\leftrightarrow \bm{k}_3).
\label{loop1-411}
\end{align}
\end{widetext}
for the bispectrum. In deriving the equations above, we assume the Gaussianity of linear density field $\delta_{\rm L}$,  and denote its power spectrum by $P_{\rm L}$. These expressions can be reduced to the real-space power spectrum and bispectrum of density field if we replace the kernels $Z_n$ with $F_n$. A notable point is that the redshift-space kernel implicitly depends on the line-of-sight direction. As a result, the statistical isotropy is broken in each term of power spectrum and bispectrum, and we need one and two more additional variables to characterize the redshift-space power spectrum and bispectrum, respectively. Another important point is that the kernels $Z_n$ at $n\geq2$ includes the mode-coupling contributions from the velocity fields, which basically come from the perturbative expansion of the exponential factor in Eq.~(\ref{fredels}). This implies that the Finger-of-God damping effect cannot be reproduced in a naive standard PT treatment, and we need to resum the infinite series of PT expansions. As it has been shown in Ref.~\cite{Taruya:2010mx}, the standard PT prediction in redshift space largely overestimates the power spectrum amplitude at one-loop order, and cannot accurately describe the BAO in redshift space. The applicable range of one-loop prediction thus becomes narrower than that of the real-space prediction. Since the expression of bispectrum also uses the redshift PT kernels and it even includes the higher-order,  the situation must be similar or rather worse than that in the power spectrum case. We will see in Sec.~\ref{sec:comparison} that without accounting the Finger-of-God damping, standard PT prediction of bispectrum starts to deviate from simulation even at large scales. A proper account of the damping effect is essential, and we will consider how to incorporates the damping effect into the PT calculation in next section.

\section{An improved model prescription for bispectrum at one-loop order}
\label{ourmodel}

In this section, we present the PT model of redshift-space bispectrum which keeps the non-perturbative damping effect. Our strategy is to decompose the contributions into non-perturbative part and the terms which can be evaluated with PT calculation, starting with the exact expression, Eq.~(\ref{bis_first}). For this purpose, we follow the treatment by Ref.~\cite{Taruya:2010mx}, and rewrite the exact expression in terms of cumulants. We then identify the non-perturbative part responsible for the FoG damping. Based on the simple proposition similarly made by Ref.~\cite{Taruya:2010mx}, the non-perturbative damping term is separated out from the rest of the contributions, for which we apply the PT calculation. We derive the expression valid at one-loop order.

Let us begin by rewriting Eq.~(\ref{bis_first}) in the form
\begin{align}
B^{\rm (s)}(\bm{k}_1,\bm{k}_2,\bm{k}_3)&=\int d\bm{r}_{13}d\bm{r}_{23}e^{i(\bm{k}_1\bm{r}_{13}+\bm{k}_2\bm{r}_{23})}\langle A_1A_2A_3e^{j_4A_4+j_5A_5}\rangle,
\label{bibi}
\end{align}
where the quantities $A_i$, $j_i$ are respectively defined by
\begin{align}
A_1&=\delta(\bm{r}_1)+f\nabla_z u_z(\bm{r}_1),\\
A_2&=\delta(\bm{r}_2)+f\nabla_z u_z(\bm{r}_2),\\
A_3&=\delta(\bm{r}_3)+f\nabla_z u_z(\bm{r}_3),\\
A_4&=u_z(\bm{r}_1)-u_z(\bm{r}_3),\\
A_5&=u_z(\bm{r}_2)-u_z(\bm{r}_3),\\
j_4&=-ik_1\mu_1 f,\\
j_5&=-ik_2\mu_2 f,
\end{align}
with $\mu_i=\bm{k}_i\cdot \hat{\bm{z}}/k_i$. To express the moment given above in terms of the cumulants, we use the relation between moment and cumulant generating functions (e.g., \cite{Taruya:2010mx, Scoccimarro:2004tg,Matsubara:2007wj}). For the stochastic vector field $\bm{A}$, we have 
\begin{align}
\langle e^{\bm{j}\cdot\bm{A}}\rangle=\exp{\left\{\langle e^{\bm{j}\cdot\bm{A}}\rangle_{\rm c}\right\}},
\label{cumulant}
\end{align}
with $j$ being an arbitrary constant vector, $\bm{j}$. To be specific, we assume that the vector fields given above are five components, i.e., $\bm{A}=\{A_1, A_2, A_3, A_4, A_5\}$ and $\bm{j}=\{j_1,j_2,j_3,j_4,j_5\}$. Then, taking the derivative three times with respect to $j_1$, $j_2$ and $j_3$, we set $j_1=j_2=j_3=0$. We obtain
\begin{widetext}
\begin{align}
\langle A_1A_2A_3e^{j_4A_4+j_5A_5}\rangle&=
\exp\left\{\langle e^{j_4A_4+j_5A_5}\rangle_{\rm c}\right\}\biggl[\langle A_1A_2A_3e^{j_4A_4+j_5A_5}\rangle_{\rm c}+\langle A_1A_2e^{j_4A_4+j_5A_5}\rangle_{\rm c}\langle A_3e^{j_4A_4+j_5A_5}\rangle_{\rm c}\nonumber\\
&+\langle A_2e^{j_4A_4+j_5A_5}\rangle_{\rm c}\langle A_1A_3e^{j_4A_4+j_5A_5}\rangle_{\rm c}+\langle A_1e^{j_4A_4+j_5A_5}\rangle_{\rm c}\langle A_2A_3e^{j_4A_4+j_5A_5}\rangle_{\rm c}\nonumber\\
&+\langle A_1e^{j_4A_4+j_5A_5}\rangle_{\rm c}\langle A_2e^{j_4A_4+j_5A_5}\rangle_{\rm c}\langle A_3e^{j_4A_4+j_5A_5}\rangle_{\rm c}\biggr].
\label{bis_sec}
\end{align}
\end{widetext}
This equation is indeed what we want to derive, and the left hand side is exactly the same one as in the integrand of Eq.~(\ref{bibi}).

The above expression shows that the pairwise velocity fields $A_4$ and $A_5$ in the exponent produce non-trivial correlations with density and velocity gradient fields. Further, these fields appear in the overall prefactor. This is indeed the same structure as seen in the expression of power spectrum \cite{Taruya:2010mx}. In power spectrum, the exponential prefactor is known to give a suppression of the amplitude due to the large-scale coherent and small-scale virialized motions, and at the large-scale of our interest, it mainly affects the broadband shape of the power spectrum. We expect that the overall prefactor in Eq.~(\ref{bis_sec}) similarly behaves like the one in the power spectrum, and it can alter the broadband shape of the bispectrum. Because of its functional form, the prefactor is likely to be affected by the small-scale nonlinearity even at large scales, and we may thus take it as non-perturbative part. On the other hand, the terms in the square bracket of Eq.~(\ref{bis_sec}) include the density fields and are responsible for reproducing the real-space bispectrum in the absence of redshift-space distortions [this is simply obtained by setting all the velocity fields in Eq.~(\ref{bis_sec}) to zero]. Thus, these terms basically carry the cosmological information, and imprints the acoustic feature of BAO. Although each term in the square bracket contains the exponential factor $e^{j_4A_4+j_5A_5}$, the contribution can be small as long as we consider the BAO scales, and the perturbative expansion may work well.

Based on the considerations,  we adopt the proposition by Ref.~\cite{Taruya:2010mx} to derive the PT model of redshift-space bispectrum valid at weakly nonlinear scales. That is, 

\begin{description}
\item[(i)]  The overall prefactor, $\exp\left\{\langle e^{j_4A_4+j_5A_5}\rangle_{\rm c}\right\}$, is kept as a non-perturbative contribution, and is replaced with general functional form $D_{\rm FoG}$,  which is assumed to be given as a function of $k_1\mu_1$, $k_2\mu_2$ and $k_3\mu_3$, ignoring the spatial correlation of $A_4$ and $A_5$. This means that the zero-lag correlation is the only dominant contribution. The relevant functional form of $D_{\rm FoG}$ will be discussed in Sec.~\ref{sec:comparison}. 

\item[(ii)]  The terms in the square bracket are treated perturbatively, and regarding the variables $j_4$ and $j_5$ as expansion parameters, we collect the contributions valid at one-loop order in standard PT. 
\end{description}

From the proposition (i), 
the overall exponential factor is factorized outside the integral over $\bm{r}_{13}$ and $\bm{r}_{23}$. We have
\begin{align}
&B^{\rm (s)}(\bm{k}_1,\bm{k}_2,\bm{k}_3)
\longrightarrow D_{\rm FoG}(k_1\mu_1,k_2\mu_2,k_3\mu_3)\,\int d\bm{r}_{13}d\bm{r}_{23}
\nonumber\\
&\quad \times\,e^{i(\bm{k}_1\bm{r}_{13}+\bm{k}_2\bm{r}_{23})}\,
\biggl[\langle A_1A_2A_3e^{j_4A_4+j_5A_5}\rangle_{\rm c}
\nonumber\\
&\qquad +\langle A_1A_2e^{j_4A_4+j_5A_5}\rangle_{\rm c}\langle A_3e^{j_4A_4+j_5A_5}\rangle_{\rm c}
\nonumber\\
&\qquad +\langle A_2e^{j_4A_4+j_5A_5}\rangle_{\rm c}\langle A_1A_3e^{j_4A_4+j_5A_5}\rangle_{\rm c}
\nonumber\\
&\qquad +\langle A_1e^{j_4A_4+j_5A_5}\rangle_{\rm c}\langle A_2A_3e^{j_4A_4+j_5A_5}\rangle_{\rm c}\nonumber\\
&\qquad +\langle A_1e^{j_4A_4+j_5A_5}\rangle_{\rm c}\langle A_2e^{j_4A_4+j_5A_5}\rangle_{\rm c}\langle A_3e^{j_4A_4+j_5A_5}\rangle_{\rm c}\biggr],
\label{bibi2}
\end{align}
We then expand the terms in the square bracket. 
Up to the third order in $j_n$, we obtain
\begin{align}
&B^{\rm (s)}(\bm{k}_1,\bm{k}_2,\bm{k}_3)
\longrightarrow 
D_{\rm FoG}(k_1\mu_1,k_2\mu_2,k_3\mu_3)\,
\sum_{n=1}^{11}C_n(\bm{k}_1,\bm{k}_2,\bm{k}_3),
\label{bis_naive}
\end{align} 
where the functions $C_n$ are defined by
\begin{align}
C_n(\bm{k}_1,\bm{k}_2,\bm{k}_3) \equiv
\int d\bm{r}_{13}d\bm{r}_{23}\,e^{i(\bm{k}_1\bm{r}_{13}+\bm{k}_2\bm{r}_{23})} \,\,S_n
\label{eq:def_C_n}
\end{align}
with the integrands $S_n$ given below:
\begin{align}
&S_1=\langle A_1A_2A_3\rangle_{\rm c},\quad
\label{eq:S_1}\\
&S_2=\langle A_1A_2\rangle_{\rm c}\langle(j_4A_4+j_5A_5)A_3\rangle_{\rm c}+{\rm cyc},
\label{eq:S_2}\\
&S_3=\langle(j_4A_4+j_5A_5)A_1A_2A_3\rangle_{\rm c},
\label{eq:S_3}\\
&S_4=\langle (j_4A_4+j_5A_5)A_1A_2\rangle_{\rm c}\langle(j_4A_4+j_5A_5)A_3\rangle_{\rm c}+{\rm cyc},
\label{eq:S_4}\\
&S_5=\langle A_1A_2\rangle_{\rm c}\langle(j_4A_4+j_5A_5)^2A_3\rangle_{\rm c}+{\rm cyc},
\label{eq:S_5}\\
&S_6=\langle(j_4A_4+j_5A_5)A_1\rangle\langle(j_4A_4+j_5A_5)A_2\rangle\langle(j_4A_4+j_5A_5)A_3\rangle,
\label{eq:S_6}\\
&S_7=\langle(j_4A_4+j_5A_5)^2A_1A_2A_3\rangle_{\rm c},
\label{eq:S_7}\\
&S_8=\langle(j_4A_4+j_5A_5)^3A_1A_2A_3\rangle_{\rm c},
\label{eq:S_8}\\
&S_{9}=\langle (j_4A_4+j_5A_5)^2A_1A_2\rangle_{\rm c}\langle(j_4A_4+j_5A_5)A_3\rangle_{\rm c}+{\rm cyc},
\label{eq:S_10}\\
&S_{10}=\langle (j_4A_4+j_5A_5)A_1A_2\rangle_{\rm c}\langle(j_4A_4+j_5A_5)^2A_3\rangle_{\rm c}+{\rm cyc},
\label{eq:S_11}\\
&S_{11}=\langle A_1A_2\rangle_{\rm c}\langle(j_4A_4+j_5A_5)^3A_3\rangle_{\rm c}+{\rm cyc}.
\label{eq:S_12}
\end{align}
In the above, the relevant terms for one-loop PT calculations, which are of the order of $\mathcal{O}(P_{\rm L}^3)$, appear at $n\leq6$, and the rest of the terms turns out to be higher-order. Hence, keeping the terms valid at the one-loop level, we model the redshift-space bispectrum as
\begin{align}
&B^{\rm (s)}_{\rm model}(\bm{k}_1,\bm{k}_2,\bm{k}_3)=
D_{\rm FoG}(k_1\mu_1,k_2\mu_2,k_3\mu_3)\,
\sum_{n=1}^{6}C_n(\bm{k}_1,\bm{k}_2,\bm{k}_3).
\label{bis_model}
\end{align} 

\begin{figure}[tbp] 
  \includegraphics[width=60mm,angle=270]{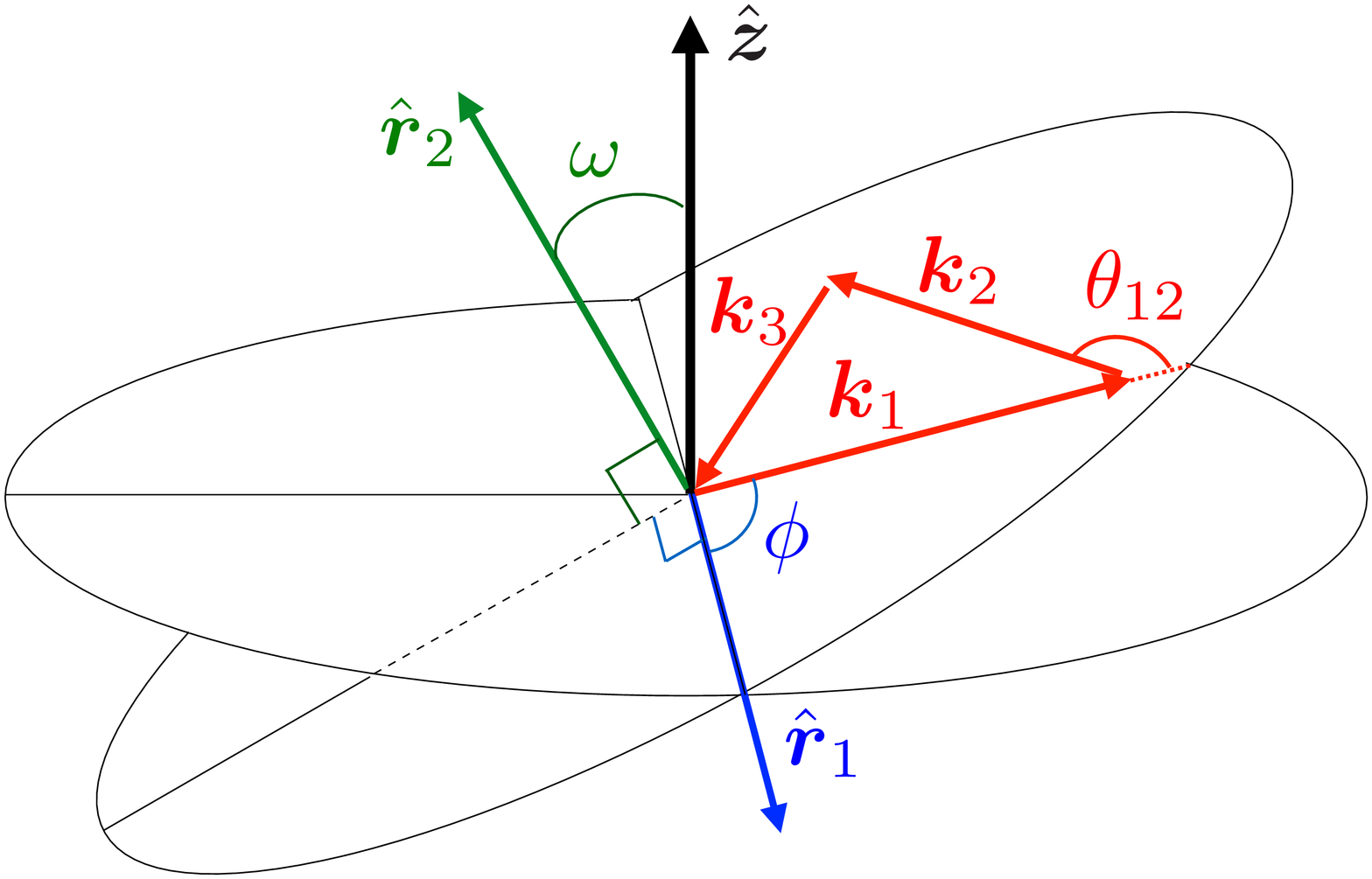}
\caption{Definition of the angles, $\omega$ and $\phi$, given in Eqs.~(\ref{eq:new_def_mu}) and (\ref{eq:new_def_phi}), which characterize the orientation of the triangle against the line-of-sight direction, $\hat{\bm{z}}$. Here, the unit vectors, $\hat{\bm{r}}_1$ and $\hat{\bm{r}}_2$, are expressed in terms of the quantities indicated in the figure by $\hat{\bm{r}}_1=\hat{\bm{z}}\times(\bm{k}_1\times\bm{k}_2)/(k_1k_2\sin{\theta_{12}}\sin{\omega})$ and $\hat{\bm{r}}_2=\bm{k}_1\times\bm{k}_2/(k_1k_2\sin{\theta_{12}})$.}
\label{fig:def_bispec_multipoles}
\end{figure}

\begin{figure*}[tb] 
  \includegraphics[width=8cm,angle=270]{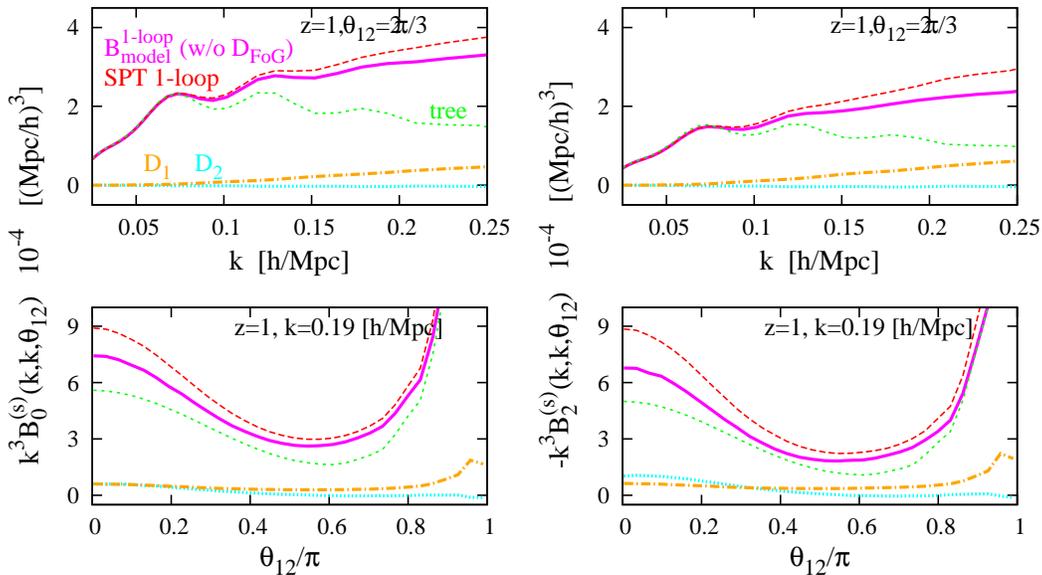}
\vspace*{-4pt}
\caption{Monopole (left) and quadrupole (right) moments of redshift-space bispectrum from PT. The results at $z=1$ are shown. While top panels show the scale-dependence of the equilateral configuration (i.e., plotted as function of $k\equiv k_1=k_2=k_3$ fixing $\theta_{12}$ to $2\pi/3$), bottom panels plot the shape dependence for isosceles configuration fixing the wave numbers to $k1=k2=k=0.19 h$\,Mpc$^{-1}$. In each panel, green dotted and red dashed lines are respectively the standard PT predictions at tree-level ($B^{\rm(s)}_{\rm SPT,tree}$) and one-loop order ($B^{\rm(s)}_{\rm SPT,1\mbox{-}loop}$). Magenta solid lines are the prediction based on Eq.~(\ref{bis_new}), $B^{\rm(s)}_{\rm model}$, for which the term $D_{\rm FoG}$ is set to $1$. This is essentially $B^{\rm(s)}_{\rm SPT,1\mbox{-}loop}$, but the terms, $D_1$ (orange) and $D_2$ (cyan), are subtracted. }
\label{equil_comp}
\end{figure*} 

The above model is compared with the standard PT results in Sec.~\ref{1-loop_sPT}, which can be also derived from Eqs.~(\ref{bibi}) and (\ref{bis_sec}) by a naive expansion of all the exponential factors, $e^{i(j_4A_4+j_5A_5)}$. Collecting the relevant contributions at one-loop order, we have
\begin{align}
&B^{\rm (s)}_{\rm SPT,1\mbox{-}loop}(\bm{k}_1,\bm{k}_2,\bm{k}_3)=
\sum_{n=1}^{6}C_n(\bm{k}_1,\bm{k}_2,\bm{k}_3) + \sum_{n=1}^2 D_n(\bm{k}_1,\bm{k}_2,\bm{k}_3).
\end{align}
with the functions $D_n$ given by 
\begin{align}
D_n(\bm{k}_1,\bm{k}_2,\bm{k}_3) &=\frac{1}{2}
\int d\bm{r}_{13}d\bm{r}_{23}\,e^{i(\bm{k}_1\bm{r}_{13}+\bm{k}_2\bm{r}_{23})} \,
\nonumber\\
&\qquad\qquad \times\,S_n\,\langle (j_4A_4+j_5A_5)^2\rangle_{\rm c},
\label{eq:expression_D_n}
\end{align}
where the function $S_n$ $(n=1,2)$ is defined by Eqs.~(\ref{eq:S_1}) and (\ref{eq:S_2}). These terms come from the expansion of the overall prefactor $\exp\left\{\langle e^{j_4A_4+j_5A_5}\rangle_{\rm c}\right\}$. Hence, at one-loop order, Eq.~(\ref{bis_model}) 
is recast as
\begin{align}
&B_{\rm model}^{\rm (s)}(\bm{k}_1,\bm{k}_2,\bm{k}_3)
= D_{\rm FoG}(k_1\mu_1,k_2\mu_2,k_3\mu_3)
\nonumber\\
&\qquad\times
\Bigl\{B^{\rm (s)}_{\rm SPT,1\mbox{-}loop}(\bm{k}_1,\bm{k}_2,\bm{k}_3)-\sum_{n=1}^2D_n(\bm{k}_1,\bm{k}_2,\bm{k}_3)\Bigr\}.
\label{bis_new}
\end{align}

In what follows, we use Eq.~(\ref{bis_new}) to compute the PT model of redshift-space bispectrum, and compare the predictions with $N$-body simulations. To be precise, we first compute $B^{\rm (s)}_{\rm SPT,1\mbox{-}loop}$ based on the standard PT calculations summarized in Sec.~\ref{1-loop_sPT} [Eqs.~(\ref{eq:bkred}) with (\ref{tree}), (\ref{loop1-222}), (\ref{loop1-321-I}), (\ref{loop1-321-II}), and (\ref{loop1-411})]. Then, we subtract $D_1$ and $D_2$ terms from the standard PT bispectrum and, we take into account Fingers-of-God effect. The explicit expressions for $D_1$ and $D_2$ relevant for the one-loop calculations are presented in Appendix \ref{A_pert}.

Before closing this section, we look at the significance of the difference between the standard PT bispectrum and the model given in Eq.~(\ref{bis_new}) or (\ref{bis_model}). In Fig.~\ref{equil_comp}, ignoring the $D_{\rm FoG}$ contribution, the monopole and quadrupole moments of the bispectrum are computed at $z=1$ for equilateral (top) and isosceles (bottom) configurations, and the results are plotted as function of $k$ and angle $\theta_{12}\equiv\cos^{-1}(\hat{\bm{k}}_1\cdot\hat{\bm{k}}_2)$, respectively. Here, the multipole moments of the bispectrum, $B^{\rm(s)}_\ell$, are defined by:
\begin{align}
B^{\rm (s)}_{\ell}(k_1,k_2,\theta_{12})
=\int_0^{2\pi}\frac{d\phi}{2\pi}\int_0^1 d\mu B^{\rm (s)}(\bm{k}_1,\bm{k}_2,\bm{k}_3)\mathcal{P}_{\ell}(\mu),
\label{eq:bispec_multipole}
\end{align}
where the function $\mathcal{P}_{\ell}(\mu)$ is the Legendre polynomials with $\mu$ being the directional cosine given by $\mu=\cos\omega$.
The angles $\omega$ and $\phi$ characterize the orientation of the triangles (i.e., $\bm{k}_1$, $\bm{k}_2$, and $\bm{k}_3$) with respect to the line-of-sight direction (see Fig.~\ref{fig:def_bispec_multipoles}). Precise definition of the angles $\omega$ and $\phi$, together with the properties of multipole expansion, is described in Appendix \ref{app:multipole}. Note that our definition differs from the one frequently used in the literature (e.g., Refs.~\cite{Scoccimarro:1999ed,2015PhRvD..92h3532S,Yamamoto:2016anp,2017MNRAS.tmp..138G}), but a nice property is that the bispectrum multipoles become fully symmetric under the permutation of the order of $k_1$, $k_2$ and $k_3$.

Fig.~\ref{equil_comp} shows that both the one-loop bispectra, $B^{\rm(s)}_{\rm SPT,1\mbox{-}loop}$ (red) and $B_{\rm model}^{\rm(s)}$ (magenta), have a larger amplitude than the tree-level prediction. While these predictions lead to similar scale and shape dependencies, differences appear manifest at smaller scales due to the the $D_1$ and $D_2$ terms. Though these two terms are basically small and weakly depend on scales, we will see below that subtracting these from $B^{\rm(s)}_{\rm SPT,1\mbox{-}loop}$ indeed plays an important role to get a consistent damping behavior of $D_{\rm FoG}$ from both power spectrum and bispectrum (see Sec.~\ref{subsec:validity_D_FoG}).

\begin{table}[tb]
\begin{ruledtabular}
  \begin{tabular}{l|c|c}
     &HR&LR\\ \hline 
    Set  &B  & C\\ 
    Number of realizations&$96$&$512$\\
    Box size [$h^{-1}$${\rm Mpc}$]&$656.25$&$1312.5$ \\ 
    Number of particles &$1024^3$&$512^3$\\
    Particle mass [$h^{-1}$${\rm M_{\odot}}$]&$1.88 \times 10^{10}$&$1.20 \times 10^{12}$\\
    Initial redshift ($z_{\rm ini}$)&$190$&$106$\\
    Output redshifts&$z=1,~0$&$z=1,~0.5,~0$\\
    \end{tabular}
  \caption{Parameters of $N$-body simulation data sets used in the paper.}
  \label{table:simucharacteristics}
  \end{ruledtabular}
\end{table}

\section{Comparison with N-body simulations}
\label{sec:comparison}

In this section, we present a detailed comparison of the redshift-space bispectrum between PT predictions and $N$-body simulations. After briefly describing the $N$-body data set used in the analysis in Sec.~\ref{exp_Nbody}, we first compare the real-space results in Sec~\ref{subsec:real_space} to see the applicable range of PT as well as the quality of $N$-body data. We then move to the redshift space, and compare the monopole and quadrupole bispectra from $N$-body simulations with those obtained by PT calculations in Sec.~\ref{subsec:redshift-space}. Sec.~\ref{subsec:validity_D_FoG} discusses the validity and consistency of the ansatz for the damping function $D_{\rm FoG}$ in our PT model of bispectrum. 

\subsection{N-body simulations and measurement of the bispectrum}
\label{exp_Nbody}

\vspace*{0.5cm}

We use the simulation ensembles Set B and Set C from Dark Energy Universe Simulation - Parallel Universe Runs (DEUS-PUR) introduced by \cite{Blot:2015}
\footnote{Set C was not directly presented in \cite{Blot:2015} but it was performed at the same time for the same project}. 
Each simulation started from a given realization of the initial matter density field. The initial conditions were generated with an improved version of MPGRAFIC \cite{Prunet:2008} while the particles evolution were computed with an optimized version of the RAMSES N-body code \cite{Teyssier:2002}. The main characteristics of the two ensembles of simulations are summarized in Table~\ref{table:simucharacteristics}. Set B consists in 96 simulations with $1024^3$ particles in a cosmological volume of ($656.25$~$h^{-1}$Mpc)$^3$. The total effective volume is 27~($h^{-1}$Gpc)$^3$ and the mass resolution is $1.88\times10^{10}$~$h^{-1}$M$_{\odot}$. We call this set HR (``High Resolution''). Set C consists in 512 simulations with $512^3$ particles in a cosmological volume of ($1312.5$~$h^{-1}$Mpc)$^3$. The total effective volume of 1158~($h^{-1}$Gpc)$^3$ is larger but the mass resolution of $1.20\times10^{12}$~$h^{-1}$M$_{\odot}$ is more limited. We call this set LR (``Low Resolution''). The two sets are complementary because they are affected at different level by numerical effects such as sample variance, finite volume and mass resolution effects (see Refs~\cite{Blot:2015,Rasera:2014} for a study of these effects). 

The matter bispectrum is estimated using the BISP\_MES code \cite{Scoccimarro:1997st} kindly provided by S.Colombi.  The code has been updated to take into account redshift space distortions (RSD) and projections onto multipoles. We provide here a short summary of the numerical methods, for more details see \cite{Scoccimarro:1997st}. Particles position and velocity from a given snapshot are provided as an input of the code. Particles position are then displaced along the $\hat{\bm{z}}$-direction of the box using Eq.~(\ref{eq:red_real_spaces}). When a particle falls outside of the simulation box, periodic boundary conditions are assumed to ensure conservation of the total number of particles. The density in Fourier-space is then computed using Cloud-In-Cell (CIC) mass assignment followed by a Fast Fourier Transform (FFT). The density field is further deconvolved with the CIC window function. We call the resulting density field from a given snapshot of the j-th simulation of a set, $\delta_{\rm sim,j}$. For triangles with sides $k_1$, $k_2$ and $k_3$, the multipole moments of the bispectrum are computed by a projection onto Legendre polynomials followed by an averaging over an homogeneous sample of modes within a bin of size $\Delta k_1$, $\Delta k_2$ and $\Delta k_3$ centered on $k_1$, $k_2$ and $k_3$. The averaging procedure is performed by randomly picking $N_{\rm mode}$ possible orientations and sizes of triangles within this interval (Monte-Carlo method). The estimated multipole projection of the bispectrum for a given snapshot of the j-th simulation is then given by
\begin{align}
&B^{(s)}_{\ell,{\rm sim,j}}(k_1,k_2,k_3)= \frac{1}{N_{\rm mode}}
\nonumber\\
&\qquad\times
 \sum_{i=1}^{N_{\rm mode}} \delta_{\rm sim,j}^{(s)}(\bm{k}_1^i) \delta_{\rm sim,j}^{(s)}(\bm{k}_2^i) \delta_{\rm sim,j}^{(s)}(\bm{k}_3^i) \mathcal{P}_{\ell}(\mu^i),
\label{eq:bispsim}
\end{align}
where the subscript $i$ indicates the rank of the orientation/size of the triangle.

The number of orientations/sizes $N_{\rm mode}$ used for the averaging procedure is set to $10^7$. With such a large value, the results are insensitive to the exact choice of $N_{\rm mode}$. The size of the bin is chosen to be equal to the fundamental frequency of the box. We  also use an FFT grid with $512^3$ elements. The corresponding Nyquist frequency is 2.45~$h$~Mpc$^{-1}$ (1.23~$h$~Mpc$^{-1}$) for the HR (LR) simulation. As a consequence, we do not apply any shot noise corrections nor any aliasing corrections since such effects are negligible at the scale of interest in this paper ($k<0.3$~$h$~Mpc$^{-1}$). However, at the very large scales of interest for comparison to perturbation theory ($k=0.01-0.1$~$h$~Mpc$^{-1}$), HR simulations might suffer from non-negligible finite-mode sampling \cite{Takahashi:2008wn}. In the linear regime, the bispectrum of the density field should be zero for perfect ensemble average. Because the total number of independent modes in the simulation is finite, the resulting bispectrum is non-zero. To mitigate this effect at large scales, we compute the initial bispectrum of each snapshot $B^{(s)}_{\ell,{\rm ini,j}}$. The linearly evolved bispectrum is then subtracted from the snapshot bispectrum to obtain a corrected bispectrum $B^{(s)}_{\ell,{\rm corr,j}}$. The linearly evolved bispectrum is computed using Eq.~(\ref{eq:bispsim}) but instead of using the snapshot density $\delta_{\rm sim,j}^{(s)}(\bm{k}^i)$, we use the linearly evolved density field
\begin{align}
\delta_{\rm linevol,j}^{(s)}(\bm{k}^i,z)= \frac{1+f(z)\mu^2}{1+f(z_{\rm ini})\mu^2}\frac{D(z)}{D(z_{\rm ini})} \times \delta_{\rm sim,j}^{(s)}(\bm{k}^i,z_{\rm ini}),
\label{eq:lineardensity}
\end{align}
where $f(z)$ is the linear growth rate and $D(z)$ is the linear growth factor. This correction plays a role at the percent level at small k ($k<0.1$~$h$~Mpc$^{-1}$) for the HR run.

 For each snapshot we consider isosceles triangles ($k_1=k_2$) and scalene triangles ($k_1=2 k_2$). For each type of triangle, we explore the scale dependence by fixing $\theta_{12}$ and varying $k_1$ and the shape dependence by fixing $k_1$ and varying $\theta_{12}$. Once the bispectrum is computed for all simulations of a given set, we perform an ensemble average of $B^{(s)}_{\ell,{\rm corr,j}}(k_1,k_2,k_3)$ to get the mean bispectrum of the set 
\begin{align}
B^{(s)}_{\ell,{\rm sim}}(k_1,k_2,k_3)=\frac{1}{N_{\rm sim}} \sum_{j=1}^{N_{\rm sim}} B^{(s)}_{\ell,{\rm corr,j}}(k_1,k_2,k_3), 
\end{align}
where $N_{\rm sim}$ is the number of simulations of the set. This bispectrum estimate is the one used in the rest of the paper. We also compute the standard deviation of the bispectrum and we estimate statistical error bars assuming 
\begin{align}
\Delta B^{(s)}_{\ell,{\rm stat}}=\frac{\sqrt{\frac{1}{N_{\rm sim}} \sum_{j=1}^{N_{\rm sim}} B^{(s)}_{\ell,{\rm corr,j}}(k_1,k_2,k_3)^2  -B^{(s)}_{\ell,{\rm sim}}(k_1,k_2,k_3)^2} }{\sqrt{N_{\rm sim}}}. 
\end{align}
The same analysis is repeated in comoving space (i.e. no redshift space distortions) by setting the velocity field and linear growth rate to zero.

\vspace*{0.5cm}

\begin{figure*}[htbp] 
  \includegraphics[width=80mm,angle=270]{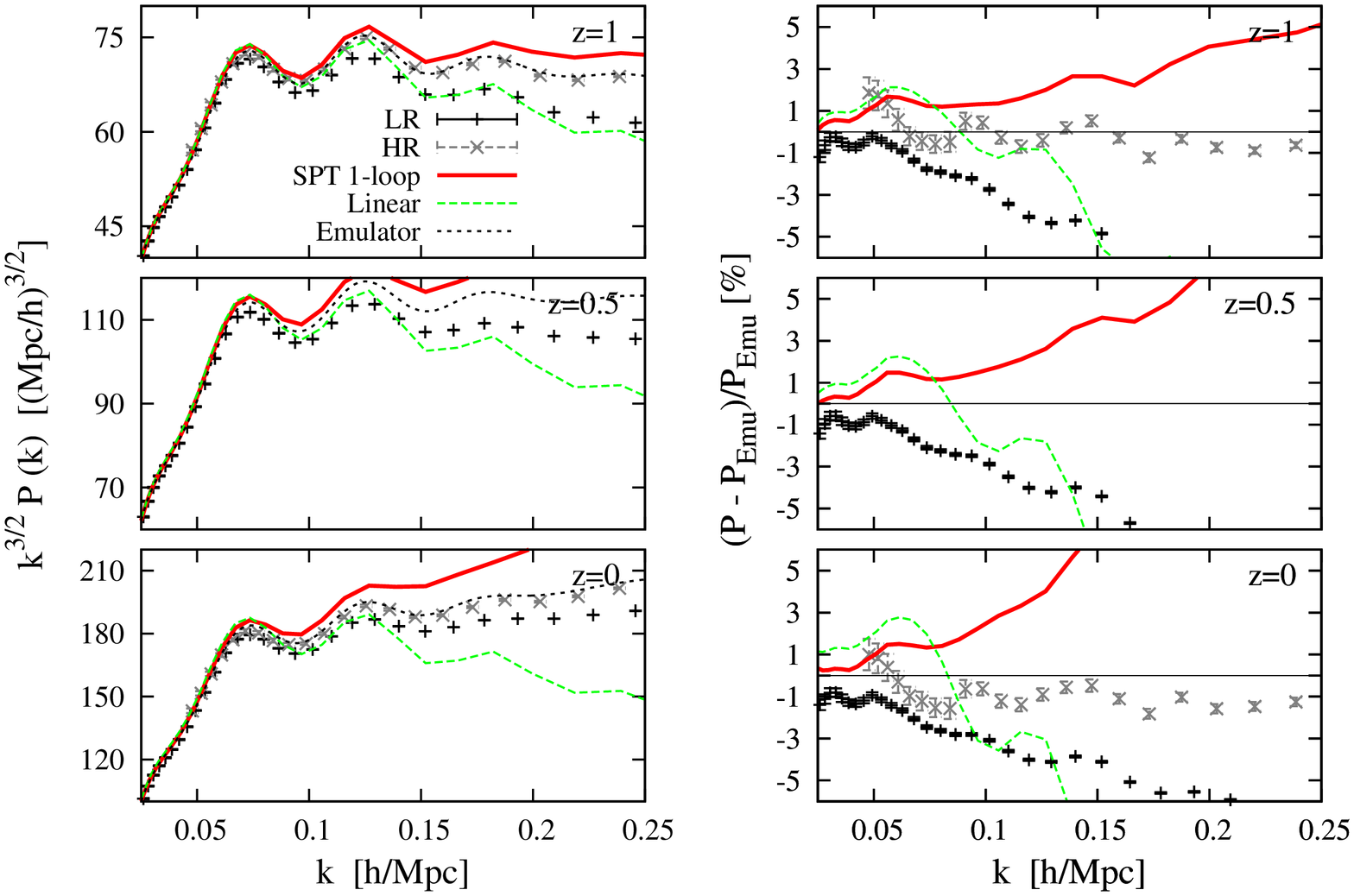}
\vspace*{-2pt}
\caption{Real-space power spectrum as function of wave number $k$ at $z=1$ (top), $0.5$ (middle) and $1$ (bottom). The left panel plots the power spectrum multiplied by $k^{3/2}$, while the right panel shows the fractional difference between the emulator and other predictions, $(P-P_{\rm Emu})/P_{\rm Emu}$. In each panel, black and gray points with errorbars indicate the results of LR and HR simulations, respectively. The green dashed and red solid lines are the linear theory and the standard PT one-loop predictions. The black dashed lines in the left panels are the prediction based on the cosmic emulator \cite{Heitmann:2008eq,Heitmann:2009cu,2010ApJ...713.1322L,Heitmann:2013bra}. }
\label{power_real}
\hspace*{-0.5cm}
 \includegraphics[width=6.4cm,angle=270]{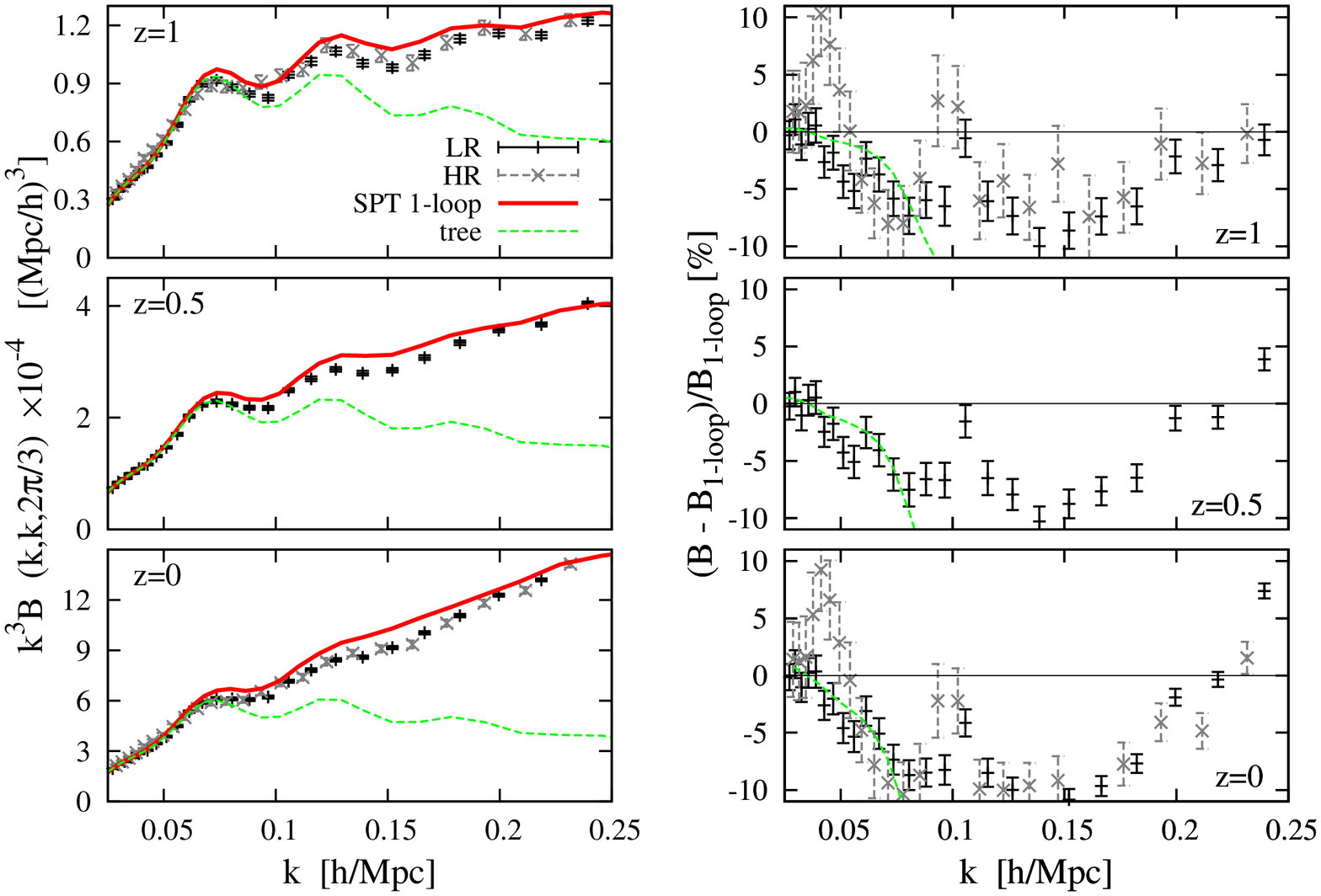}
\hspace*{0.cm}
  \includegraphics[width=6.4cm,angle=270]{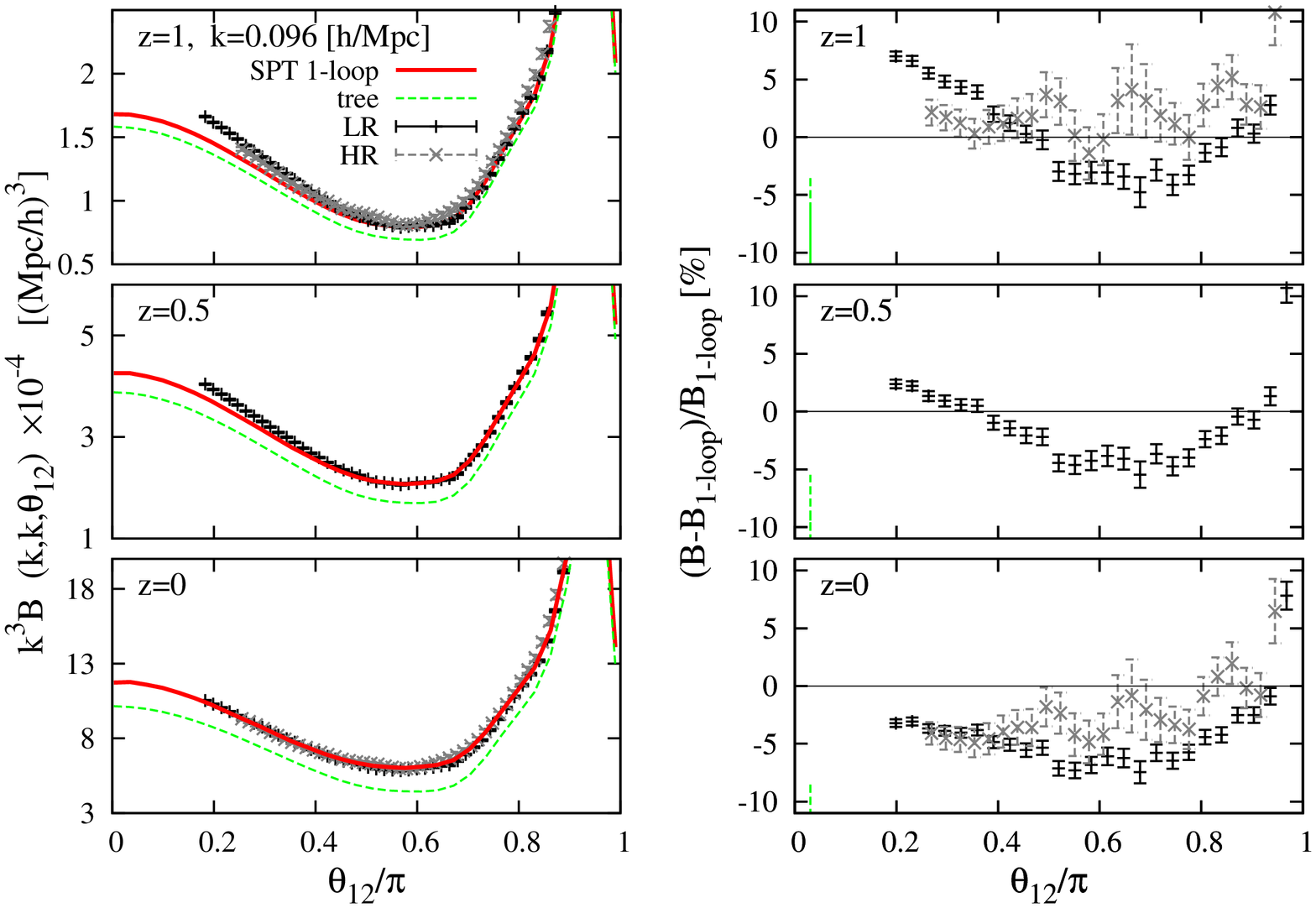}
\caption{Real-space bispectrum in real space at $z=1$ (top), $0.5$ (middle), and $0$ (bottom). The first two panels from the left show the results for equilateral triangular shape (i.e., $k_1=k_2=k_3=k$), plotted as function of wave number $k$. The leftmost panel is the bispectrum amplitude multiplied by $k^3$, while the second panel plots the fractional difference between standard PT one-loop predictions and others, i.e., $(B-B_{\rm 1\mbox{-}loop})/B_{\rm 1\mbox{-}loop}$. The right two panels also present the results similarly to the left two panels, but the cases with isosceles triangular shape ($k_1=k_2$) are plotted as function of the angle $\theta_{12}\equiv\cos^{-1}(\hat{\bm{k}}_1\cdot\hat{\bm{k}}_2)$, fixing the scale $k_1=k_2$ to $0.096\,h$\,Mpc$^{-1}$. In each panel,  meanings of the symbols and line types are the same as in Fig.~\ref{power_real}, except for the green dashed lines, which indicate the standard PT prediction at tree-level order.}
\label{realB_kdep_tdep}
\end{figure*}

\subsection{Results in real space}
\label{subsec:real_space}

Let us first look at the results in real space and check the applicable range of PT as well as the quality of $N$-body data. 

Fig.~\ref{power_real} shows the power spectra at $z=1$ (top), $0.5$ (middle) and $0$ (bottom). Left panel plots the power spectra multiplied by $k^{3/2}$, while the right panel summarizes their fractional difference, for which we take the predictions by the Emulator code \cite{Heitmann:2008eq,Heitmann:2009cu,2010ApJ...713.1322L,Heitmann:2013bra} as the base value, and evaluate $[P(k)-P_{\rm Emu}(k)]/P_{\rm Emu}(k)$ with $P_{\rm Emu}$ being the Emulator power spectrum. Note here that the error bars in the $N$-body results indicate the standard error of the averaged power spectrum over the number of realizations. The claimed error bars of the Emulator are $1\%$.

The PT predictions at one-loop order (red) reasonably agree with those of the emulator code, and the agreement is at the 3 percent level for the scales of $k\lesssim0.18$, $0.12$ and $0.1\,h$\,Mpc$^{-1}$ at $z=1$, $0.5$ and $0$, respectively. This is consistent with what has been found in the literature. The high-resolution $N$-body data (HR) also shows a reasonable agreement with the emulator (1 percent level over all the studied range of wavenumber) and one-loop PT predictions, but the low-resolution data (LR) systematically deviates from others at small scales. The deviation gradually increases from about 0.5 percent near $k=0.05\,h$\,Mpc$^{-1}$) to 5 percent near $k=0.2\,h$\,Mpc$^{-1}$. This is a well known mass-resolution effect which tends to decrease the power at small scale \cite{Heitmann:2008eq,Rasera:2014}. Indeed, in the case of high-resolution setup, we found that the simulations starting at a lower initial redshift gives a better agreement with PT predictions.

On the other hand, turning to look at the real-space bispectrum, we do not clearly see the systematic difference between HR and LR simulations. Fig.~\ref{realB_kdep_tdep} shows the results for the equilateral triangular shapes plotted as function of $k$ (left two panels), and those for the isosceles triangles plotted as function of $\theta_{12} \equiv\cos^{-1}(\hat{\bm{k}}_1\cdot\hat{\bm{k}}_2)$ (right two panels). Note that in panels showing the amplitude of bispectrum,  the results are all multiplied by $k^3$.  Also, in plotting the fractional difference (second left and rightmost panels), we take the one-loop PT predictions as the base model, and evaluate the ratio, $(B-B_{\rm 1\mbox{-}loop})/B_{\rm 1\mbox{-}loop}$.

Compared to the power spectrum case, the statistical errors in $N$-body simulations is larger, and the difference between the two data set can be seen only at large scales $k\lesssim0.05\,h$\,Mpc$^{-1}$. Rather, there seems to be a systematic difference between simulations and PT prediction, and because of this, the agreement between simulations and PT prediction look somewhat worse, and is at $5\%$ level even at the scales where the reasonable agreement at a few percent level can be seen in the power spectrum. Perhaps, these results might be partly ascribed to the setup of initial conditions (early starting redshift) or to the overestimate of non-linear effects on the bispectrum by standard PT, but without any reference,  it is difficult to further clarify the origin of systematics. We will leave it to future investigation. Nevertheless, one important point is that the one-loop PT predictions can capture the major trend in the $N$-body simulations at small scales; scale-dependent enhancement of the bispectrum amplitude at $k\gtrsim0.05\,h$\,Mpc$^{-1}$. Because of this, the agreement still remains at the $5-10$\% level even at small scales at all redshifts.

Keeping the systematics and a level of agreement in the real space in mind, we will move to the redshift space, and continue the comparison in next subsection.

\begin{figure*}[tbp] 
  \includegraphics[width=110mm,angle=270]{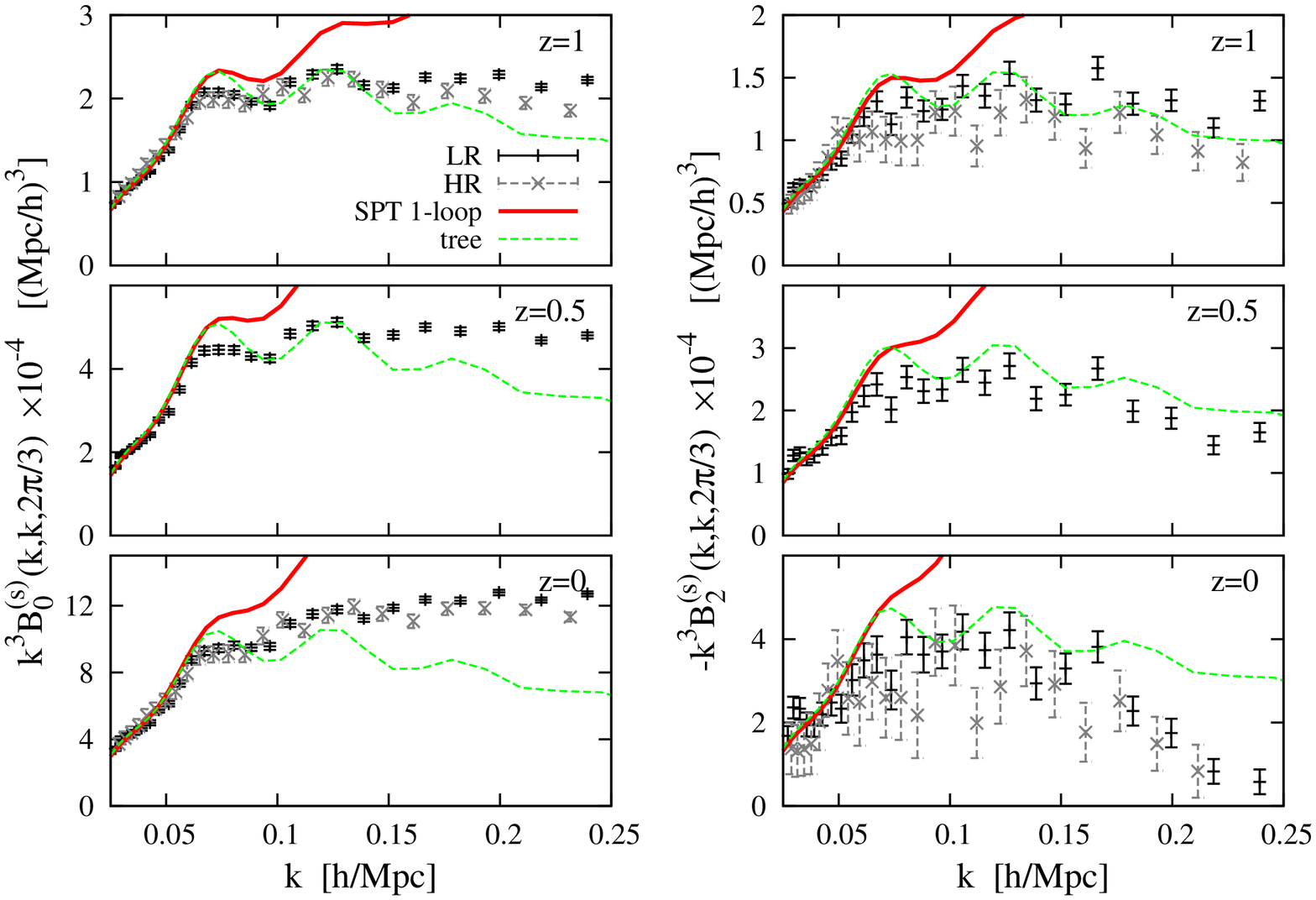}
\vspace{-1pt}
\caption{Monopole (left) and quadrupole (right) moments of the redshift-space bispectrum for the equilateral triangles, plotted as function of wave number $k$ at $z=1$ (top), $0.5$ (middle), and $0$ (bottom).  The standard PT predictions at tree-level (green dashed) and one-loop (red solid) order are compared with the measured bispectrum from LR (black) and HR (gray) data of $N$-body simulations. The results for equilateral shape triangles are particularly shown. }
\label{red_bis_kdep_SPT}
\vspace*{-1pt}
  \includegraphics[width=110mm,angle=270]{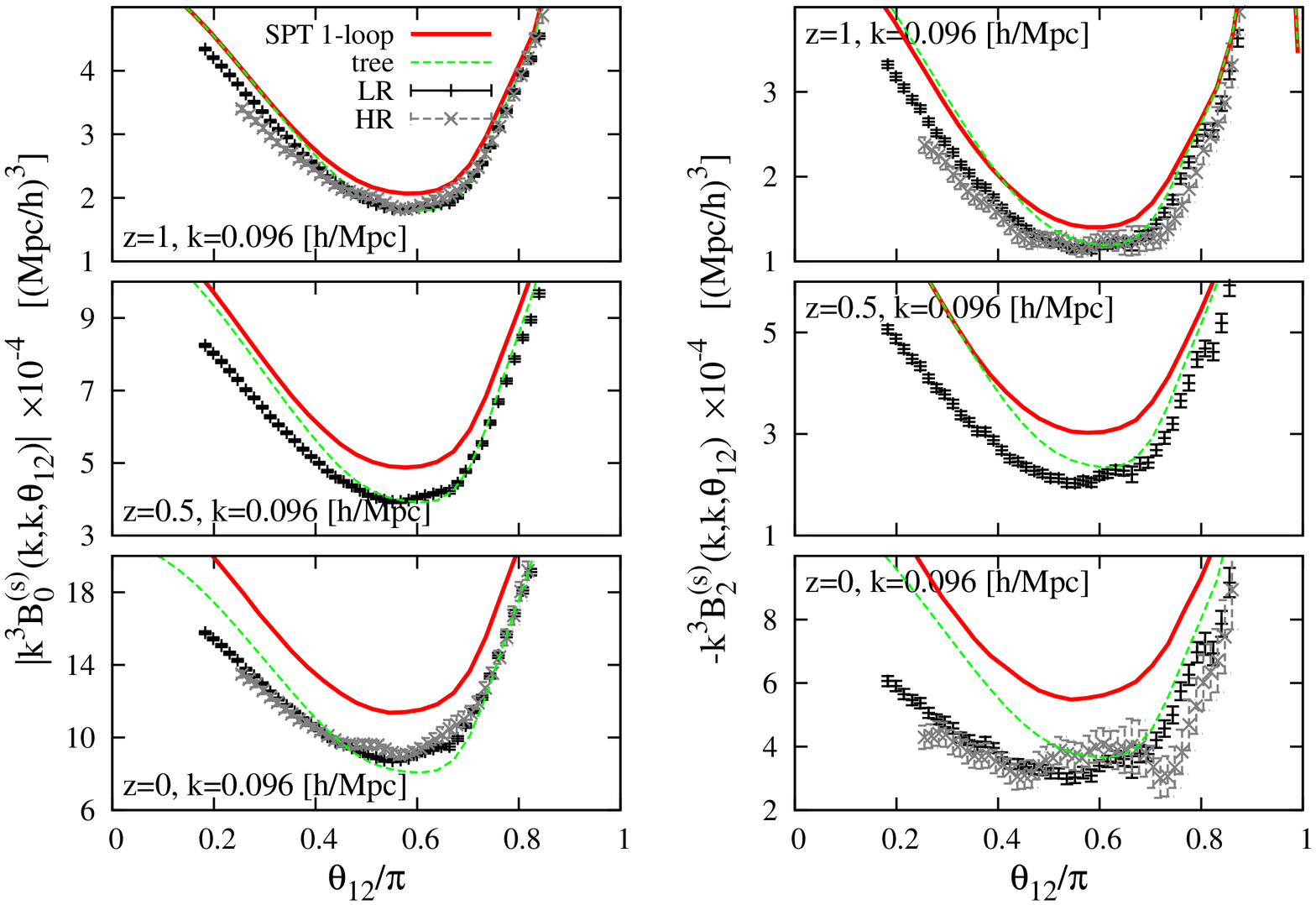}
\vspace{-1pt}
\caption{Same as Fig.~\ref{red_bis_kdep_SPT}, but the shape-dependence of the bispectrum is plotted as function of $\theta_{12}=\cos^{-1}(\hat{\bm k}_1\cdot\hat{\bm k}_2)$, fixing both $k_1$ and $k_2$ to $0.096\,h$\,Mpc$^{-1}$. The meanings of the line types are the same as in Fig.~\ref{red_bis_kdep_SPT}. }
\label{red_bis_tdep_SPT}
\end{figure*}

\subsection{Results in redshift space}
\label{subsec:redshift-space}

To see the impact of FoG damping on the bispectrum, we first compare the standard PT predictions with results of $N$-body simulations. 

Figs.~\ref{red_bis_kdep_SPT} and \ref{red_bis_tdep_SPT} respectively show the scale- and shape-dependencies of the bispectrum for equilateral and isosceles configurations. Based on the definition of multipole expansion in Eq.~(\ref{eq:bispec_multipole}) and Appendix \ref{app:multipole}, we compute and measure the monopole and quadrupole moments of the bispectrum, which are respectively plotted in left and right panels. Overall, the $N$-body results both from the LR and HR data agree well each other. A closer look at $k\gtrsim0.1\,h$\,Mpc$^{-1}$ reveals that the amplitude of the LR data is systematically larger than that of the HR data, but within the errorbars there is no significant discrepancy in both monopole and quadrupole moments. Rather, a discrepancy between the $N$-body results and PT predictions is manifest, and compared to the results in real space, the range of agreement is fairly narrower for standard PT one-loop, and is restricted to a low-$k$ region. This is rather manifest at higher redshifts, and the prediction generally overestimates the simulations. On the other hand, shifting the overall amplitude, the shape dependence predicted by standard PT one-loop seems to reasonably match the measured results of $N$-body simulations (see Fig.~\ref{red_bis_tdep_SPT}). The results clearly manifest that a naive standard PT fails to describe the damping behaviors seen in the $N$-body simulations, and an appropriate prescription for the damping effect is important for prediction even at large scales.

Let us then see how the PT model presented in Sec.~\ref{ourmodel} works well. In doing so, an appropriate functional form of $D_{\rm FoG}$ needs to be first specified. The function $D_{\rm FoG}$ is generally expressed as function of $k_1\mu_1$, $k_2\mu_2$, and $k_3\mu_3$, and it contains
the non-perturbative damping behavior arising from the exponential factor, $\exp\{\langle e^{j_4A_4+j_5A_5} \rangle_c\}$ [see Eq.~(\ref{bis_sec}) and proposition (i) in Sec.~\ref{ourmodel}]. Here, we adopt the following Gaussian form as a simple and relevant ansatz, similar to what has been frequently used in the power spectrum cases \cite{Taruya:2010mx}:
\begin{align}
  D_{\rm FoG}= \exp\left[-\frac{1}{2}(f\,\sigmav)^2\,
    \Bigl\{(k_1\mu_1)^2+(k_2\mu_2)^2+(k_3\mu_3)^2\Bigr\}\,\right].
  \label{eq:dfog}
\end{align}
The function $f$ is the linear growth rate, and $\sigma_v$ is the constant parameter corresponding to the one-dimensional velocity dispersion, which controls the strength of the FoG damping. Note that similar functional form is obtained by expanding the exponential pre-factor and truncating it at the second-order in cumulants, just ignoring the spatial correlation. A non-trivial point may be whether $D_{\rm FoG}$ is still expressed as univariate function of $(k_1\mu_1)^2+(k_2\mu_2)^2+(k_3\mu_3)^2$ or not beyond the scales relevant for tree-level predictions. We will discuss and check it in Sec.~\ref{subsec:validity_D_FoG}.

\begin{figure*}[tbp]
  \begin{center}
        \begin{center}
          \includegraphics[width=105mm,angle=270]{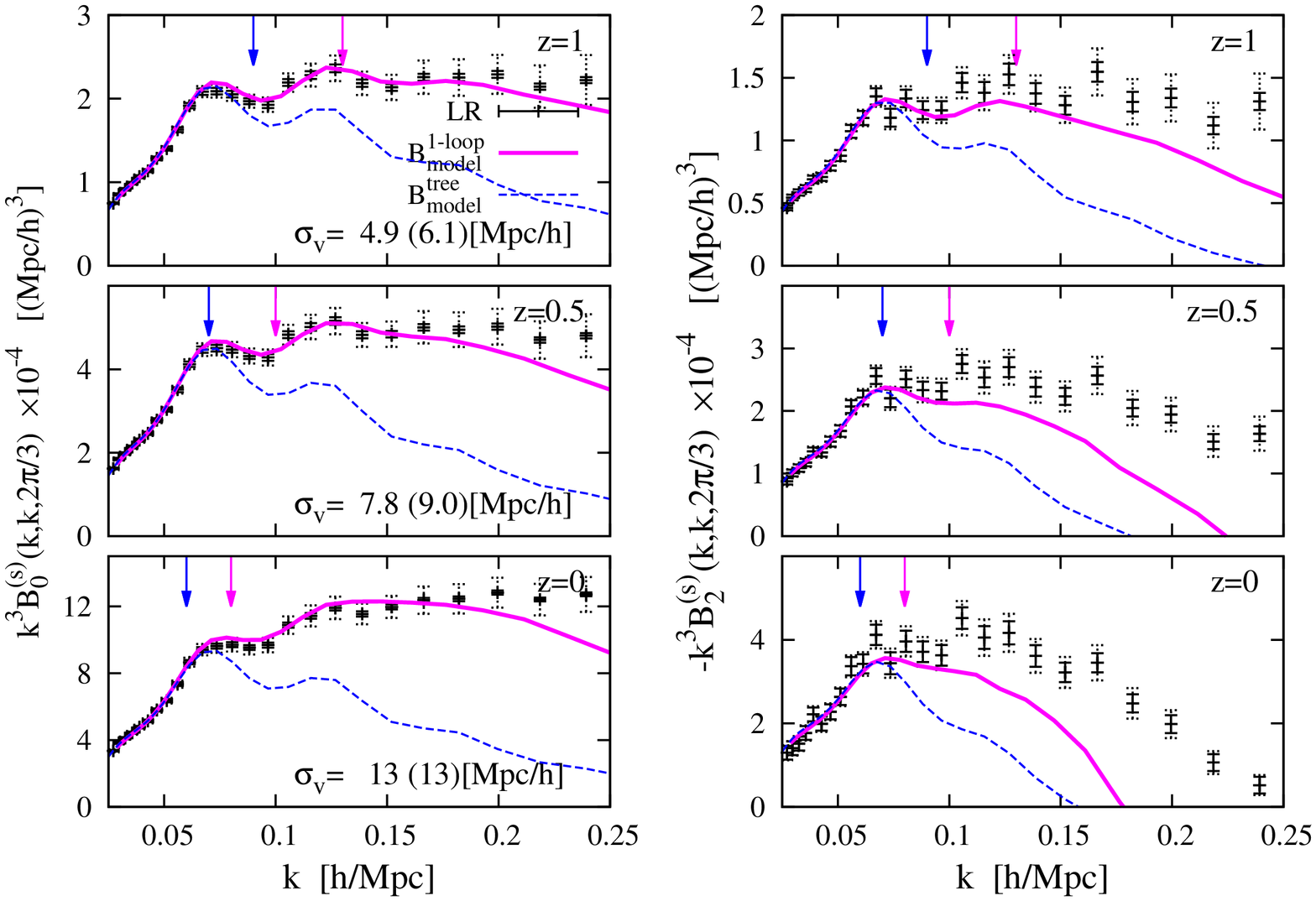}
        \end{center}
        \vspace*{-1pt}
     \caption{Monopole (left) and quadrupole (right) moments of the redshift-space bispectrum for the equilateral triangles, plotted as function of wave number $k$ at $z=1$ (top), $0.5$ (middle), and $0$ (bottom). Here, the PT models involving the damping function $D_{\rm FoG}$ are compared with the measured results obtained from the LR data of $N$-body simulations. The predictions of tree- and one-loop PT model are depicted as blue dashed and magenta solid lines, respectively. In plotting the predictions, the free parameter $\sigmav$ in the damping function is determined by fitting the monopole and quadrupole moments to the $N$-body data at $0.05\,h$\,Mpc$^{-1}$$\leq k\leq k_{\rm max}$, with $k_{\rm max}$ indicated by the vertical arrows in each panel. Note that the error bars depicted as solid and dotted lines respectively represent the statistical error averaged over the number of realizations, and the one including both the statistical and systematic errors (see Sec.~\ref{subsec:validity_D_FoG}). 
}
    \label{equil_kdep}
  \end{center}
 \vspace*{-15pt}
  \begin{center}
          \includegraphics[width=105mm,angle=270]{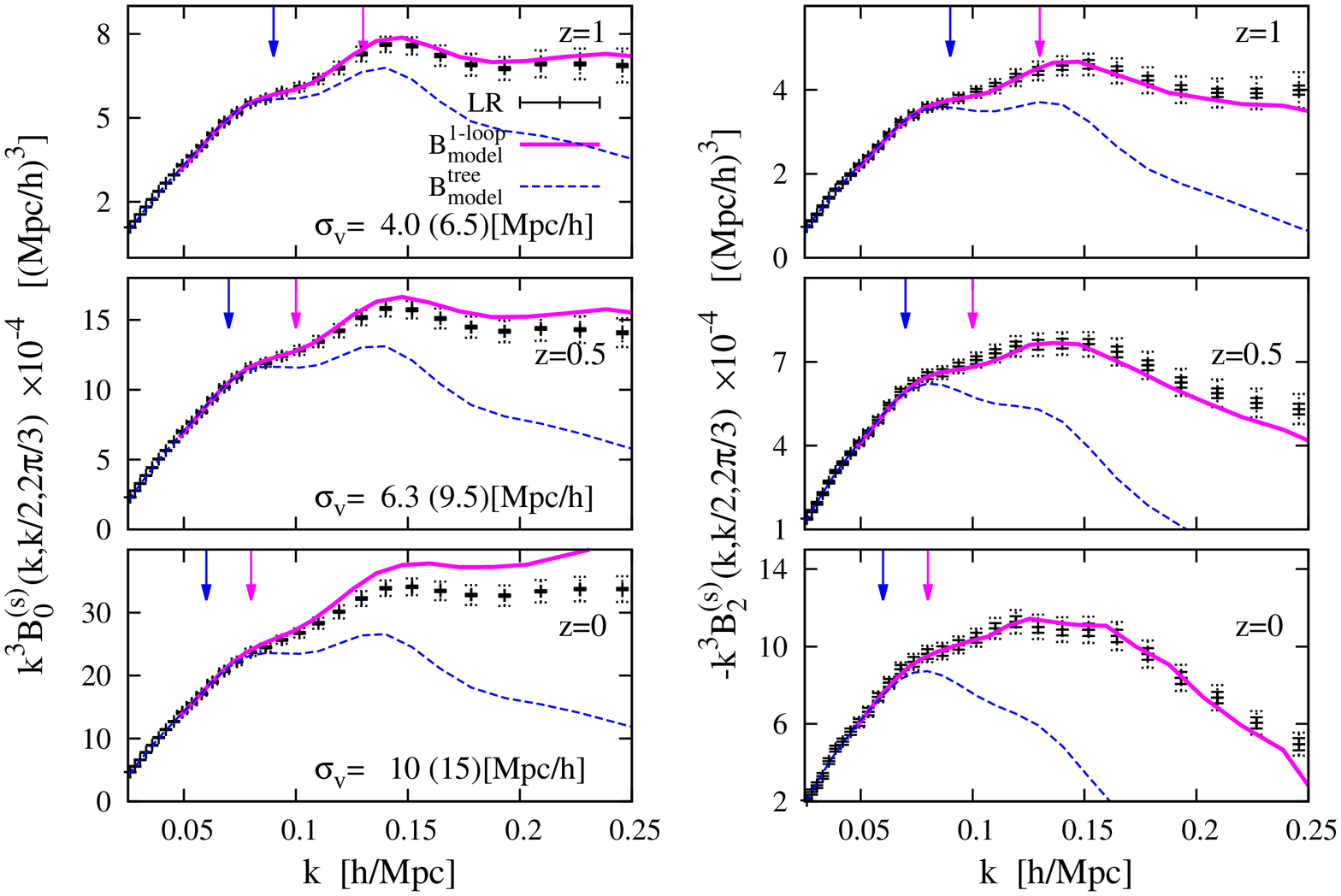}
          \end{center}
 \vspace*{-10pt}
      \caption{Same as Fig.\ref{equil_kdep}, but the results for scalene triangle of $k_1=2k_2=(2/\sqrt{3})k_3=k$ are shown. }
    \label{iso_kdep}
  \end{figure*}

\begin{figure*}[tbp]
  \begin{center}
          \includegraphics[width=105mm,angle=270]{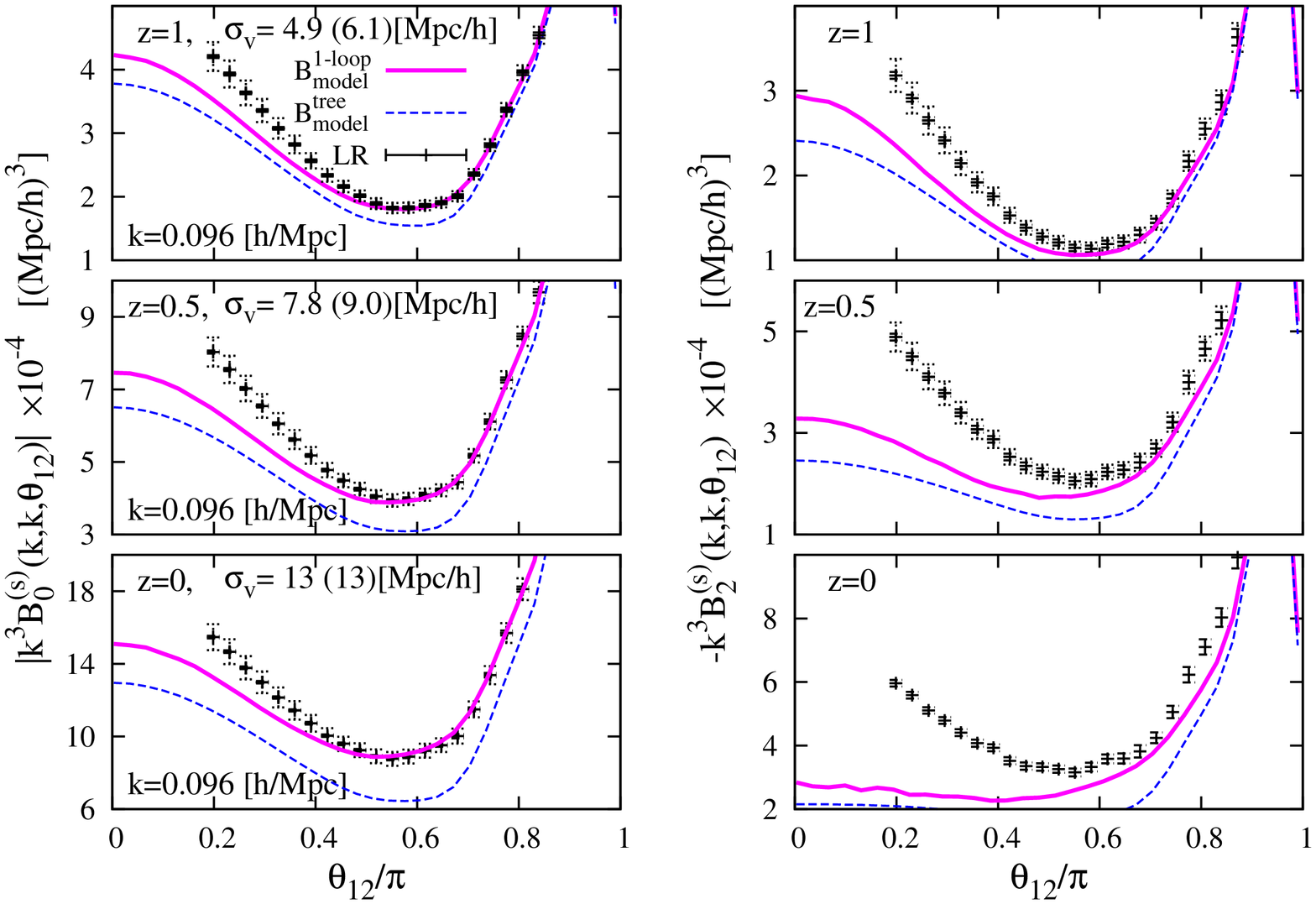}
          \end{center}
          \vspace*{-5pt}
      \caption{Same as Fig.~\ref{equil_kdep}, but here the shape dependence of the bispectrum for isosceles triangle ($k_1=k_2$) is plotted as function of $\theta_{12}=\cos^{-1}(\hat{\bm k}_1\cdot\hat{\bm k}_2)$, fixing the scale of $k_1$ and $k_2$ to $0.096\,h$\,Mpc$^{-1}$. Here, in plotting the predictions of PT models, we adopt the fitted results of $\sigmav$ obtained from Fig.~\ref{equil_kdep}. }\label{equil_tdep}
 \vspace*{-10pt}
  \begin{center}
   \includegraphics[width=105mm,angle=270]{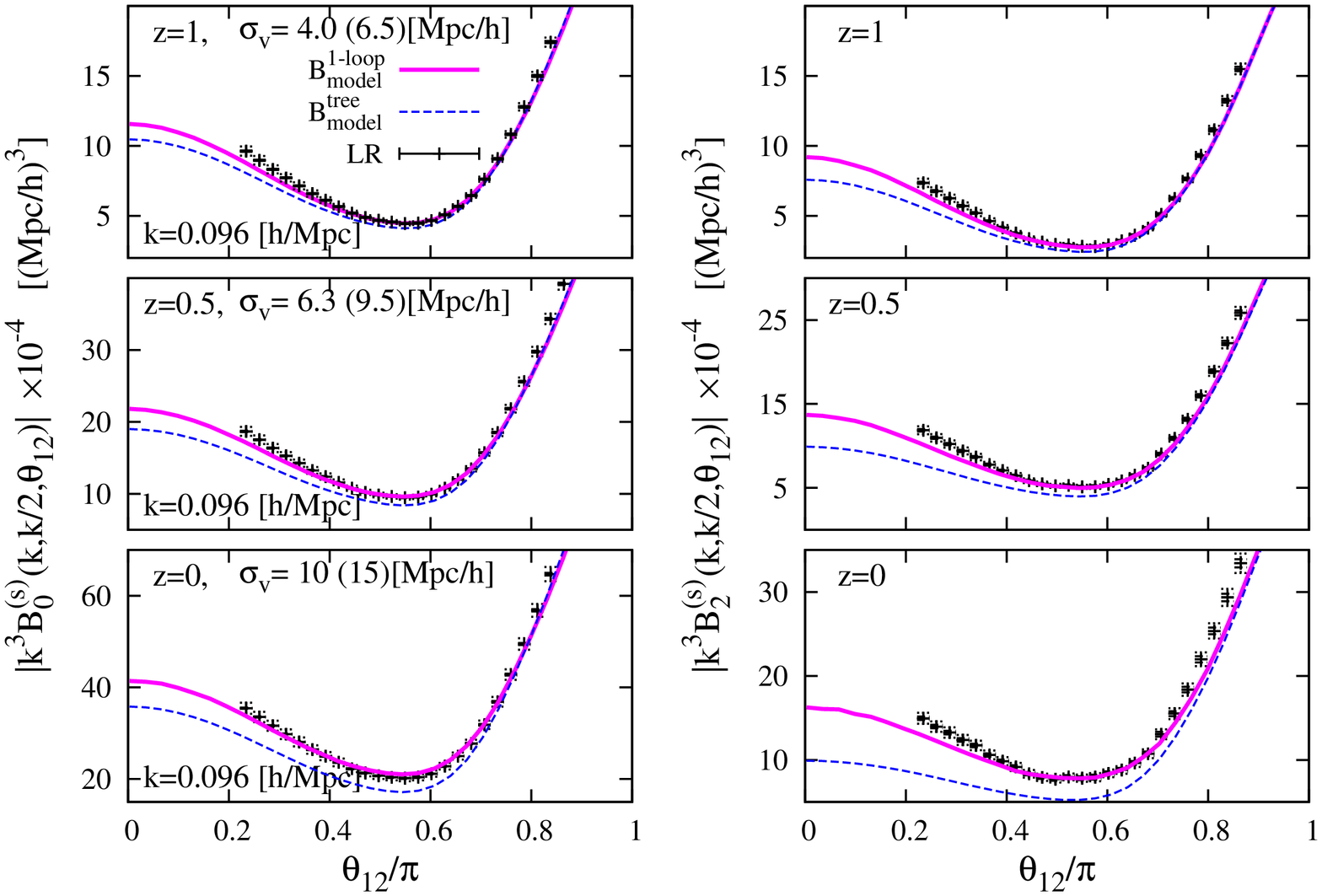}
          \end{center}
          \vspace*{-5pt}
      \caption{Same as Fig.~\ref{equil_tdep}, but the results for another triangular shape with $k_1=2k_2$ are shown, again fixing $k_1$ and $2k_2$ to $0.096\,h$\,Mpc$^{-1}$. Here, in plotting the predictions of PT models, we adopt the fitted results of $\sigmav$ obtained from Fig.~\ref{iso_kdep}. }
    \label{iso_tdep}
  \end{figure*}

Adopting Eq.~(\ref{eq:dfog}), Figs.~\ref{equil_kdep}-\ref{iso_tdep} compare the prediction of the PT model with $N$-body simulations. While Figs.~\ref{equil_kdep} and \ref{iso_kdep} show the scale-dependence of the bispectrum amplitudes for equilateral ($k_1=k_2=k_3$) and scalene triangular configuration with $k_1=2k_2=(2/\sqrt{3})k_3$, Figs.~\ref{equil_tdep} and \ref{iso_tdep} respectively plot the shape-dependence of the bispectrum as function of $\theta_{12}=\cos^{-1}(\hat{\bm{k}}_1\cdot\hat{\bm{k}}_2)$ for the triangles of $k_1=k_2$ and $k_1=2k_2$. In each figure, the monopole and quadrupole moments of the bispectrum are computed/measured according to the definition in Appendix \ref{app:multipole}, and the results are presented in left and right panels, respectively. Here, the measured results of the bispectra are shown only for LR data, since no notable difference has been found in both LR and HR data.

\begin{figure*}[tbp] 
  \includegraphics[width=8cm,angle=270]{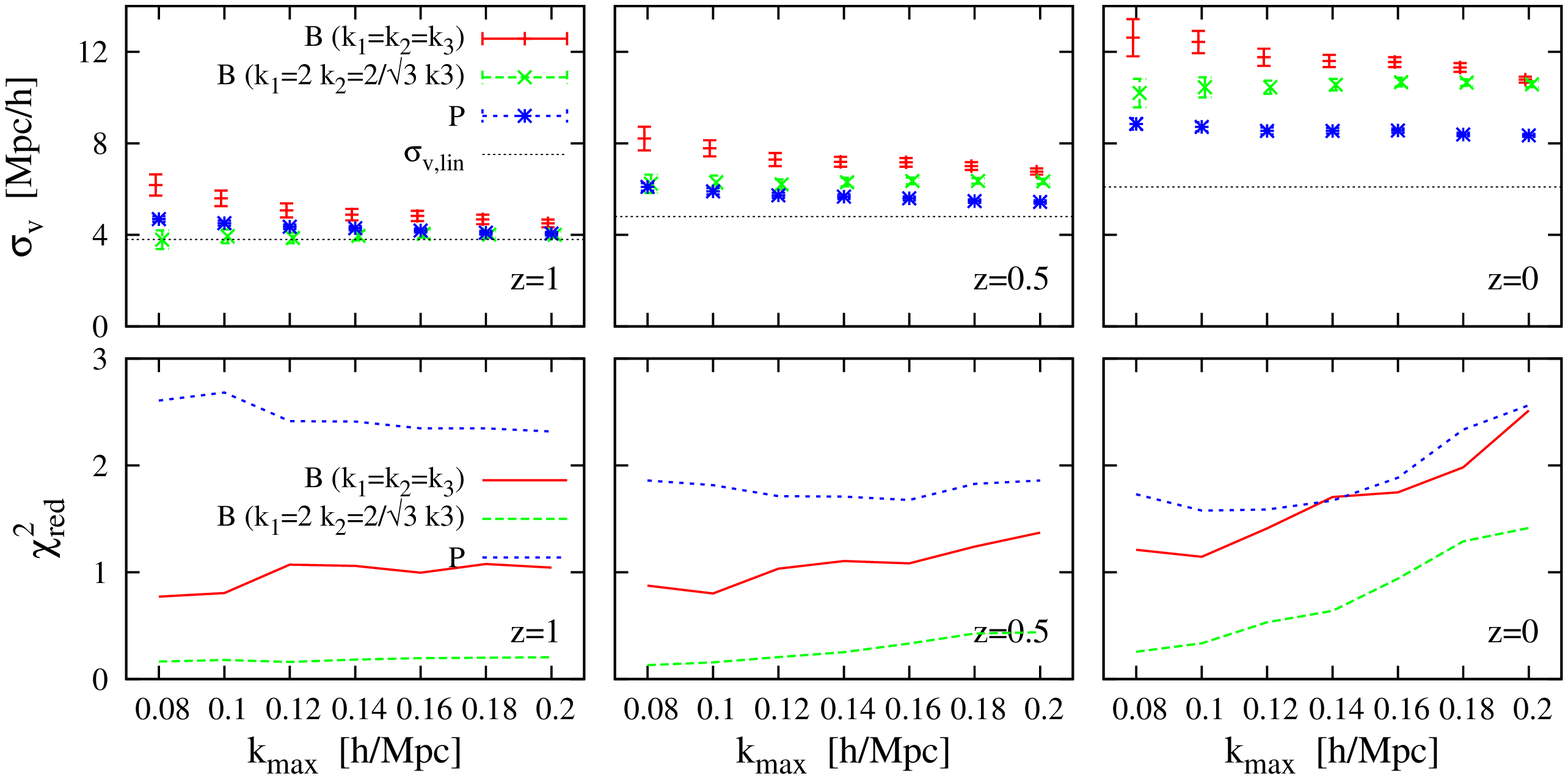}
\caption{Dependence of the fitted results of parameter $\sigma_v$ (top), and the reduced chi-square (bottom) on the maximum wavenumber $k_{\rm max}$ at $z=1$ (left), $0.5$ (middle), and $0$ (right). Red and green symbols respectively represent the results from the bispectrum for equilateral and scalene shape with $k_1=2k_2=(2/\sqrt{3})k_3$. On the other hand, blue symbols are obtained from power spectrum, based on the PT model by Ref.~\cite{Taruya:2010mx}. Note that all the results are obtained by fitting the one-loop PT model predictions to the measured results from LR data of $N$-body simulations.} 
\label{sigv_kmaxdep}
  \includegraphics[width=8cm,angle=270]{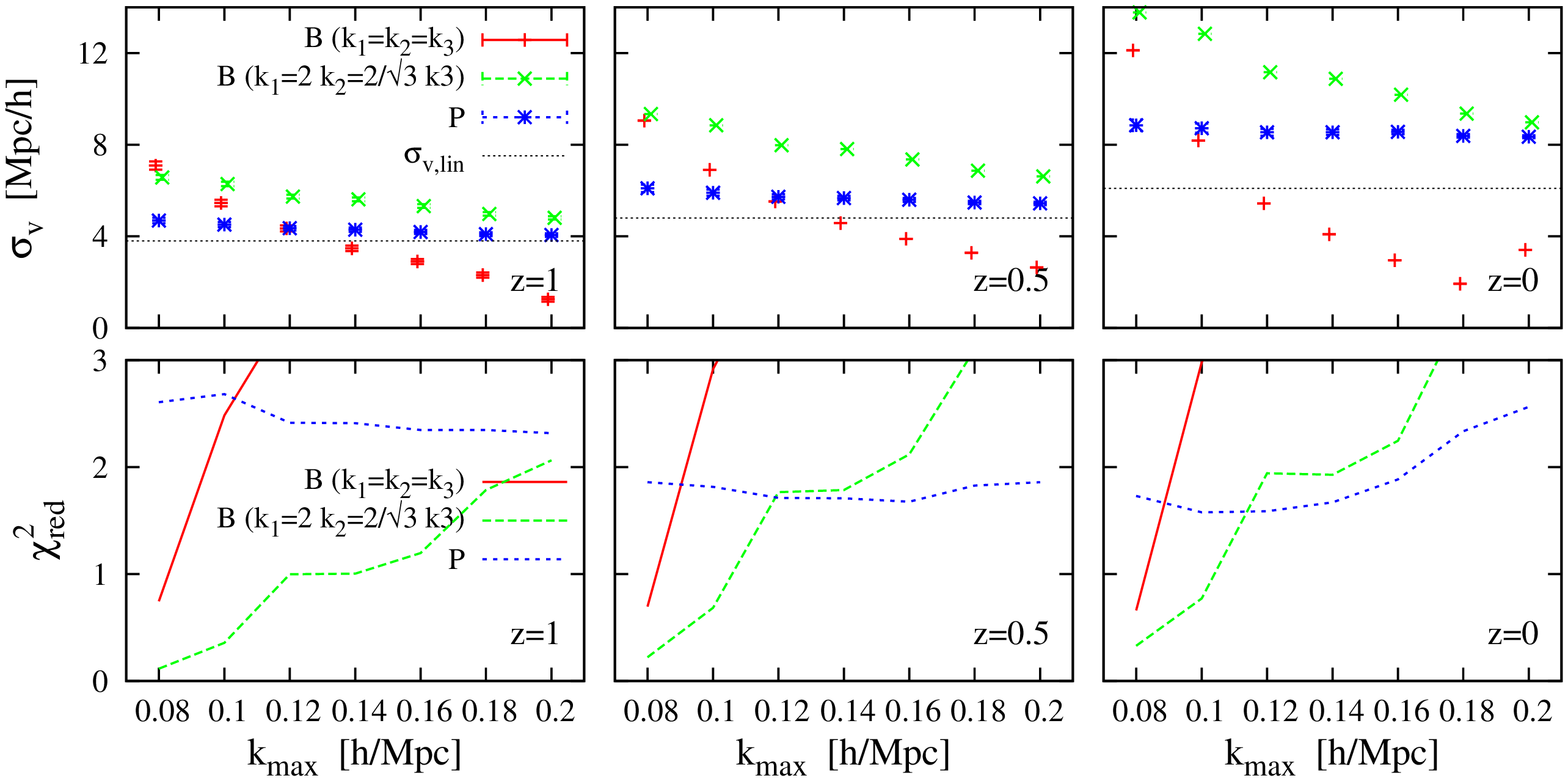}
\caption{Same as in Fig.~\ref{sigv_kmaxdep}, but the results obtained by fitting the tree-level PT model predictions to the $N$-body data are shown. Meanings of the lines and symbols are the same as in Fig.~\ref{sigv_kmaxdep}. }
\label{sigv_kmaxdep_tree}
\end{figure*}

Overall, the one-loop PT model depicted as magenta solid lines better agree with simulations over a wider range of $k$ as well as for a wide range of shapes. Note that the free parameter $\sigmav$ in Eq.~(\ref{eq:dfog}) is determined at each redshift by fitting the predicted monopole and quadrupole moments with measured results of $N$-body simulations at the range $[k_{\rm min},\,k_{\rm max}]$. While $k_{\rm min}$ is set to $0.05\,h$\,Mpc$^{-1}$, we adopt $k_{1\%}$ defined by Ref.~\cite{Nishimichi:2008ry} as the maximum wavenumber $k_{\rm max}$, indicated by the vertical arrow in each panel of Figs.~\ref{equil_kdep} and \ref{iso_kdep} (blue for tree-level PT and red for one-loop PT). The $k_{1\%}$ indicates the maximum wavenumber below which the predicted power spectrum is shown to well reproduce the N-body result within $1\%$ accuracy in the real space, and from Fig.~\ref{realB_kdep_tdep}, we see that the $k_{1\%}$ also gives a good indicator for the applicable range of one-loop bispectrum in real space. Because the deviations between PT and simulations do not behave as a monotonous function it is however hard to extract an exact scale. This is why we rely on more accurate power spectrum measurements to define $k_{1\%}$. The fitted result of the parameter $\sigmav$ is given in left panel of each figure, which are close to the linear theory prediction of one-dimensional velocity dispersion (see also Table \ref{kmax_del_v}, and Figs.~\ref{sigv_kmaxdep} and \ref{sigv_kmaxdep_tree}).

A notable point may be that the one-loop PT model reproduces the $N$-body results even beyond the fitting range of $\sigmav$. For comparison, in Figs.~\ref{equil_kdep}-\ref{iso_tdep}, we have also plotted the tree-level standard PT predictions multiplied by the damping function of Eq.~(\ref{eq:dfog})\footnote{In terms of the descriptions given in Sec.~\ref{ourmodel}, the tree-level standard PT multiplied by $D_{\rm FoG}$ corresponds to the leading-order PT calculations of Eq.~(\ref{bis_model}) with the functions $C_n$ summing up to $n=2$. }, depicted as blue short-dashed lines, but the agreement with $N$-body simulations is restricted to the fitting range indicated by the blue vertical arrow. In this respect, the one-loop corrections play an important role, together with damping function, to better describe the redshift-space bispectrum at weakly non-linear regime.  A closer look at the equilateral case in Fig.~\ref{equil_kdep} reveals that the agreement of the one-loop model is bit degraded compared to the scalene case (Fig.~\ref{iso_kdep}), especially for the quadrupole moment. Similar trend is also found for the shape dependence in Figs.~\ref{equil_tdep} and \ref{iso_tdep}. This is partly because for a fixed wavenumber $k$, one of the side length for the triangle becomes smaller or larger than $k$, and the results can be less or more sensitive to the nonlinearity of the gravitational clustering and RSD. Indeed, for the cases shown in Figs.~\ref{iso_kdep} and \ref{iso_tdep},  the tree-level PT predictions get closer to the one-loop PT results at large scales. 

\begin{table*}[tbp]
\begin{ruledtabular}
  \begin{tabular}{c|c|c||c|c|c|c|c|c}
     &\multicolumn{2}{c||}{$k_{\rm max}~ [h\,{\rm Mpc}^{-1}]$}& $P_{\rm model}^{(s)}$: 1-loop &\multicolumn{2}{c|}{$B_{\rm model}^{(s)}$: 1-loop }&\multicolumn{2}{c|}{$B_{\rm model}^{(s)}$: tree}& $\sigma_{\rm v,lin}$ \\ \cline{2-9}
     &1-loop&tree&& $k_1=k_2=k_3$ & $k_1=2k_2=(2/\sqrt{3})k_3$& $k_1=k_2=k_3$& $k_1=2k_2=(2/\sqrt{3})k_3$ & \\ \hline
    $z=1$&0.13&0.09&$4.3\pm 0.1$~($4.2\pm 0.04$)&$4.9\pm 0.3$~($5.9\pm 0.3$)&$4.0\pm 0.2$~($4.3\pm 0.3$)&$6.1\pm 0.2$&$6.5\pm 0.1$ & $3.8$\\ 
    $z=0.5$&0.1&0.07&$5.9\pm 0.1$&$7.8\pm 0.4$&$6.3\pm 0.3$&$9.0\pm 0.3$&$9.5\pm 0.2$ & $4.8$ \\ 
  $z=0$&0.08&0.06&$8.8\pm 0.3$~($9.8\pm 0.3$)&$13\pm 0.8$~($15\pm 1.0$)&$10\pm 0.6$~($9.1\pm 1.1$)&$13\pm 0.6$&$15\pm 0.4$ & $6.1$\\   
    \end{tabular}
  \caption{Fitted values of $\sigmav$ given in Eq.~(\ref{eq:dfog}) at various redshifts and triangular shapes. The results are obtained by fitting the monopole and quadrupole predictions of PT models to the measured results from the LR (HR) data of $N$-body simulations, and are listed in units of $h\,$\,Mpc$^{-1}$. For reference, the rightmost column shows the linear theory predictions. }
  \label{kmax_del_v}
 \end{ruledtabular}
  \end{table*}

\subsection{Testing the ansatz of damping function}
 \label{subsec:validity_D_FoG}

\begin{figure*}[tbp] 
  \includegraphics[width=85mm,angle=270]{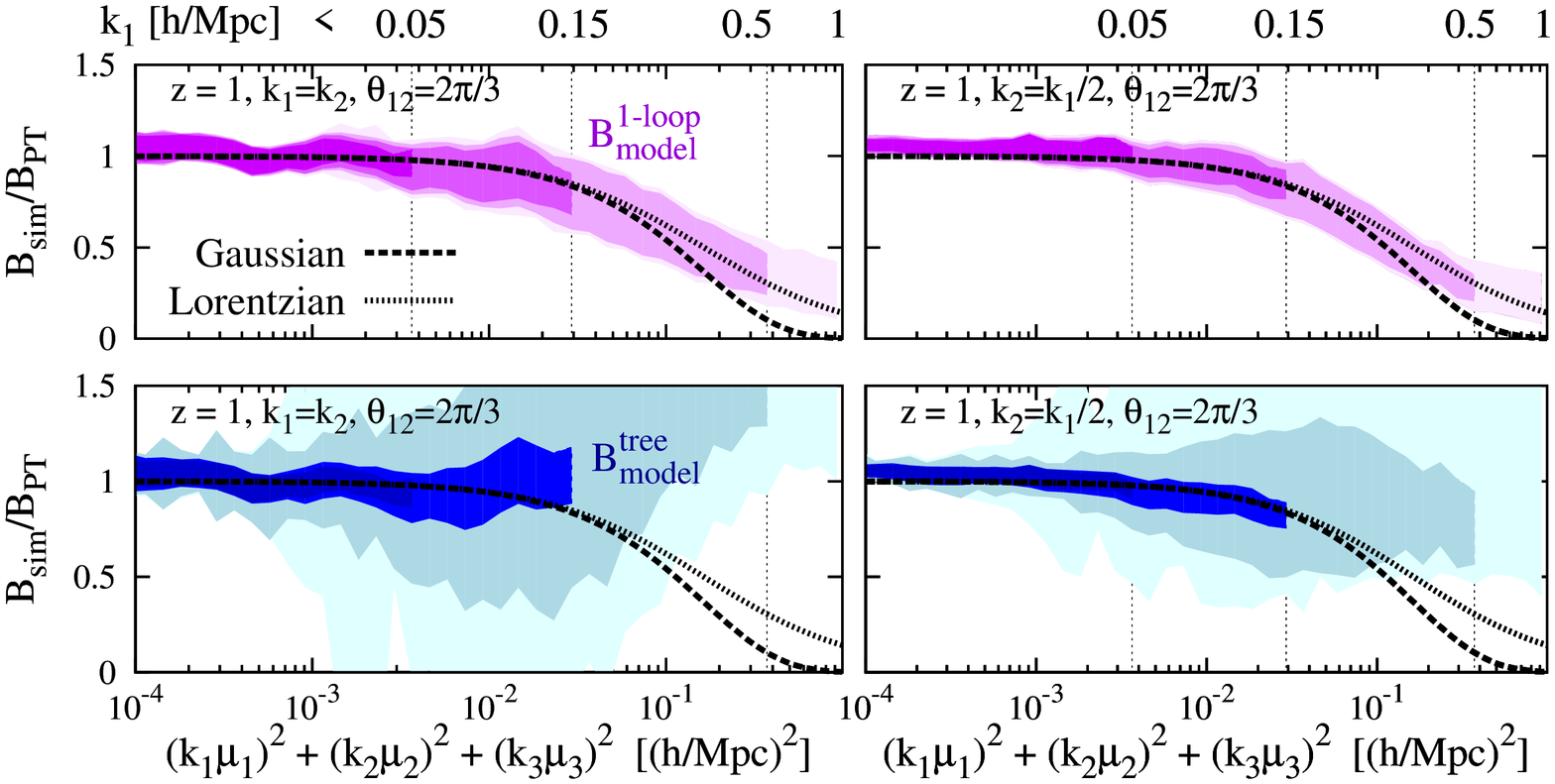}
\caption{Ratio of the redshift-space bispectrum measured from the LR data and to  the one-loop (top) and tree-level (bottom) PT models, setting $D_{\rm FoG}$ to unity (or equivalently, taking the limit $\sigmav\to0$). Here, without applying multipole expansion, the ratio is evaluated for various orientation of the triangle, and the scatters of the estimated results are shown as color shaded regions, fixing the equilateral shape (left) and scalene shape with $k_1=2k_2=(2/\sqrt{3})k_3$ (right). The resultant ratio is then plotted against $(k_1\mu_1)^2+(k_2\mu_2)^2+(k_3\mu_3)^2$. Note that the four different color strengths imply the length of $k_1$ used to estimate the ratio, indicated at the upper horizontal axis. For reference, Gaussian (dashed) and Lorentzian (dotted) forms of the damping function are also plotted. 
}
\label{fig:test_ansatz}
\end{figure*}

Adopting the Gaussian form of the non-perturbative damping function, we have seen that the PT model successfully describes the measured results of the bispectrum. However, the validity and consistency of the treatment have to be checked and/or verified, at least in the following two aspects. One is the consistency of the fitted values of $\sigmav$. Since the present PT model is constructed in similar manner to the model of power spectrum in Ref.~\cite{Taruya:2010mx}, the fitted values of the parameter $\sigmav$ derived from bispectrum have to be consistent with those from the power spectrum. The other aspect is the functional form of the damping function. We have assumed in Eq.~(\ref{eq:dfog}) that the damping function is  expressed as the univariate function of $(k_1\mu_1)^2+(k_2\mu_2)^2+(k_3\mu_3)^2$, but this can be verified only at the leading-order, and have to be checked at the scales relevant for the one-loop PT.

Let us first discuss the consistency of the fitted value, $\sigmav$. Table \ref{kmax_del_v} summarizes the results derived both from the bispectrum and power spectrum. The estimated results are based on the $N$-body simulations of the LR (HR) data, and we used the one-loop model for power spectrum, while for the bispectrum, the fitted results from two different configurations are presented in both tree-level and one-loop PT cases. The best-fitted values both from the power spectrum and bispectrum one-loop models reasonably agree with each other at $z=1$. Although a deviation is manifest at lower redshifts, this is small compared to the cases with tree-level predictions of bispectrum.

To see the robustness of the fitted values, we extend the analysis in Table~\ref{kmax_del_v}, and using the LR data, we examine the fitting in various range of $k$. The results are shown in Figs.~\ref{sigv_kmaxdep} and \ref{sigv_kmaxdep_tree}. In each figure, top and bottom panels respectively plot the fitted values $\sigmav$ and the reduced $\chi^2$ as function of $k_{\rm max}$. Note that in estimating $\chi^2_{\rm red}$ and the errors of the fitted $\sigmav$, we took account of the systematics in the $N$-body simulations. That is, at each data point, we added the systematic errors $\Delta P_{\ell,{\rm sys}}^{\rm(s)}$ and $\Delta B_{\ell,{\rm sys}}^{\rm(s)}$ to the statistical errors of the power spectrum and bispectrum multipoles, $\Delta P_{\ell,{\rm stat}}^{\rm(s)}$ and $\Delta B_{\ell,{\rm stat}}^{\rm(s)}$, as shown in Figs.~\ref{equil_kdep}-\ref{iso_tdep} for the bispectrum case (solid error bars: $\Delta B^{\rm(s)}_{\ell,{\rm stat}}$, dotted error bars: $\Delta B^{\rm(s)}_{\ell,{\rm stat}}+\Delta B^{\rm(s)}_{\ell,{\rm sys}}$). The size of the systematics is estimated from the real-space results in Figs.~\ref{power_real} and \ref{realB_kdep_tdep}. We assume that the relative systematics in redshift space for the power spectrum and bispectrum are proportional to the relative systematics in real space for the matter power spectrum, $ \Delta P_{\rm sys}/P_{\rm sim} = |P_{\rm sim}-P_{\rm ref}|/P_{\rm sim}$, where $P_{\rm sim}$ represents the measured real-space power spectrum in $N$-body simulations, and $P_{\rm ref}$ is our reference for the real-space power spectrum. For the reference, we choose the emulator power spectrum (although the estimated systematics is of order 1 percent, we neglect this contribution). The systematics for the power spectrum and bispectrum in redshift space are estimated as,
\begin{align}
 \frac{\Delta P^{\rm(s)}_{\ell,{\rm sys}}(k_a)}{P^{\rm(s)}_{\ell,{\rm sim}}(k_a)}&=\alpha  \frac{\Delta P_{\rm sys}(k_a)}{P_{\rm sim}(k_a)},\\
\frac{\Delta B^{\rm(s)}_{\ell,{\rm sys}}(k_1,k_2,k_3)}{B^{\rm(s)}_{\ell,{\rm sim}}(k_1,k_2,k_3)} &= \beta \sqrt{\frac{1}{3} \sum_{a=1}^3
\left( \frac{\Delta P_{\rm sys}(k_a)}{P_{\rm sim}(k_a)}\right)^2},
\label{eq:syst}
\end{align}
where $\alpha$ and $\beta$ are two fudge factors that we have fixed to $\alpha \sim 1$ and $\beta \sim 1$. We have checked that at large-k (where statistical error bars are small compared to systematics), Eq.~(\ref{eq:syst}) reproduces the order of magnitude of the relative difference between the bispectrum of the HR set and the LR set (which is affected by mass resolution effect).   

The rough systematic errors adopted here may result in a rather crude estimate of the goodness of fit, and thus the derived $\chi_{\rm red}^2$ can only be used for a comparison between tree-level and one-loop results in Fig.~\ref{sigv_kmaxdep} and \ref{sigv_kmaxdep_tree}. Nevertheless, we see that the fitting results in one-loop PT cases are basically stable against the variation of $k_{\rm max}$, and the estimated values of $\chi_{\rm red}^2$ are smaller than those in the tree-level cases. Further, we checked that the (best-)fitted values of $\sigmav$ are robust against the systematic errors, and the resultant values in one-loop PT reasonably agree well with each other, especially at $z=1$. Although the deviation becomes manifest at lower redshift, this would be probably due to the break down of the one-loop predictions. In fact, the $\chi_{\rm red}^2$ systematically increases with $k_{\rm max}$, indicating that the fitting starts to fail. Thus, at least at the redshift $z=1$, the one-loop PT models work fine, and the FoG damping is described with a single parameter.

Next consider the validity of the ansatz for $D_{\rm FoG}$ at Eq.~(\ref{eq:dfog}). To clarify whether the non-perturbative part is described by the univariate function or not, we directly measure the bispectrum, not applying the multipole expansion. In this case, the bispectrum in redshift space is described by the five variables. For each shape and orientation of the bispectrum, we compute the corresponding PT prediction based on Eq.~(\ref{bis_model}) or (\ref{bis_new}), but setting $D_{\rm FoG}$ to $1$. Taking the ratio gives
\begin{align}
&\frac{B_{\rm sim}^{\rm (s)}(k_1,k_2,\theta_{12},\omega,\phi)}{B_{\rm model}^{\rm(s)}(k_1,k_2,\theta_{12},\omega,\phi)\Bigr|_{D_{\rm FoG}=1}}.
\label{eq:ratio_DFoG}
\end{align}
At the scales where the one-loop PT is applicable, this ratio directly quantifies the functional form of $D_{\rm FoG}$, and thus we can check whether it is expressed as univariate function of $(k_1\mu_1)^2+(k_2\mu_2)^2+(k_3\mu_3)^2$ or not. 

Fig.~\ref{fig:test_ansatz} shows the measured results of the ratio, Eq.~(\ref{eq:ratio_DFoG}), for various orientations at $z=1$, plotted against $(k_1\mu_1)^2+(k_2\mu_2)^2+(k_3\mu_3)^2$. To be precise, what is shown here is the dispersion of the measured ratio depicted as shaded color region, and the four different color strengths imply the length of $k_1$ used to estimate the ratio: $k_1\leq0.05$, $0.15$, $0.5$, and $1\,h$\,Mpc$^{-1}$ from dark to light.  The results are compared with the univariate damping function of the Gaussian (dashed) and Lorentzian (dotted) form. Clearly, the scatter of the ratio for the one-loop PT model, given in top panel, is small, and its mean values fairly trace the univariate damping function. This is in marked contrast to the results for tree-level PT shown in bottom panel, where we see a large scatter. Further, the results seem robust irrespective of the shape of the bispectrum triangle, as seen in both left and right panels, where we respectively show the results for the equilateral case ($k_1=k_2=k_3$) and the scalene triangle with $k_1=2k_2=(2/\sqrt{3})k_3$. A closer look at results suggests that Lorentzian form describes the measured ratio reasonably well at the high-$k$ tail, although it is mostly the boundary where we can apply one-loop PT prediction. Hence, we conclude that the univariate ansatz for $D_{\rm FoG}$ is validated at least in the applicable range of one-loop predictions.

\section{Discussion and conclusion\label{sec:sum}}

In this paper, we have studied the matter bispectrum in redshift space, and presented a perturbation theory (PT) model that can keep the non-perturbative damping effect of the redshift-space distortions (RSD) under control. Starting with the exact formula for redshift-space bispectrum, we rewrite the expression in terms of the cumulants to identify the non-perturbative term. Separating the non-perturbative term responsible for the so-called Fingers-of-God (FoG) damping effect, we derive the perturbative expressions for bispectrum valid at one-loop order. The resultant model has been constructed similarly to the power spectrum model in Ref.~\cite{Taruya:2010mx}, and it incorporates the non-perturbative damping term on top of the terms that can be computed with standard PT.

Adopting the Gaussian form of damping function, we have performed a detailed comparison between the predictions of PT model with the measured results of the bispectrum from a suite of cosmological $N$-body simulations. Incorporating a single free parameter into the damping function, the one-loop PT model reproduces the simulation results fairly well at weakly nonlinear scales at $z=0-1$. The fitted results of the parameter $\sigmav$ are found to agree well with those obtained from the power spectrum, and the agreement generally holds irrespective of the shape of the triangles. On the other hand, even if we incorporate the damping function into the model, the tree-level PT predictions start to deviate from $N$-body results at rather low-$k$. Also, the fitted value $\sigmav$ does not match the one obtained from the power spectrum, and varies with triangular shapes.

We have further examined the validity of the ansatz imposed in the functional form of the damping function [Eq.~(\ref{eq:dfog})]. Combining the simulation data with standard PT results, we confirmed that the univariate ansatz for damping function $D_{\rm FoG}$ indeed hold for one-loop PT model, and its functional form is shown to be very close to the Gaussian, although the Lorentzian form looks slightly better. Note, however, that in the case of the tree-level PT model, univariate ansatz does not give a good description, and this can be a part of the reason why the failure of tree-level PT prediction appears at rather low-$k$. Hence, even at large scales, a careful modeling with one-loop correction is essential, and together with the model of power spectrum proposed by Ref.~\cite{Taruya:2010mx}, the present one-loop model of bispectrum gives a coherent description for RSD.

Finally, we note that toward the practical application to real applications, there still remain several issues to be addressed. One is the improvement of the PT prediction. Including the higher-order (two-loop) corrections or applying the resummation technique, the applicable range of PT is expected to become wider, and a more tighter test of gravity will be made. Effective field theory approach may also help to improve the prediction (e.g., Refs.~\cite{2012JCAP...07..051B,2012JHEP...09..082C,2014PhRvD..89d3521H,Baldauf:2015aha}). Another issue is the estimation of statistical error covariance of bispectrum, which is crucial and necessary for unbiased and robust cosmological data analysis. The $N$-body measurement of the covariance is, however, known to be computationally extensive (e.g., Refs.~\cite{2013MNRAS.429..344K,2013PhRvD..87l3538S} for weak lensing case), and a clever approach, involving the analytic treatment, may have to be developed, for instance. One final big issue is the galaxy biasing. Throughout the paper, we have focused on the matter bispectrum, but the real observable is the biased object. Incorporating the prescription of galaxy biasing is thus very crucial. Toward the practically useful model, the study with halo/subhalo catalogs may help a lot, and we will tackle this issue in the near future.

\begin{acknowledgments}
This work is in part supported by MEXT/JSPS KAKENHI Grant Number JP15H05899 and JP16H03977. It is also supported by JSPS Grant L16519. This work was granted access to HPC resources of IDRIS through allocations made by GENCI (Grand Equipement National de Calcul Intensif) under the allocations 2015-042287 and 2016-042287. We acknowledge support from the DIM ACAV of the Region Ile-de-France. We also thanks Stéphane Colombi for providing his bispectrum code.
\end{acknowledgments}

\appendix

\section{Redshift-space perturbation theory kernels}
\label{app:kernel}

In this Appendix, we summarize the explicit expressions for redshift-space kernels $Z_n$ defined in Eq.~(\ref{RSD_fluc}). Substituting the higher-order PT solutions given by Eqs.~(\ref{eq:delta_n_SPT}) and (\ref{eq:theta_n_SPT}) into the Taylor-expanded form of the redshift-space density field in Eq.~(\ref{teylar}), the kernel $Z_n$ can be read off from the reorganized perturbative expansion in powers of $\delta_{\rm L}$.  The kernels up to fourth order become (e.g., \cite{Scoccimarro:1999ed}):
\begin{widetext}
\begin{align}
&Z_1(\bm{k})=1+f\mu^2,
\label{eq:Z_1}
\\
&Z_2(\bm{k}_1,\bm{k}_2)=F_2(\bm{k}_1,\bm{k}_2)+f\mu^2G_2(\bm{k}_1,\bm{k}_2)+\frac{f\mu k}{2}\left[\frac{\mu_1}{k_1}(1+f\mu_2^2)+\frac{\mu_2}{k_2}(1+f\mu_1^2)\right].
\label{eq:Z_2}
\\
&Z_3(\bm{k}_1,\bm{k}_2,\bm{k}_3)=F_3(\bm{k}_1,\bm{k}_2,\bm{k}_3)+f\mu^2G_3(\bm{k}_1,\bm{k}_2,\bm{k}_3)+f\mu k\left[F_2(\bm{k}_2,\bm{k}_3)+f\mu_{23}^2G_2(\bm{k}_2,\bm{k}_3)\right]\frac{\mu_1}{k_1}
\nonumber\\
&\qquad\qquad+f\mu k(1+f\mu_1^2)\frac{\mu_{23}}{k_{23}}G_2(\bm{k}_2,\bm{k}_3)+\frac{(f\mu k)^2}{2}(1+f\mu_1^2)\frac{\mu_2}{k_2}\frac{\mu_3}{k_3},
\label{eq:Z_3}
\\
&Z_4(\bm{k}_1,\bm{k}_2,\bm{k}_3,\bm{k}_4)=Z_{4{\rm a}}(\bm{k}_1,\bm{k}_2,\bm{k}_3,\bm{k}_4)+Z_{4{\rm b}}(\bm{k}_1,\bm{k}_2,\bm{k}_3,\bm{k}_4).
\label{eq:Z_4}
\end{align}
Here, the fourth-order kernels $Z_{\rm 4a}$ and $Z_{\rm 4b}$ are respectively given by
\begin{align}
&Z_{4{\rm a}}(\bm{k}_1,\bm{k}_2,\bm{k}_3,\bm{k}_4)=F_4(\bm{k}_1,\bm{k}_2,\bm{k}_3,\bm{k}_4)+f\mu^2G_4(\bm{k}_1,\bm{k}_2,\bm{k}_3,\bm{k}_4)+f\mu k\left[F_3(\bm{k}_2,\bm{k}_3,\bm{k}_4)+f\mu_{234}^2G_3(\bm{k}_2,\bm{k}_3,\bm{k}_4)\right]\frac{\mu_1}{k_1}
\nonumber\\
&\qquad\qquad
+f\mu k(1+f\mu_1^2)\frac{\mu_{234}}{k_{234}}G_3(\bm{k}_2,\bm{k}_3,\bm{k}_4)+\frac{(f\mu k)^3}{6}(1+f\mu_1^2)\frac{\mu_2}{k_2}\frac{\mu_3}{k_3}\frac{\mu_4}{k_4},
\label{eq:Z_4a}
\\
&Z_{4{\rm b}}(\bm{k}_1,\bm{k}_2,\bm{k}_3,\bm{k}_4)=f\mu k\left[F_2(\bm{k}_1,\bm{k}_2)+f\mu_{12}^2G_2(\bm{k}_1,\bm{k}_2)\right]\frac{\mu_{34}}{k_{34}}G_2(\bm{k}_3,\bm{k}_4)
\nonumber\\
&\qquad\qquad+\frac{(f\mu k)^2}{2}\left[F_2(\bm{k}_1,\bm{k}_2)+f\mu_{12}^2G_2(\bm{k}_1,\bm{k}_2)\right]\frac{\mu_{3}}{k_{3}}\frac{\mu_{4}}{k_{4}}+\frac{(f\mu k)^2}{2}\frac{\mu_{12}}{k_{12}}G_2(\bm{k}_1,\bm{k}_2)\left[(1+f\mu_3^2)\frac{\mu_4}{k_4}+(1+f\mu_4^2)\frac{\mu_3}{k_3}\right].
\label{eq:Z_4b}
\end{align}
In the above, the vector $\bm{k}$ in the $n$-th order kernel implies 
$\bm{k}=\bm{k}_1+\cdots+\bm{k}_n$. The quantities $\mu$, $\mu_i$, $\mu_{ij}$, and $\mu_{ijk}$ are defined by: 
\begin{align}
\mu\equiv \frac{\bm{k}\cdot\hat{\bm{z}}}{k}, \quad
\mu_i\equiv \frac{\bm{k}_i\cdot\hat{\bm{z}}}{k_i}, \quad
\mu_{ij}\equiv \frac{(\bm{k}_i+\bm{k}_j)\cdot\hat{\bm{z}}}{|\bm{k}_i+\bm{k}_j|},\quad
\mu_{ijk}\equiv \frac{(\bm{k}_i+\bm{k}_j+\bm{k}_k)\cdot\hat{\bm{z}}}{|\bm{k}_i+\bm{k}_j+\bm{k}_k|}.
\end{align}
\end{widetext}

In applying these results to the statistical calculations described in Sec.~\ref{1-loop_sPT}, the kernels $Z_n$ have to be symmetrized under the exchange of each argument. One important remark is that even with fully symmetrized kernels $F_n$ and $G_n$, the resultant redshift-space kernels $Z_n$ at $n\geq3$ only preserve partial symmetry. For instance, while the kernel $Z_3$ given at Eq.~(\ref{eq:Z_3}) is symmetric under $\bm{k}_1\leftrightarrow\bm{k}_3$, the expression for $Z_{4a}$ preserve the symmetry of $\bm{k}_2\leftrightarrow\bm{k}_3\leftrightarrow\bm{k}_4$. These kernels become fully symmetrized if we take the cyclic permutations.

\section{Perturbative calculations for $D_1$ and $D_2$ terms}
\label{A_pert}

In this Appendix, we present the perturbative expressions for the $D_1$ and $D_2$ terms defined in Eq.~(\ref{eq:expression_D_n}), relevant at one-loop order calculations.

First note that Eq.~(\ref{eq:expression_D_n}) is expressed in terms of the real-space quantities, including the auto- and cross-power spectra and bispectra of density and velocity fields. For convenience, we introduce the two-component multiplet: 
\begin{align}
\Psi_a(\bm{k})=\Bigl(\delta(\bm{k}),\,\theta(\bm{k})\Bigr),
\end{align}
with $\theta$ being the dimensionless velocity divergence defined in real space by $\theta(\bm{x})=-\nabla\cdot\bm{v}/(f\,a H)$. Then the auto- and cross-power spectra and bispectra are given by 
\begin{align}
 &(2\pi)^3\delta_{\rm D}(\bm{k}_1+\bm{k}_2)\,P_{ab}(k_1)\equiv\langle\Psi_a(\bm{k}_1)\Psi_b(\bm{k}_2)\rangle,
\label{pk_real}
\\
 &(2\pi)^3\delta_{\rm D}(\bm{k}_1+\bm{k}_2+\bm{k}_3)\,B_{abc}(\bm{k}_1,\bm{k}_2,\bm{k}_3) 
\nonumber\\
&\qquad\qquad\qquad\qquad
\equiv\langle\Psi_a(\bm{k}_1)\Psi_b(\bm{k}_2)\Psi_c(\bm{k}_3)\rangle, 
\label{bis_real}
\end{align}
where subscripts $a,b,c$ run from $1$ to $2$. 

Since the integrand of the $D_1$ and $D_2$ terms respectively involves $B_{abc}P_{de}$ and $P_{ab}P_{cd}P_{ef}$, the leading-order non-vanishing contributions become $\mathcal{O}(P_{\rm L}^3)$. This is the same order as in the one-loop redshift-space bispectrum. Hence, the tree-level calculations of $P_{ab}$ and $B_{abc}$ are sufficient for a consistent one-loop treatment of redshift-space bispectrum. That is, 
\begin{align}
& P_{ab}(k) \simeq P_{\rm L}(k),
\label{eq:P_ab_linear}
\\
& B_{abc}(\bm{k}_1,\bm{k}_2,\bm{k}_3) \simeq 2\Bigl\{
{\mathcal F}_a(\bm{k}_2,\bm{k}_3)\,P_{\rm L}(k_2)P_{\rm L}(k_3)
\nonumber\\
&+{\mathcal F}_b(\bm{k}_1,\bm{k}_3)\,P_{\rm L}(k_1)P_{\rm L}(k_3)+
{\mathcal F}_c(\bm{k}_1,\bm{k}_2)\,P_{\rm L}(k_1)P_{\rm L}(k_2)
\Bigr\},
\label{eq:B_abc_tree}
\end{align}
where the kernel ${\mathcal F}_a$ is the standard PT kernels at second-order, given by ${\mathcal F}_a=(F_2,\,G_2)$. Below, we will separately present the explicit expressions for $D_1$  and $D_2$ terms. In doing so, we use the following expressions for the cumulants: 
\begin{align}
&\langle A_1A_2A_3\rangle_c =\int\frac{d^3\bm{p}_1d^3\bm{p}_2}{(2\pi)^6}\,
e^{i\left\{\bm{p}_1\cdot\bm{r}_{13}+\bm{p}_2\cdot\bm{r}_{23}\right\}}
\sum_{a,b,c=1}^2 f^{a+b+c-3}
\nonumber\\
&\qquad\quad \times
\,\mu_{p_1}^{2(a-1)}\mu_{p_2}^{2(b-1)}\mu_{p_3}^{2(c-1)}\,
B_{abc}(\bm{p}_1,\bm{p}_2,\bm{p}_3),
\label{eq:A1A2A3}
\\
&\langle (j_4A_4+j_5A_5)^2\rangle_c =-f^2\int\frac{d^3\bm{p}}{(2\pi)^3}\,
\frac{\mu_p^2}{p^2}\,P_{22}(p)
\nonumber\\
&\qquad \quad\times \Bigl\{(k_1\mu_1)^2+(k_2\mu_2)^2+(k_3\mu_3)^2 
+2\bigl( k_1k_2\mu_1\mu_2 e^{i\bm{p}\cdot\bm{r}_{12}}
\nonumber\\
&\qquad\quad
+ k_1k_3\mu_1\mu_3 e^{i\bm{p}\cdot\bm{r}_{13}}
+ k_2k_3\mu_2\mu_3 e^{i\bm{p}\cdot\bm{r}_{23}} \bigr)
\Bigr\},
\label{eq:(j4A4+j5A5)^2}
\\
&\langle A_1A_2\rangle_c =\int\frac{d^3\bm{p}}{(2\pi)^3}\,e^{i\bm{p}\cdot r_{12}}
\nonumber\\
&\qquad\quad\times
\Bigl\{P_{11}(p)+2f\mu_p^2\,P_{12}(p)+f^2\mu_p^4P_{22}(p)\Bigr\},
\label{eq:A1A2}
\\
&\langle (j_4A_4+j_5A_5)A_3)\rangle_c =-f\int\frac{d^3\bm{p}}{(2\pi)^3}\,
\frac{\mu_p}{p}
\nonumber\\
&
\qquad\quad\times
\Bigl\{P_{12}(p)+f\mu_p^2P_{22}(p)\Bigr\}
\nonumber\\
&
\qquad\quad\times
\Bigl\{k_1\mu_1\,e^{i\bm{p}\cdot\bm{r}_{13}}+
k_2\mu_2\,e^{i\bm{p}\cdot\bm{r}_{23}}+k_3\mu_3
\Bigr\}
\label{eq:(j4A4+j5A5)A3}
\end{align}
with the quantities $\mu_i$ and $\mu_p$ respectively defined by $\mu_i=(\bm{k}_i\cdot\hat{\bm{z}})/k_i$ and $\mu_p=(\bm{p}\cdot\hat{\bm{z}})/p$.



\subsection{$D_1$ term}
\label{subsec:D1}

According to Eq.~(\ref{eq:expression_D_n}), the $D_1$ term is explicitly written as
\begin{align}
& D_1(\bm{k}_1,\bm{k}_2,\bm{k}_3)=\frac{1}{2}\int d^3\bm{r}_{13}d^3\bm{r}_{23}
\,e^{i(\bm{k}_1\cdot\bm{r}_{13}+\bm{k}_2\cdot\bm{r}_{23})}\,
\nonumber\\
&\qquad\qquad\times
\langle A_1A_2A_3\rangle_c\,\langle (j_4A_4+j_5A_5))^2\rangle_c.
\end{align}
Substituting Eq.~(\ref{eq:A1A2A3}) and (\ref{eq:(j4A4+j5A5)^2}) into the above, we obtain
\begin{align}
&D_1(\bm{k}_1,\bm{k}_2,\bm{k}_3)
=-\frac{f^2}{2}\Bigl\{(k_1\mu_1)^2+(k_2\mu_2)^2+(k_3\mu_3)^2\Bigr\}
\nonumber\\
&\quad\times\sigma_{\rm v,lin}^2 \,C_1(\bm{k}_1,\bm{k}_2,\bm{k}_3)
 -f^2\int\frac{d^3\bm{p}}{(2\pi)^3}\frac{\mu_p^2}{p^2}P_{22}(p)
\nonumber\\
&\quad\times
\Bigl\{(k_1k_2\mu_1\mu_2)\,C_1(\bm{k}_1+\bm{p},\,\bm{k}_2-\bm{p},\,\bm{k}_3) 
\nonumber\\
&\quad\quad
+(k_1k_3\mu_1\mu_3)\,C_1(\bm{k}_1+\bm{p},\,\bm{k}_2,\,\bm{k}_3-\bm{p}) 
\nonumber\\
&\quad\quad
+(k_2k_3\mu_2\mu_3)\,C_1(\bm{k}_1,\,\bm{k}_2+\bm{p},\,\bm{k}_3-\bm{p}) 
\Bigr\}
\end{align}
with the quantity $\sigma_{\rm v,lin}$ being the linear theory estimate of 
the one-dimensional velocity dispersion, given by
\begin{align}
 \sigma_{\rm v,lin}^2\equiv\int\frac{d^3\bm{p}}{(2\pi)^3}\frac{\mu_p^2}{p^2}\,P_{22}(p)\simeq \int\frac{dp}{6\pi^2}\,P_{\rm L}(p).
\label{eq:sigma_v,lin}
\end{align}
Here, the function $C_1$ is the same one as defined in Eq.~(\ref{eq:def_C_n} ) with (\ref{eq:S_1}), and is explicitly given by 
\begin{align}
C_1(\bm{k}_1,\bm{k}_2,\bm{k}_3)=\sum_{a,b,c=1}^2\mu_1^{2(a-1)}\mu_2^{2(b-1)}\mu_3^{2(c-1)}\,B_{abc}(\bm{k}_1,\bm{k}_2,\bm{k}_3).
\label{eq:C1_term}
\end{align}

\subsection{$D_2$ term}
\label{subsec:D2}

From Eq.~(\ref{eq:expression_D_n}), 
the $D_2$ term is expressed in terms of the cumulants below:
\begin{align}
& D_2(\bm{k}_1,\bm{k}_2,\bm{k}_3)=\frac{1}{2}\int d^3\bm{r}_{13}d^3\bm{r}_{23}
\,e^{i(\bm{k}_1\cdot\bm{r}_{13}+\bm{k}_2\cdot\bm{r}_{23})}\,
\nonumber\\
&\qquad\qquad\times
\Bigl[
  \langle A_1A_2\rangle_c\,\langle (j_4A_4+j_5A_5))A_3\rangle_c 
\nonumber\\
&~\qquad\qquad + \langle A_1A_3\rangle_c\,\langle (j_4A_4+j_5A_5))A_2\rangle_c 
\nonumber\\
&~\qquad\qquad + \langle A_2A_3\rangle_c\,\langle (j_4A_4+j_5A_5))A_1\rangle_c 
\Bigr]\,
\nonumber\\
&\qquad\qquad \times\,
\langle (j_4A_4+j_5A_5))^2\rangle_c.
\end{align}
Substituting Eqs.~(\ref{eq:(j4A4+j5A5)^2})-(\ref{eq:(j4A4+j5A5)A3}) into the above expression, after lengthy calculation, we obtain
\begin{align}
D_2(\bm{k}_1,\bm{k}_2,\bm{k}_3)&=
-\frac{f^2}{2}\{(k_1\mu_1)^2+(k_2\mu_2)^2+(k_3\mu_3)^2\}\,
\nonumber\\
&\qquad \times\,\sigma_{\rm v,lin}^2\,C_2(\bm{k}_1,\bm{k}_2,\bm{k}_3)
\nonumber\\
&+f^3(k_1\mu_1)(k_2\mu_2)(k_3\mu_3)\,\,J_2(\bm{k}_1,\bm{k}_2,\bm{k}_3)
\nonumber\\
&+f^3\,\,K_2(\bm{k}_1,\bm{k}_2,\bm{k}_3).
\end{align}
Here, the function $C_2$ is the same one as defined in Eq.~(\ref{eq:def_C_n}) with (\ref{eq:S_2}). The explicit expressions for the functions $C_2$, $J_2$ and $K_2$ are given below:
\begin{align}
 C_2(\bm{k}_1,\bm{k}_2,\bm{k}_3)&=
-f\,(k_1\mu_1)\Bigl\{ X(\bm{k}_2)Y(\bm{k}_3) + X(\bm{k}_3)Y(\bm{k}_2)\Bigr\} 
\nonumber
\\
&+ \mbox{cyc.}
\label{eq:def_C2}
\\
 J_2(\bm{k}_1,\bm{k}_2,\bm{k}_3)&=\int\frac{d^3\bm{p}}{(2\pi)^3}\,\frac{\mu_p^2}{p^2}P_{22}(p)
\Bigl\{X(\bm{k}_2-\bm{p})Y(\bm{k}_1+\bm{p})
\nonumber
\\
&+ X(\bm{k}_1+\bm{p})Y(\bm{k}_2-\bm{p})\Bigr\}+ \mbox{cyc.}
\label{eq:def_J2}
\\
 K_2(\bm{k}_1,\bm{k}_2,\bm{k}_3)&=\int\frac{d^3\bm{p}}{(2\pi)^3}\,\frac{\mu_p^2}{p^2}P_{22}(p)
\Bigl[\Bigl\{(k_1\mu_1)^2(k_2\mu_2)\,X(\bm{k}_2-\bm{p}) 
\nonumber
\\
&+ (k_1\mu_1)(k_2\mu_2)^2\,X(\bm{k}_1-\bm{p})\Bigr\}\,Y(\bm{k}_3) 
\nonumber
\\
&
+(X\,\, \leftrightarrow\,\, Y)\,\,\Bigr]+\mbox{cyc}.
\label{eq:def_K2}
\end{align}
with the functions $X$ and $Y$ defined by
\begin{align}
&X(\bm{p})=P_{11}(p)+2f\,\mu_p^2\,P_{12}(p) + f^2\,\mu_p^4\,P_{22}(p),
\label{eq:def_func_X}
\\
&Y(\bm{p})=\frac{\mu_p}{p}\,\Bigl\{ P_{12}(p)+f\,\mu_p^2\,P_{22}(p)\Bigr\}.
\label{eq:def_func_Y}
\end{align}

\section{Multipole expansion of redshift-space bispectrum }
\label{app:multipole}

In Sec.~\ref{ourmodel} and \ref{sec:comparison}, we have applied the multipole expansion to the redshift-space bispectrum, and evaluated its monopole and quadrupole moments. In this Appendix, we present the definition of our multipole expansion which differs from the one frequently used in the literature (e.g., \cite{Scoccimarro:1999ed,2015PhRvD..92h3532S,Yamamoto:2016anp,2017MNRAS.tmp..138G}). The newly defined bispectrum multipoles have several nice properties, which we will discuss below.  

The redshift-space bispectrum is characterized as a function of five variables. Three of them characterize the shape of triangle, i.e., the length of two wave vectors $\bm{k}_1$ and $\bm{k}_2$, and the angle between them, $\theta_{12}\equiv\cos^{-1}(\hat{\bm{k}}_1 \cdot \hat{\bm{k}}_2)$. The two remaining variables describe the orientation of the triangle with respect to the line-of-sight. We denote them by $\omega$ and $\phi$. In Ref.~\cite{Scoccimarro:1999ed}, the dependance of the orientation of the triangle are conveniently described by decomposing into spherical harmonics: 
\begin{align}
B_{\rm s}(\bm{k}_1,\bm{k}_2,\bm{k}_3)=\sum^\infty_{\ell=0}\sum^\ell_{m=-\ell} B_{\rm s}^{(\ell,m)}(k_1,k_2,\theta_{12})Y_{\ell m}(\omega,\phi).
\end{align}  
Similar to Ref.~\cite{Scoccimarro:1999ed}, we focus on the $m=0$ multipoles, which correspond to averaging over $\phi$. Then, the above equation leads to  
\begin{align}
\int_0^{2\pi}\frac{d\phi}{2\pi}B_{\rm s}(\bm{k}_1,\bm{k}_2,\bm{k}_3)=\sum^\infty_{\ell=0} B_{\rm s}^{(\ell)}(k_1,k_2,\theta_{12})P_{\ell}(\mu), 
\label{eq:multipole_expansion}
\end{align}
were $P_{\ell}(\mu)$ is the Legendre polynomials. The variable $\mu$ represents the directional cosine of the orientation, $\mu=\cos{\omega}$. 
This is rewritten with 
\begin{align}
B_{\rm s}^{(\ell)}(k_1,k_2,\theta_{12}) = 
\frac{2\ell+1}{2}\int_{-1}^1\,d\mu\,P_{\ell}(\mu)\,
\int_0^{2\pi}\frac{d\phi}{2\pi}\,
B_{\rm s}(\bm{k}_1,\bm{k}_2,\bm{k}_3).
\label{eq:multipole}
\end{align}

In the above, the bispectrum multipole, $B_{\rm s}^{(\ell)}$, is the quantity of our interest,  but at this moment, Eq.~(\ref{eq:multipole}) is ambiguous because we do not yet specify what is $\omega$ and $\phi$. To describe the orientation of the triangle, a simple way is to choose $\bm{k}_1$ specifically, and define the orientation angle as $\mu=\cos{\omega}=\hat{\bm{k}}_1\cdot\hat{\bm{z}}$ \cite{Scoccimarro:1999ed}. Then, we set $\phi$ to the azimuthal angle around $\bm{k}_1$. In this case, however, the resultant bispectrum multipoles $B_{\rm s}^{(\ell)}(k_1,k_2,\theta_{12})$ are not fully symmetric under the permutation of order, $k_1$, $k_2$ and $k_3$.

Here, we give alternative definitions of $\omega$ and $\phi$ to preserve the symmetry of bispectrum multipoles. This is shown in Fig.~\ref{fig:def_bispec_multipoles}. Given two vectors $\bm{k}_1$ and $\bm{k}_2$, we define the orientation of the triangle by the angle between line-of-sight direction and the vector normal to the triangle, i.e., $\hat{\bm{k}}_1\times\hat{\bm{k}}_2$. That is, 
\begin{align}
\mu=\cos\omega=\frac{(\hat{\bm{k}}_1\times\hat{\bm{k}}_2)\cdot\hat{\bm{z}}}{\sin{\theta_{12}}}.
\label{eq:new_def_mu}
\end{align}
The remaining angle $\phi$ may be defined as the azimuthal angle around 
the vector $\hat{\bm{k}}_1\times\hat{\bm{k}}_2$. Since this is 
perpendicular to the plane of the triangle, we have
\begin{align}
\cos\phi =\frac{\left\{\hat{\bm{z}}\times(\hat{\bm{k}}_1\times\hat{\bm{k}}_2)\right\}\cdot\hat{\bm{k}}_1}{\sin{\omega}}.
\label{eq:new_def_phi}
\end{align}

One may suspect that the above definition still breaks 
the symmetry. Indeed, the vectors $\bm{k}_1$ and $\bm{k}_2$ can be arbitrarily chosen among three vectors, and we can even 
exchange $\hat{\bm{k}}_1\,\longleftrightarrow\,\hat{\bm{k}}_2$. Then the orientation angle is changed to $\omega\to \pi-\omega$. Nevertheless, the bispectrum 
multipole $B_{\rm s}^{(\ell)}$ is invariant. This is because only the $\ell=$ even modes of the bispectrum become non-vanishing, and we have $P_\ell(\cos\omega)=P_\ell(\cos(\pi-\omega))$ for $\ell=$ even.

Once accepting the new definition, the measurement of the bispectrum multipoles is straightforward: 

\begin{itemize}
\item In harmonic space, we first pick up the three density fields, $\delta(\bm{k}_1)$, $\delta(\bm{k}_2)$, and $\delta(\bm{k}_3)$, with the vectors satisfying 
$\bm{k}_1+\bm{k}_2+\bm{k}_3=\bm{0}$. Here, labels of $\bm{k}_1$, $\bm{k}_2$ and $\bm{k}_3$ are arbitrary. 
\item Choosing the two vectors among the three, we calculate the directional cosine $\mu$ according to Eq.~(\ref{eq:new_def_mu}). 
\item Multiplying the bispectrum estimator by the Legendre polynomial $P_\ell(\mu)$.  
\item Repeating the above three steps for the same triangle configuration but with different $\mu$, we average the {\it weighted} bispectrum estimator over $\mu$. Further multiplying the averaged bispectrum by the factor $(2\ell+1)/2$, we finally obtain the bispectrum multipole, 
$B_{\rm s}^{(\ell)}$, which is characterized by the length of two vectors, and the angle between them [see Eq.~(\ref{eq:multipole})]. 
\end{itemize}

\bibliographystyle{apsrev4-1}
\bibliography{bib2}

\begin{thebibliography}{64}%
\makeatletter
\providecommand \@ifxundefined [1]{%
 \@ifx{#1\undefined}
}%
\providecommand \@ifnum [1]{%
 \ifnum #1\expandafter \@firstoftwo
 \else \expandafter \@secondoftwo
 \fi
}%
\providecommand \@ifx [1]{%
 \ifx #1\expandafter \@firstoftwo
 \else \expandafter \@secondoftwo
 \fi
}%
\providecommand \natexlab [1]{#1}%
\providecommand \enquote  [1]{``#1''}%
\providecommand \bibnamefont  [1]{#1}%
\providecommand \bibfnamefont [1]{#1}%
\providecommand \citenamefont [1]{#1}%
\providecommand \href@noop [0]{\@secondoftwo}%
\providecommand \href [0]{\begingroup \@sanitize@url \@href}%
\providecommand \@href[1]{\@@startlink{#1}\@@href}%
\providecommand \@@href[1]{\endgroup#1\@@endlink}%
\providecommand \@sanitize@url [0]{\catcode `\\12\catcode `\$12\catcode
  `\&12\catcode `\#12\catcode `\^12\catcode `\_12\catcode `\%12\relax}%
\providecommand \@@startlink[1]{}%
\providecommand \@@endlink[0]{}%
\providecommand \url  [0]{\begingroup\@sanitize@url \@url }%
\providecommand \@url [1]{\endgroup\@href {#1}{\urlprefix }}%
\providecommand \urlprefix  [0]{URL }%
\providecommand \Eprint [0]{\href }%
\providecommand \doibase [0]{http://dx.doi.org/}%
\providecommand \selectlanguage [0]{\@gobble}%
\providecommand \bibinfo  [0]{\@secondoftwo}%
\providecommand \bibfield  [0]{\@secondoftwo}%
\providecommand \translation [1]{[#1]}%
\providecommand \BibitemOpen [0]{}%
\providecommand \bibitemStop [0]{}%
\providecommand \bibitemNoStop [0]{.\EOS\space}%
\providecommand \EOS [0]{\spacefactor3000\relax}%
\providecommand \BibitemShut  [1]{\csname bibitem#1\endcsname}%
\let\auto@bib@innerbib\@empty
\bibitem [{\citenamefont {{Davis}}\ and\ \citenamefont
  {{Peebles}}(1983)}]{1983ApJ...267..465D}%
  \BibitemOpen
  \bibfield  {author} {\bibinfo {author} {\bibfnamefont {M.}~\bibnamefont
  {{Davis}}}\ and\ \bibinfo {author} {\bibfnamefont {P.~J.~E.}\ \bibnamefont
  {{Peebles}}},\ }\href {\doibase 10.1086/160884} {\bibfield  {journal}
  {\bibinfo  {journal} {Astrophys. J.}\ }\textbf {\bibinfo {volume} {267}},\
  \bibinfo {pages} {465} (\bibinfo {year} {1983})}\BibitemShut {NoStop}%
\bibitem [{\citenamefont {Kaiser}(1987)}]{Kaiser:1987qv}%
  \BibitemOpen
  \bibfield  {author} {\bibinfo {author} {\bibfnamefont {N.}~\bibnamefont
  {Kaiser}},\ }\href@noop {} {\bibfield  {journal} {\bibinfo  {journal} {Mon.
  Not. Roy. Astron. Soc.}\ }\textbf {\bibinfo {volume} {227}},\ \bibinfo
  {pages} {1} (\bibinfo {year} {1987})}\BibitemShut {NoStop}%
\bibitem [{\citenamefont {Hamilton}(1992)}]{Hamilton:1992zz}%
  \BibitemOpen
  \bibfield  {author} {\bibinfo {author} {\bibfnamefont {A.~J.~S.}\
  \bibnamefont {Hamilton}},\ }\href {\doibase 10.1086/186264} {\bibfield
  {journal} {\bibinfo  {journal} {Astrophys. J.}\ }\textbf {\bibinfo {volume}
  {385}},\ \bibinfo {pages} {L5} (\bibinfo {year} {1992})}\BibitemShut
  {NoStop}%
\bibitem [{\citenamefont {Scoccimarro}(2004)}]{Scoccimarro:2004tg}%
  \BibitemOpen
  \bibfield  {author} {\bibinfo {author} {\bibfnamefont {R.}~\bibnamefont
  {Scoccimarro}},\ }\href {\doibase 10.1103/PhysRevD.70.083007} {\bibfield
  {journal} {\bibinfo  {journal} {Phys. Rev.}\ }\textbf {\bibinfo {volume}
  {D70}},\ \bibinfo {pages} {083007} (\bibinfo {year} {2004})},\ \Eprint
  {http://arxiv.org/abs/astro-ph/0407214} {arXiv:astro-ph/0407214 [astro-ph]}
  \BibitemShut {NoStop}%
\bibitem [{\citenamefont {Percival}\ and\ \citenamefont
  {White}(2009)}]{Percival:2008sh}%
  \BibitemOpen
  \bibfield  {author} {\bibinfo {author} {\bibfnamefont {W.~J.}\ \bibnamefont
  {Percival}}\ and\ \bibinfo {author} {\bibfnamefont {M.}~\bibnamefont
  {White}},\ }\href {\doibase 10.1111/j.1365-2966.2008.14211.x} {\bibfield
  {journal} {\bibinfo  {journal} {Mon. Not. Roy. Astron. Soc.}\ }\textbf
  {\bibinfo {volume} {393}},\ \bibinfo {pages} {297} (\bibinfo {year}
  {2009})},\ \Eprint {http://arxiv.org/abs/0808.0003} {arXiv:0808.0003
  [astro-ph]} \BibitemShut {NoStop}%
\bibitem [{\citenamefont {Song}\ and\ \citenamefont
  {Percival}(2009)}]{Song:2008qt}%
  \BibitemOpen
  \bibfield  {author} {\bibinfo {author} {\bibfnamefont {Y.-S.}\ \bibnamefont
  {Song}}\ and\ \bibinfo {author} {\bibfnamefont {W.~J.}\ \bibnamefont
  {Percival}},\ }\href {\doibase 10.1088/1475-7516/2009/10/004} {\bibfield
  {journal} {\bibinfo  {journal} {JCAP}\ }\textbf {\bibinfo {volume} {0910}},\
  \bibinfo {pages} {004} (\bibinfo {year} {2009})},\ \Eprint
  {http://arxiv.org/abs/0807.0810} {arXiv:0807.0810 [astro-ph]} \BibitemShut
  {NoStop}%
\bibitem [{\citenamefont {Yamamoto}\ \emph {et~al.}(2008)\citenamefont
  {Yamamoto}, \citenamefont {Sato},\ and\ \citenamefont
  {Huetsi}}]{Yamamoto:2008gr}%
  \BibitemOpen
  \bibfield  {author} {\bibinfo {author} {\bibfnamefont {K.}~\bibnamefont
  {Yamamoto}}, \bibinfo {author} {\bibfnamefont {T.}~\bibnamefont {Sato}}, \
  and\ \bibinfo {author} {\bibfnamefont {G.}~\bibnamefont {Huetsi}},\ }\href
  {\doibase 10.1143/PTP.120.609} {\bibfield  {journal} {\bibinfo  {journal}
  {Prog. Theor. Phys.}\ }\textbf {\bibinfo {volume} {120}},\ \bibinfo {pages}
  {609} (\bibinfo {year} {2008})},\ \Eprint {http://arxiv.org/abs/0805.4789}
  {arXiv:0805.4789 [astro-ph]} \BibitemShut {NoStop}%
\bibitem [{\citenamefont {{Eisenstein}}\ and\ \citenamefont
  {{Hu}}(1998)}]{EisensteinHu1998}%
  \BibitemOpen
  \bibfield  {author} {\bibinfo {author} {\bibfnamefont {D.~J.}\ \bibnamefont
  {{Eisenstein}}}\ and\ \bibinfo {author} {\bibfnamefont {W.}~\bibnamefont
  {{Hu}}},\ }\href {\doibase 10.1086/305424} {\bibfield  {journal} {\bibinfo
  {journal} {\apj}\ }\textbf {\bibinfo {volume} {496}},\ \bibinfo {pages} {605}
  (\bibinfo {year} {1998})},\ \Eprint {http://arxiv.org/abs/astro-ph/9709112}
  {astro-ph/9709112} \BibitemShut {NoStop}%
\bibitem [{\citenamefont {Meiksin}\ \emph {et~al.}(1998)\citenamefont
  {Meiksin}, \citenamefont {White},\ and\ \citenamefont
  {Peacock}}]{Meiksin:1998ra}%
  \BibitemOpen
  \bibfield  {author} {\bibinfo {author} {\bibfnamefont {A.}~\bibnamefont
  {Meiksin}}, \bibinfo {author} {\bibfnamefont {M.~J.}\ \bibnamefont {White}},
  \ and\ \bibinfo {author} {\bibfnamefont {J.~A.}\ \bibnamefont {Peacock}},\
  }\href@noop {} {\  (\bibinfo {year} {1998})},\ \Eprint
  {http://arxiv.org/abs/astro-ph/9812214} {arXiv:astro-ph/9812214} \BibitemShut
  {NoStop}%
\bibitem [{\citenamefont {Eisenstein}\ \emph {et~al.}(2005)\citenamefont
  {Eisenstein} \emph {et~al.}}]{Eisenstein:2005su}%
  \BibitemOpen
  \bibfield  {author} {\bibinfo {author} {\bibfnamefont {D.~J.}\ \bibnamefont
  {Eisenstein}} \emph {et~al.} (\bibinfo {collaboration} {SDSS}),\ }\href
  {\doibase 10.1086/466512} {\bibfield  {journal} {\bibinfo  {journal}
  {Astrophys. J.}\ }\textbf {\bibinfo {volume} {633}},\ \bibinfo {pages} {560}
  (\bibinfo {year} {2005})},\ \Eprint {http://arxiv.org/abs/astro-ph/0501171}
  {arXiv:astro-ph/0501171} \BibitemShut {NoStop}%
\bibitem [{\citenamefont {Taruya}\ \emph {et~al.}(2010)\citenamefont {Taruya},
  \citenamefont {Nishimichi},\ and\ \citenamefont {Saito}}]{Taruya:2010mx}%
  \BibitemOpen
  \bibfield  {author} {\bibinfo {author} {\bibfnamefont {A.}~\bibnamefont
  {Taruya}}, \bibinfo {author} {\bibfnamefont {T.}~\bibnamefont {Nishimichi}},
  \ and\ \bibinfo {author} {\bibfnamefont {S.}~\bibnamefont {Saito}},\ }\href
  {\doibase 10.1103/PhysRevD.82.063522} {\bibfield  {journal} {\bibinfo
  {journal} {Phys. Rev.}\ }\textbf {\bibinfo {volume} {D82}},\ \bibinfo {pages}
  {063522} (\bibinfo {year} {2010})},\ \Eprint {http://arxiv.org/abs/1006.0699}
  {arXiv:1006.0699 [astro-ph.CO]} \BibitemShut {NoStop}%
\bibitem [{\citenamefont {{Reid}}\ \emph {et~al.}(2012)\citenamefont {{Reid}},
  \citenamefont {{Samushia}}, \citenamefont {{White}}, \citenamefont
  {{Percival}}, \citenamefont {{Manera}}, \citenamefont {{Padmanabhan}},
  \citenamefont {{Ross}}, \citenamefont {{S{\'a}nchez}}, \citenamefont
  {{Bailey}}, \citenamefont {{Bizyaev}}, \citenamefont {{Bolton}},
  \citenamefont {{Brewington}}, \citenamefont {{Brinkmann}}, \citenamefont
  {{Brownstein}}, \citenamefont {{Cuesta}}, \citenamefont {{Eisenstein}},
  \citenamefont {{Gunn}}, \citenamefont {{Honscheid}}, \citenamefont
  {{Malanushenko}}, \citenamefont {{Malanushenko}}, \citenamefont {{Maraston}},
  \citenamefont {{McBride}}, \citenamefont {{Muna}}, \citenamefont {{Nichol}},
  \citenamefont {{Oravetz}}, \citenamefont {{Pan}}, \citenamefont {{de
  Putter}}, \citenamefont {{Roe}}, \citenamefont {{Ross}}, \citenamefont
  {{Schlegel}}, \citenamefont {{Schneider}}, \citenamefont {{Seo}},
  \citenamefont {{Shelden}}, \citenamefont {{Sheldon}}, \citenamefont
  {{Simmons}}, \citenamefont {{Skibba}}, \citenamefont {{Snedden}},
  \citenamefont {{Swanson}}, \citenamefont {{Thomas}}, \citenamefont
  {{Tinker}}, \citenamefont {{Tojeiro}}, \citenamefont {{Verde}}, \citenamefont
  {{Wake}}, \citenamefont {{Weaver}}, \citenamefont {{Weinberg}}, \citenamefont
  {{Zehavi}},\ and\ \citenamefont {{Zhao}}}]{2012MNRAS.426.2719R}%
  \BibitemOpen
  \bibfield  {author} {\bibinfo {author} {\bibfnamefont {B.~A.}\ \bibnamefont
  {{Reid}}}, \bibinfo {author} {\bibfnamefont {L.}~\bibnamefont {{Samushia}}},
  \bibinfo {author} {\bibfnamefont {M.}~\bibnamefont {{White}}}, \bibinfo
  {author} {\bibfnamefont {W.~J.}\ \bibnamefont {{Percival}}}, \bibinfo
  {author} {\bibfnamefont {M.}~\bibnamefont {{Manera}}}, \bibinfo {author}
  {\bibfnamefont {N.}~\bibnamefont {{Padmanabhan}}}, \bibinfo {author}
  {\bibfnamefont {A.~J.}\ \bibnamefont {{Ross}}}, \bibinfo {author}
  {\bibfnamefont {A.~G.}\ \bibnamefont {{S{\'a}nchez}}}, \bibinfo {author}
  {\bibfnamefont {S.}~\bibnamefont {{Bailey}}}, \bibinfo {author}
  {\bibfnamefont {D.}~\bibnamefont {{Bizyaev}}}, \bibinfo {author}
  {\bibfnamefont {A.~S.}\ \bibnamefont {{Bolton}}}, \bibinfo {author}
  {\bibfnamefont {H.}~\bibnamefont {{Brewington}}}, \bibinfo {author}
  {\bibfnamefont {J.}~\bibnamefont {{Brinkmann}}}, \bibinfo {author}
  {\bibfnamefont {J.~R.}\ \bibnamefont {{Brownstein}}}, \bibinfo {author}
  {\bibfnamefont {A.~J.}\ \bibnamefont {{Cuesta}}}, \bibinfo {author}
  {\bibfnamefont {D.~J.}\ \bibnamefont {{Eisenstein}}}, \bibinfo {author}
  {\bibfnamefont {J.~E.}\ \bibnamefont {{Gunn}}}, \bibinfo {author}
  {\bibfnamefont {K.}~\bibnamefont {{Honscheid}}}, \bibinfo {author}
  {\bibfnamefont {E.}~\bibnamefont {{Malanushenko}}}, \bibinfo {author}
  {\bibfnamefont {V.}~\bibnamefont {{Malanushenko}}}, \bibinfo {author}
  {\bibfnamefont {C.}~\bibnamefont {{Maraston}}}, \bibinfo {author}
  {\bibfnamefont {C.~K.}\ \bibnamefont {{McBride}}}, \bibinfo {author}
  {\bibfnamefont {D.}~\bibnamefont {{Muna}}}, \bibinfo {author} {\bibfnamefont
  {R.~C.}\ \bibnamefont {{Nichol}}}, \bibinfo {author} {\bibfnamefont
  {D.}~\bibnamefont {{Oravetz}}}, \bibinfo {author} {\bibfnamefont
  {K.}~\bibnamefont {{Pan}}}, \bibinfo {author} {\bibfnamefont
  {R.}~\bibnamefont {{de Putter}}}, \bibinfo {author} {\bibfnamefont {N.~A.}\
  \bibnamefont {{Roe}}}, \bibinfo {author} {\bibfnamefont {N.~P.}\ \bibnamefont
  {{Ross}}}, \bibinfo {author} {\bibfnamefont {D.~J.}\ \bibnamefont
  {{Schlegel}}}, \bibinfo {author} {\bibfnamefont {D.~P.}\ \bibnamefont
  {{Schneider}}}, \bibinfo {author} {\bibfnamefont {H.-J.}\ \bibnamefont
  {{Seo}}}, \bibinfo {author} {\bibfnamefont {A.}~\bibnamefont {{Shelden}}},
  \bibinfo {author} {\bibfnamefont {E.~S.}\ \bibnamefont {{Sheldon}}}, \bibinfo
  {author} {\bibfnamefont {A.}~\bibnamefont {{Simmons}}}, \bibinfo {author}
  {\bibfnamefont {R.~A.}\ \bibnamefont {{Skibba}}}, \bibinfo {author}
  {\bibfnamefont {S.}~\bibnamefont {{Snedden}}}, \bibinfo {author}
  {\bibfnamefont {M.~E.~C.}\ \bibnamefont {{Swanson}}}, \bibinfo {author}
  {\bibfnamefont {D.}~\bibnamefont {{Thomas}}}, \bibinfo {author}
  {\bibfnamefont {J.}~\bibnamefont {{Tinker}}}, \bibinfo {author}
  {\bibfnamefont {R.}~\bibnamefont {{Tojeiro}}}, \bibinfo {author}
  {\bibfnamefont {L.}~\bibnamefont {{Verde}}}, \bibinfo {author} {\bibfnamefont
  {D.~A.}\ \bibnamefont {{Wake}}}, \bibinfo {author} {\bibfnamefont {B.~A.}\
  \bibnamefont {{Weaver}}}, \bibinfo {author} {\bibfnamefont {D.~H.}\
  \bibnamefont {{Weinberg}}}, \bibinfo {author} {\bibfnamefont
  {I.}~\bibnamefont {{Zehavi}}}, \ and\ \bibinfo {author} {\bibfnamefont
  {G.-B.}\ \bibnamefont {{Zhao}}},\ }\href {\doibase
  10.1111/j.1365-2966.2012.21779.x} {\bibfield  {journal} {\bibinfo  {journal}
  {Mon. Not. Roy. Astron. Soc.}\ }\textbf {\bibinfo {volume} {426}},\ \bibinfo
  {pages} {2719} (\bibinfo {year} {2012})},\ \Eprint
  {http://arxiv.org/abs/1203.6641} {arXiv:1203.6641 [astro-ph.CO]} \BibitemShut
  {NoStop}%
\bibitem [{\citenamefont {Oka}\ \emph {et~al.}(2014)\citenamefont {Oka},
  \citenamefont {Saito}, \citenamefont {Nishimichi}, \citenamefont {Taruya},\
  and\ \citenamefont {Yamamoto}}]{Oka:2013cba}%
  \BibitemOpen
  \bibfield  {author} {\bibinfo {author} {\bibfnamefont {A.}~\bibnamefont
  {Oka}}, \bibinfo {author} {\bibfnamefont {S.}~\bibnamefont {Saito}}, \bibinfo
  {author} {\bibfnamefont {T.}~\bibnamefont {Nishimichi}}, \bibinfo {author}
  {\bibfnamefont {A.}~\bibnamefont {Taruya}}, \ and\ \bibinfo {author}
  {\bibfnamefont {K.}~\bibnamefont {Yamamoto}},\ }\href {\doibase
  10.1093/mnras/stu111} {\bibfield  {journal} {\bibinfo  {journal} {Mon. Not.
  Roy. Astron. Soc.}\ }\textbf {\bibinfo {volume} {439}},\ \bibinfo {pages}
  {2515} (\bibinfo {year} {2014})},\ \Eprint {http://arxiv.org/abs/1310.2820}
  {arXiv:1310.2820 [astro-ph.CO]} \BibitemShut {NoStop}%
\bibitem [{\citenamefont {Beutler}\ \emph {et~al.}(2013)\citenamefont {Beutler}
  \emph {et~al.}}]{Beutler:2013yhm}%
  \BibitemOpen
  \bibfield  {author} {\bibinfo {author} {\bibfnamefont {F.}~\bibnamefont
  {Beutler}} \emph {et~al.} (\bibinfo {collaboration} {BOSS Collaboration}),\
  }\href@noop {} {\  (\bibinfo {year} {2013})},\ \Eprint
  {http://arxiv.org/abs/1312.4611} {arXiv:1312.4611 [astro-ph.CO]} \BibitemShut
  {NoStop}%
\bibitem [{\citenamefont {Linder}(2008)}]{Linder:2007nu}%
  \BibitemOpen
  \bibfield  {author} {\bibinfo {author} {\bibfnamefont {E.~V.}\ \bibnamefont
  {Linder}},\ }\href {\doibase 10.1016/j.astropartphys.2008.03.002} {\bibfield
  {journal} {\bibinfo  {journal} {Astropart. Phys.}\ }\textbf {\bibinfo
  {volume} {29}},\ \bibinfo {pages} {336} (\bibinfo {year} {2008})},\ \Eprint
  {http://arxiv.org/abs/0709.1113} {arXiv:0709.1113 [astro-ph]} \BibitemShut
  {NoStop}%
\bibitem [{\citenamefont {Seo}\ and\ \citenamefont
  {Eisenstein}(2003)}]{Seo:2003pu}%
  \BibitemOpen
  \bibfield  {author} {\bibinfo {author} {\bibfnamefont {H.-J.}\ \bibnamefont
  {Seo}}\ and\ \bibinfo {author} {\bibfnamefont {D.~J.}\ \bibnamefont
  {Eisenstein}},\ }\href {\doibase 10.1086/379122} {\bibfield  {journal}
  {\bibinfo  {journal} {Astrophys. J.}\ }\textbf {\bibinfo {volume} {598}},\
  \bibinfo {pages} {720} (\bibinfo {year} {2003})},\ \Eprint
  {http://arxiv.org/abs/astro-ph/0307460} {arXiv:astro-ph/0307460 [astro-ph]}
  \BibitemShut {NoStop}%
\bibitem [{\citenamefont {Blake}\ and\ \citenamefont
  {Glazebrook}(2003)}]{Blake:2003rh}%
  \BibitemOpen
  \bibfield  {author} {\bibinfo {author} {\bibfnamefont {C.}~\bibnamefont
  {Blake}}\ and\ \bibinfo {author} {\bibfnamefont {K.}~\bibnamefont
  {Glazebrook}},\ }\href {\doibase 10.1086/376983} {\bibfield  {journal}
  {\bibinfo  {journal} {Astrophys. J.}\ }\textbf {\bibinfo {volume} {594}},\
  \bibinfo {pages} {665} (\bibinfo {year} {2003})},\ \Eprint
  {http://arxiv.org/abs/astro-ph/0301632} {arXiv:astro-ph/0301632 [astro-ph]}
  \BibitemShut {NoStop}%
\bibitem [{\citenamefont {Glazebrook}\ and\ \citenamefont
  {Blake}(2005)}]{Glazebrook:2005mb}%
  \BibitemOpen
  \bibfield  {author} {\bibinfo {author} {\bibfnamefont {K.}~\bibnamefont
  {Glazebrook}}\ and\ \bibinfo {author} {\bibfnamefont {C.}~\bibnamefont
  {Blake}},\ }\href {\doibase 10.1086/432497} {\bibfield  {journal} {\bibinfo
  {journal} {Astrophys. J.}\ }\textbf {\bibinfo {volume} {631}},\ \bibinfo
  {pages} {1} (\bibinfo {year} {2005})},\ \Eprint
  {http://arxiv.org/abs/astro-ph/0505608} {arXiv:astro-ph/0505608 [astro-ph]}
  \BibitemShut {NoStop}%
\bibitem [{\citenamefont {{Song}}\ \emph {et~al.}(2015)\citenamefont {{Song}},
  \citenamefont {{Taruya}}, \citenamefont {{Linder}}, \citenamefont {{Koyama}},
  \citenamefont {{Sabiu}}, \citenamefont {{Zhao}}, \citenamefont
  {{Bernardeau}}, \citenamefont {{Nishimichi}},\ and\ \citenamefont
  {{Okumura}}}]{2015PhRvD..92d3522S}%
  \BibitemOpen
  \bibfield  {author} {\bibinfo {author} {\bibfnamefont {Y.-S.}\ \bibnamefont
  {{Song}}}, \bibinfo {author} {\bibfnamefont {A.}~\bibnamefont {{Taruya}}},
  \bibinfo {author} {\bibfnamefont {E.}~\bibnamefont {{Linder}}}, \bibinfo
  {author} {\bibfnamefont {K.}~\bibnamefont {{Koyama}}}, \bibinfo {author}
  {\bibfnamefont {C.~G.}\ \bibnamefont {{Sabiu}}}, \bibinfo {author}
  {\bibfnamefont {G.-B.}\ \bibnamefont {{Zhao}}}, \bibinfo {author}
  {\bibfnamefont {F.}~\bibnamefont {{Bernardeau}}}, \bibinfo {author}
  {\bibfnamefont {T.}~\bibnamefont {{Nishimichi}}}, \ and\ \bibinfo {author}
  {\bibfnamefont {T.}~\bibnamefont {{Okumura}}},\ }\href {\doibase
  10.1103/PhysRevD.92.043522} {\bibfield  {journal} {\bibinfo  {journal}
  {\prd}\ }\textbf {\bibinfo {volume} {92}},\ \bibinfo {eid} {043522} (\bibinfo
  {year} {2015})},\ \Eprint {http://arxiv.org/abs/1507.01592}
  {arXiv:1507.01592} \BibitemShut {NoStop}%
\bibitem [{\citenamefont {Guzzo}\ \emph {et~al.}(2008)\citenamefont {Guzzo}
  \emph {et~al.}}]{Guzzo:2008ac}%
  \BibitemOpen
  \bibfield  {author} {\bibinfo {author} {\bibfnamefont {L.}~\bibnamefont
  {Guzzo}} \emph {et~al.},\ }\href {\doibase 10.1038/nature06555} {\bibfield
  {journal} {\bibinfo  {journal} {Nature}\ }\textbf {\bibinfo {volume} {451}},\
  \bibinfo {pages} {541} (\bibinfo {year} {2008})},\ \Eprint
  {http://arxiv.org/abs/0802.1944} {arXiv:0802.1944 [astro-ph]} \BibitemShut
  {NoStop}%
\bibitem [{\citenamefont {Okumura}\ \emph {et~al.}(2016)\citenamefont {Okumura}
  \emph {et~al.}}]{Okumura:2015lvp}%
  \BibitemOpen
  \bibfield  {author} {\bibinfo {author} {\bibfnamefont {T.}~\bibnamefont
  {Okumura}} \emph {et~al.},\ }\href {\doibase 10.1093/pasj/psw029} {\bibfield
  {journal} {\bibinfo  {journal} {Publ. Astron. Soc. Jap.}\ }\textbf {\bibinfo
  {volume} {68}},\ \bibinfo {pages} {24} (\bibinfo {year} {2016})},\ \Eprint
  {http://arxiv.org/abs/1511.08083} {arXiv:1511.08083 [astro-ph.CO]}
  \BibitemShut {NoStop}%
\bibitem [{\citenamefont {{de la Torre}}\ \emph {et~al.}(2013)\citenamefont
  {{de la Torre}}, \citenamefont {{Guzzo}}, \citenamefont {{Peacock}},
  \citenamefont {{Branchini}}, \citenamefont {{Iovino}}, \citenamefont
  {{Granett}}, \citenamefont {{Abbas}}, \citenamefont {{Adami}}, \citenamefont
  {{Arnouts}}, \citenamefont {{Bel}}, \citenamefont {{Bolzonella}},
  \citenamefont {{Bottini}}, \citenamefont {{Cappi}}, \citenamefont {{Coupon}},
  \citenamefont {{Cucciati}}, \citenamefont {{Davidzon}}, \citenamefont {{De
  Lucia}}, \citenamefont {{Fritz}}, \citenamefont {{Franzetti}}, \citenamefont
  {{Fumana}}, \citenamefont {{Garilli}}, \citenamefont {{Ilbert}},
  \citenamefont {{Krywult}}, \citenamefont {{Le Brun}}, \citenamefont {{Le
  F{\`e}vre}}, \citenamefont {{Maccagni}}, \citenamefont {{Ma{\l}ek}},
  \citenamefont {{Marulli}}, \citenamefont {{McCracken}}, \citenamefont
  {{Moscardini}}, \citenamefont {{Paioro}}, \citenamefont {{Percival}},
  \citenamefont {{Polletta}}, \citenamefont {{Pollo}}, \citenamefont
  {{Schlagenhaufer}}, \citenamefont {{Scodeggio}}, \citenamefont {{Tasca}},
  \citenamefont {{Tojeiro}}, \citenamefont {{Vergani}}, \citenamefont
  {{Zanichelli}}, \citenamefont {{Burden}}, \citenamefont {{Di Porto}},
  \citenamefont {{Marchetti}}, \citenamefont {{Marinoni}}, \citenamefont
  {{Mellier}}, \citenamefont {{Monaco}}, \citenamefont {{Nichol}},
  \citenamefont {{Phleps}}, \citenamefont {{Wolk}},\ and\ \citenamefont
  {{Zamorani}}}]{2013A&A...557A..54D}%
  \BibitemOpen
  \bibfield  {author} {\bibinfo {author} {\bibfnamefont {S.}~\bibnamefont {{de
  la Torre}}}, \bibinfo {author} {\bibfnamefont {L.}~\bibnamefont {{Guzzo}}},
  \bibinfo {author} {\bibfnamefont {J.~A.}\ \bibnamefont {{Peacock}}}, \bibinfo
  {author} {\bibfnamefont {E.}~\bibnamefont {{Branchini}}}, \bibinfo {author}
  {\bibfnamefont {A.}~\bibnamefont {{Iovino}}}, \bibinfo {author}
  {\bibfnamefont {B.~R.}\ \bibnamefont {{Granett}}}, \bibinfo {author}
  {\bibfnamefont {U.}~\bibnamefont {{Abbas}}}, \bibinfo {author} {\bibfnamefont
  {C.}~\bibnamefont {{Adami}}}, \bibinfo {author} {\bibfnamefont
  {S.}~\bibnamefont {{Arnouts}}}, \bibinfo {author} {\bibfnamefont
  {J.}~\bibnamefont {{Bel}}}, \bibinfo {author} {\bibfnamefont
  {M.}~\bibnamefont {{Bolzonella}}}, \bibinfo {author} {\bibfnamefont
  {D.}~\bibnamefont {{Bottini}}}, \bibinfo {author} {\bibfnamefont
  {A.}~\bibnamefont {{Cappi}}}, \bibinfo {author} {\bibfnamefont
  {J.}~\bibnamefont {{Coupon}}}, \bibinfo {author} {\bibfnamefont
  {O.}~\bibnamefont {{Cucciati}}}, \bibinfo {author} {\bibfnamefont
  {I.}~\bibnamefont {{Davidzon}}}, \bibinfo {author} {\bibfnamefont
  {G.}~\bibnamefont {{De Lucia}}}, \bibinfo {author} {\bibfnamefont
  {A.}~\bibnamefont {{Fritz}}}, \bibinfo {author} {\bibfnamefont
  {P.}~\bibnamefont {{Franzetti}}}, \bibinfo {author} {\bibfnamefont
  {M.}~\bibnamefont {{Fumana}}}, \bibinfo {author} {\bibfnamefont
  {B.}~\bibnamefont {{Garilli}}}, \bibinfo {author} {\bibfnamefont
  {O.}~\bibnamefont {{Ilbert}}}, \bibinfo {author} {\bibfnamefont
  {J.}~\bibnamefont {{Krywult}}}, \bibinfo {author} {\bibfnamefont
  {V.}~\bibnamefont {{Le Brun}}}, \bibinfo {author} {\bibfnamefont
  {O.}~\bibnamefont {{Le F{\`e}vre}}}, \bibinfo {author} {\bibfnamefont
  {D.}~\bibnamefont {{Maccagni}}}, \bibinfo {author} {\bibfnamefont
  {K.}~\bibnamefont {{Ma{\l}ek}}}, \bibinfo {author} {\bibfnamefont
  {F.}~\bibnamefont {{Marulli}}}, \bibinfo {author} {\bibfnamefont {H.~J.}\
  \bibnamefont {{McCracken}}}, \bibinfo {author} {\bibfnamefont
  {L.}~\bibnamefont {{Moscardini}}}, \bibinfo {author} {\bibfnamefont
  {L.}~\bibnamefont {{Paioro}}}, \bibinfo {author} {\bibfnamefont {W.~J.}\
  \bibnamefont {{Percival}}}, \bibinfo {author} {\bibfnamefont
  {M.}~\bibnamefont {{Polletta}}}, \bibinfo {author} {\bibfnamefont
  {A.}~\bibnamefont {{Pollo}}}, \bibinfo {author} {\bibfnamefont
  {H.}~\bibnamefont {{Schlagenhaufer}}}, \bibinfo {author} {\bibfnamefont
  {M.}~\bibnamefont {{Scodeggio}}}, \bibinfo {author} {\bibfnamefont
  {L.~A.~M.}\ \bibnamefont {{Tasca}}}, \bibinfo {author} {\bibfnamefont
  {R.}~\bibnamefont {{Tojeiro}}}, \bibinfo {author} {\bibfnamefont
  {D.}~\bibnamefont {{Vergani}}}, \bibinfo {author} {\bibfnamefont
  {A.}~\bibnamefont {{Zanichelli}}}, \bibinfo {author} {\bibfnamefont
  {A.}~\bibnamefont {{Burden}}}, \bibinfo {author} {\bibfnamefont
  {C.}~\bibnamefont {{Di Porto}}}, \bibinfo {author} {\bibfnamefont
  {A.}~\bibnamefont {{Marchetti}}}, \bibinfo {author} {\bibfnamefont
  {C.}~\bibnamefont {{Marinoni}}}, \bibinfo {author} {\bibfnamefont
  {Y.}~\bibnamefont {{Mellier}}}, \bibinfo {author} {\bibfnamefont
  {P.}~\bibnamefont {{Monaco}}}, \bibinfo {author} {\bibfnamefont {R.~C.}\
  \bibnamefont {{Nichol}}}, \bibinfo {author} {\bibfnamefont {S.}~\bibnamefont
  {{Phleps}}}, \bibinfo {author} {\bibfnamefont {M.}~\bibnamefont {{Wolk}}}, \
  and\ \bibinfo {author} {\bibfnamefont {G.}~\bibnamefont {{Zamorani}}},\
  }\href {\doibase 10.1051/0004-6361/201321463} {\bibfield  {journal} {\bibinfo
   {journal} {Astron. Astrophys.}\ }\textbf {\bibinfo {volume} {557}},\
  \bibinfo {eid} {A54} (\bibinfo {year} {2013})},\ \Eprint
  {http://arxiv.org/abs/1303.2622} {arXiv:1303.2622} \BibitemShut {NoStop}%
\bibitem [{\citenamefont {Beutler}\ \emph {et~al.}(2016)\citenamefont {Beutler}
  \emph {et~al.}}]{Beutler:2016arn}%
  \BibitemOpen
  \bibfield  {author} {\bibinfo {author} {\bibfnamefont {F.}~\bibnamefont
  {Beutler}} \emph {et~al.} (\bibinfo {collaboration} {BOSS}),\ }\href@noop {}
  {\bibfield  {journal} {\bibinfo  {journal} {Submitted to: Mon. Not. Roy.
  Astron. Soc.}\ } (\bibinfo {year} {2016})},\ \Eprint
  {http://arxiv.org/abs/1607.03150} {arXiv:1607.03150 [astro-ph.CO]}
  \BibitemShut {NoStop}%
\bibitem [{\citenamefont {Blake}\ \emph {et~al.}(2011)\citenamefont {Blake},
  \citenamefont {Brough}, \citenamefont {Colless}, \citenamefont {Contreras},
  \citenamefont {Couch} \emph {et~al.}}]{Blake:2011rj}%
  \BibitemOpen
  \bibfield  {author} {\bibinfo {author} {\bibfnamefont {C.}~\bibnamefont
  {Blake}}, \bibinfo {author} {\bibfnamefont {S.}~\bibnamefont {Brough}},
  \bibinfo {author} {\bibfnamefont {M.}~\bibnamefont {Colless}}, \bibinfo
  {author} {\bibfnamefont {C.}~\bibnamefont {Contreras}}, \bibinfo {author}
  {\bibfnamefont {W.}~\bibnamefont {Couch}},  \emph {et~al.},\ }\href {\doibase
  10.1111/j.1365-2966.2011.18903.x} {\bibfield  {journal} {\bibinfo  {journal}
  {Mon. Not. Roy. Astron. Soc.}\ }\textbf {\bibinfo {volume} {415}},\ \bibinfo
  {pages} {2876} (\bibinfo {year} {2011})},\ \Eprint
  {http://arxiv.org/abs/1104.2948} {arXiv:1104.2948 [astro-ph.CO]} \BibitemShut
  {NoStop}%
\bibitem [{\citenamefont {{Planck Collaboration}}\ \emph
  {et~al.}(2016)\citenamefont {{Planck Collaboration}}, \citenamefont {{Ade}},
  \citenamefont {{Aghanim}}, \citenamefont {{Arnaud}}, \citenamefont
  {{Ashdown}}, \citenamefont {{Aumont}}, \citenamefont {{Baccigalupi}},
  \citenamefont {{Banday}}, \citenamefont {{Barreiro}}, \citenamefont
  {{Bartlett}},\ and\ \citenamefont {et~al.}}]{Planck2015_XIII}%
  \BibitemOpen
  \bibfield  {author} {\bibinfo {author} {\bibnamefont {{Planck
  Collaboration}}}, \bibinfo {author} {\bibfnamefont {P.~A.~R.}\ \bibnamefont
  {{Ade}}}, \bibinfo {author} {\bibfnamefont {N.}~\bibnamefont {{Aghanim}}},
  \bibinfo {author} {\bibfnamefont {M.}~\bibnamefont {{Arnaud}}}, \bibinfo
  {author} {\bibfnamefont {M.}~\bibnamefont {{Ashdown}}}, \bibinfo {author}
  {\bibfnamefont {J.}~\bibnamefont {{Aumont}}}, \bibinfo {author}
  {\bibfnamefont {C.}~\bibnamefont {{Baccigalupi}}}, \bibinfo {author}
  {\bibfnamefont {A.~J.}\ \bibnamefont {{Banday}}}, \bibinfo {author}
  {\bibfnamefont {R.~B.}\ \bibnamefont {{Barreiro}}}, \bibinfo {author}
  {\bibfnamefont {J.~G.}\ \bibnamefont {{Bartlett}}}, \ and\ \bibinfo {author}
  {\bibnamefont {et~al.}},\ }\href {\doibase 10.1051/0004-6361/201525830}
  {\bibfield  {journal} {\bibinfo  {journal} {Astron. Astrophys.}\ }\textbf
  {\bibinfo {volume} {594}},\ \bibinfo {eid} {A13} (\bibinfo {year} {2016})},\
  \Eprint {http://arxiv.org/abs/1502.01589} {arXiv:1502.01589} \BibitemShut
  {NoStop}%
\bibitem [{\citenamefont {{Planck Collaboration}}\ \emph
  {et~al.}(2015)\citenamefont {{Planck Collaboration}}, \citenamefont {{Ade}},
  \citenamefont {{Aghanim}}, \citenamefont {{Arnaud}}, \citenamefont
  {{Ashdown}}, \citenamefont {{Aumont}}, \citenamefont {{Baccigalupi}},
  \citenamefont {{Banday}}, \citenamefont {{Barreiro}}, \citenamefont
  {{Bartlett}},\ and\ \citenamefont {et~al.}}]{2015arXiv150201589P}%
  \BibitemOpen
  \bibfield  {author} {\bibinfo {author} {\bibnamefont {{Planck
  Collaboration}}}, \bibinfo {author} {\bibfnamefont {P.~A.~R.}\ \bibnamefont
  {{Ade}}}, \bibinfo {author} {\bibfnamefont {N.}~\bibnamefont {{Aghanim}}},
  \bibinfo {author} {\bibfnamefont {M.}~\bibnamefont {{Arnaud}}}, \bibinfo
  {author} {\bibfnamefont {M.}~\bibnamefont {{Ashdown}}}, \bibinfo {author}
  {\bibfnamefont {J.}~\bibnamefont {{Aumont}}}, \bibinfo {author}
  {\bibfnamefont {C.}~\bibnamefont {{Baccigalupi}}}, \bibinfo {author}
  {\bibfnamefont {A.~J.}\ \bibnamefont {{Banday}}}, \bibinfo {author}
  {\bibfnamefont {R.~B.}\ \bibnamefont {{Barreiro}}}, \bibinfo {author}
  {\bibfnamefont {J.~G.}\ \bibnamefont {{Bartlett}}}, \ and\ \bibinfo {author}
  {\bibnamefont {et~al.}},\ }\href@noop {} {\bibfield  {journal} {\bibinfo
  {journal} {ArXiv e-prints}\ } (\bibinfo {year} {2015})},\ \Eprint
  {http://arxiv.org/abs/1502.01589} {arXiv:1502.01589} \BibitemShut {NoStop}%
\bibitem [{\citenamefont {Sefusatti}\ \emph {et~al.}(2006)\citenamefont
  {Sefusatti}, \citenamefont {Crocce}, \citenamefont {Pueblas},\ and\
  \citenamefont {Scoccimarro}}]{Sefusatti:2006pa}%
  \BibitemOpen
  \bibfield  {author} {\bibinfo {author} {\bibfnamefont {E.}~\bibnamefont
  {Sefusatti}}, \bibinfo {author} {\bibfnamefont {M.}~\bibnamefont {Crocce}},
  \bibinfo {author} {\bibfnamefont {S.}~\bibnamefont {Pueblas}}, \ and\
  \bibinfo {author} {\bibfnamefont {R.}~\bibnamefont {Scoccimarro}},\ }\href
  {\doibase 10.1103/PhysRevD.74.023522} {\bibfield  {journal} {\bibinfo
  {journal} {Phys. Rev.}\ }\textbf {\bibinfo {volume} {D74}},\ \bibinfo {pages}
  {023522} (\bibinfo {year} {2006})},\ \Eprint
  {http://arxiv.org/abs/astro-ph/0604505} {arXiv:astro-ph/0604505 [astro-ph]}
  \BibitemShut {NoStop}%
\bibitem [{\citenamefont {{Kayo}}\ \emph {et~al.}(2013)\citenamefont {{Kayo}},
  \citenamefont {{Takada}},\ and\ \citenamefont
  {{Jain}}}]{2013MNRAS.429..344K}%
  \BibitemOpen
  \bibfield  {author} {\bibinfo {author} {\bibfnamefont {I.}~\bibnamefont
  {{Kayo}}}, \bibinfo {author} {\bibfnamefont {M.}~\bibnamefont {{Takada}}}, \
  and\ \bibinfo {author} {\bibfnamefont {B.}~\bibnamefont {{Jain}}},\ }\href
  {\doibase 10.1093/mnras/sts340} {\bibfield  {journal} {\bibinfo  {journal}
  {Mon. Not. Roy. Astron. Soc.}\ }\textbf {\bibinfo {volume} {429}},\ \bibinfo
  {pages} {344} (\bibinfo {year} {2013})},\ \Eprint
  {http://arxiv.org/abs/1207.6322} {arXiv:1207.6322 [astro-ph.CO]} \BibitemShut
  {NoStop}%
\bibitem [{\citenamefont {Sato}\ and\ \citenamefont
  {Nishimichi}(2013)}]{Sato:2013mq}%
  \BibitemOpen
  \bibfield  {author} {\bibinfo {author} {\bibfnamefont {M.}~\bibnamefont
  {Sato}}\ and\ \bibinfo {author} {\bibfnamefont {T.}~\bibnamefont
  {Nishimichi}},\ }\href {\doibase 10.1103/PhysRevD.87.123538} {\bibfield
  {journal} {\bibinfo  {journal} {Phys. Rev.}\ }\textbf {\bibinfo {volume}
  {D87}},\ \bibinfo {pages} {123538} (\bibinfo {year} {2013})},\ \Eprint
  {http://arxiv.org/abs/1301.3588} {arXiv:1301.3588 [astro-ph.CO]} \BibitemShut
  {NoStop}%
\bibitem [{\citenamefont {Greig}\ \emph {et~al.}(2013)\citenamefont {Greig},
  \citenamefont {Komatsu},\ and\ \citenamefont {Wyithe}}]{Greig:2012zw}%
  \BibitemOpen
  \bibfield  {author} {\bibinfo {author} {\bibfnamefont {B.}~\bibnamefont
  {Greig}}, \bibinfo {author} {\bibfnamefont {E.}~\bibnamefont {Komatsu}}, \
  and\ \bibinfo {author} {\bibfnamefont {J.~S.~B.}\ \bibnamefont {Wyithe}},\
  }\href {\doibase 10.1093/mnras/stt292} {\bibfield  {journal} {\bibinfo
  {journal} {Mon. Not. Roy. Astron. Soc.}\ }\textbf {\bibinfo {volume} {431}},\
  \bibinfo {pages} {1777} (\bibinfo {year} {2013})},\ \Eprint
  {http://arxiv.org/abs/1212.0977} {arXiv:1212.0977 [astro-ph.CO]} \BibitemShut
  {NoStop}%
\bibitem [{\citenamefont {Song}\ \emph {et~al.}(2015)\citenamefont {Song},
  \citenamefont {Taruya},\ and\ \citenamefont {Oka}}]{Song:2015gca}%
  \BibitemOpen
  \bibfield  {author} {\bibinfo {author} {\bibfnamefont {Y.-S.}\ \bibnamefont
  {Song}}, \bibinfo {author} {\bibfnamefont {A.}~\bibnamefont {Taruya}}, \ and\
  \bibinfo {author} {\bibfnamefont {A.}~\bibnamefont {Oka}},\ }\href {\doibase
  10.1088/1475-7516/2015/08/007} {\bibfield  {journal} {\bibinfo  {journal}
  {JCAP}\ }\textbf {\bibinfo {volume} {1508}},\ \bibinfo {pages} {007}
  (\bibinfo {year} {2015})},\ \Eprint {http://arxiv.org/abs/1502.03099}
  {arXiv:1502.03099 [astro-ph.CO]} \BibitemShut {NoStop}%
\bibitem [{\citenamefont {{Gagrani}}\ and\ \citenamefont
  {{Samushia}}(2017)}]{2017MNRAS.tmp..138G}%
  \BibitemOpen
  \bibfield  {author} {\bibinfo {author} {\bibfnamefont {P.}~\bibnamefont
  {{Gagrani}}}\ and\ \bibinfo {author} {\bibfnamefont {L.}~\bibnamefont
  {{Samushia}}},\ }\href {\doibase 10.1093/mnras/stx135} {\bibfield  {journal}
  {\bibinfo  {journal} {Mon. Not. Roy. Astron. Soc.}\ } (\bibinfo {year}
  {2017}),\ 10.1093/mnras/stx135},\ \Eprint {http://arxiv.org/abs/1610.03488}
  {arXiv:1610.03488} \BibitemShut {NoStop}%
\bibitem [{\citenamefont {Scoccimarro}\ \emph {et~al.}(1999)\citenamefont
  {Scoccimarro}, \citenamefont {Couchman},\ and\ \citenamefont
  {Frieman}}]{Scoccimarro:1999ed}%
  \BibitemOpen
  \bibfield  {author} {\bibinfo {author} {\bibfnamefont {R.}~\bibnamefont
  {Scoccimarro}}, \bibinfo {author} {\bibfnamefont {H.~M.~P.}\ \bibnamefont
  {Couchman}}, \ and\ \bibinfo {author} {\bibfnamefont {J.~A.}\ \bibnamefont
  {Frieman}},\ }\href {\doibase 10.1086/307220} {\bibfield  {journal} {\bibinfo
   {journal} {Astrophys. J.}\ }\textbf {\bibinfo {volume} {517}},\ \bibinfo
  {pages} {531} (\bibinfo {year} {1999})},\ \Eprint
  {http://arxiv.org/abs/astro-ph/9808305} {arXiv:astro-ph/9808305 [astro-ph]}
  \BibitemShut {NoStop}%
\bibitem [{\citenamefont {Reid}\ and\ \citenamefont
  {White}(2011)}]{Reid:2011ar}%
  \BibitemOpen
  \bibfield  {author} {\bibinfo {author} {\bibfnamefont {B.~A.}\ \bibnamefont
  {Reid}}\ and\ \bibinfo {author} {\bibfnamefont {M.}~\bibnamefont {White}},\
  }\href {\doibase 10.1111/j.1365-2966.2011.19379.x} {\bibfield  {journal}
  {\bibinfo  {journal} {Mon. Not. Roy. Astron. Soc.}\ }\textbf {\bibinfo
  {volume} {417}},\ \bibinfo {pages} {1913} (\bibinfo {year} {2011})},\ \Eprint
  {http://arxiv.org/abs/1105.4165} {arXiv:1105.4165 [astro-ph.CO]} \BibitemShut
  {NoStop}%
\bibitem [{\citenamefont {Vlah}\ \emph {et~al.}(2012)\citenamefont {Vlah},
  \citenamefont {Seljak}, \citenamefont {McDonald}, \citenamefont {Okumura},\
  and\ \citenamefont {Baldauf}}]{Vlah:2012ni}%
  \BibitemOpen
  \bibfield  {author} {\bibinfo {author} {\bibfnamefont {Z.}~\bibnamefont
  {Vlah}}, \bibinfo {author} {\bibfnamefont {U.}~\bibnamefont {Seljak}},
  \bibinfo {author} {\bibfnamefont {P.}~\bibnamefont {McDonald}}, \bibinfo
  {author} {\bibfnamefont {T.}~\bibnamefont {Okumura}}, \ and\ \bibinfo
  {author} {\bibfnamefont {T.}~\bibnamefont {Baldauf}},\ }\href {\doibase
  10.1088/1475-7516/2012/11/009} {\bibfield  {journal} {\bibinfo  {journal}
  {JCAP}\ }\textbf {\bibinfo {volume} {1211}},\ \bibinfo {pages} {009}
  (\bibinfo {year} {2012})},\ \Eprint {http://arxiv.org/abs/1207.0839}
  {arXiv:1207.0839 [astro-ph.CO]} \BibitemShut {NoStop}%
\bibitem [{\citenamefont {Matsubara}(2008{\natexlab{a}})}]{Matsubara:2007wj}%
  \BibitemOpen
  \bibfield  {author} {\bibinfo {author} {\bibfnamefont {T.}~\bibnamefont
  {Matsubara}},\ }\href {\doibase 10.1103/PhysRevD.77.063530} {\bibfield
  {journal} {\bibinfo  {journal} {Phys. Rev.}\ }\textbf {\bibinfo {volume}
  {D77}},\ \bibinfo {pages} {063530} (\bibinfo {year} {2008}{\natexlab{a}})},\
  \Eprint {http://arxiv.org/abs/0711.2521} {arXiv:0711.2521 [astro-ph]}
  \BibitemShut {NoStop}%
\bibitem [{\citenamefont {Matsubara}(2008{\natexlab{b}})}]{Matsubara:2008wx}%
  \BibitemOpen
  \bibfield  {author} {\bibinfo {author} {\bibfnamefont {T.}~\bibnamefont
  {Matsubara}},\ }\href {\doibase 10.1103/PhysRevD.78.109901,
  10.1103/PhysRevD.78.083519} {\bibfield  {journal} {\bibinfo  {journal} {Phys.
  Rev.}\ }\textbf {\bibinfo {volume} {D78}},\ \bibinfo {pages} {083519}
  (\bibinfo {year} {2008}{\natexlab{b}})},\ \bibinfo {note} {[Erratum: Phys.
  Rev.D78,109901(2008)]},\ \Eprint {http://arxiv.org/abs/0807.1733}
  {arXiv:0807.1733 [astro-ph]} \BibitemShut {NoStop}%
\bibitem [{\citenamefont {Carlson}\ \emph {et~al.}(2013)\citenamefont
  {Carlson}, \citenamefont {Reid},\ and\ \citenamefont
  {White}}]{Carlson:2012bu}%
  \BibitemOpen
  \bibfield  {author} {\bibinfo {author} {\bibfnamefont {J.}~\bibnamefont
  {Carlson}}, \bibinfo {author} {\bibfnamefont {B.}~\bibnamefont {Reid}}, \
  and\ \bibinfo {author} {\bibfnamefont {M.}~\bibnamefont {White}},\ }\href
  {\doibase 10.1093/mnras/sts457} {\bibfield  {journal} {\bibinfo  {journal}
  {Mon. Not. Roy. Astron. Soc.}\ }\textbf {\bibinfo {volume} {429}},\ \bibinfo
  {pages} {1674} (\bibinfo {year} {2013})},\ \Eprint
  {http://arxiv.org/abs/1209.0780} {arXiv:1209.0780 [astro-ph.CO]} \BibitemShut
  {NoStop}%
\bibitem [{\citenamefont {Matsubara}(2014)}]{Matsubara:2013ofa}%
  \BibitemOpen
  \bibfield  {author} {\bibinfo {author} {\bibfnamefont {T.}~\bibnamefont
  {Matsubara}},\ }\href {\doibase 10.1103/PhysRevD.90.043537} {\bibfield
  {journal} {\bibinfo  {journal} {Phys. Rev.}\ }\textbf {\bibinfo {volume}
  {D90}},\ \bibinfo {pages} {043537} (\bibinfo {year} {2014})},\ \Eprint
  {http://arxiv.org/abs/1304.4226} {arXiv:1304.4226 [astro-ph.CO]} \BibitemShut
  {NoStop}%
\bibitem [{\citenamefont {{Taruya}}\ \emph {et~al.}(2013)\citenamefont
  {{Taruya}}, \citenamefont {{Nishimichi}},\ and\ \citenamefont
  {{Bernardeau}}}]{2013PhRvD..87h3509T}%
  \BibitemOpen
  \bibfield  {author} {\bibinfo {author} {\bibfnamefont {A.}~\bibnamefont
  {{Taruya}}}, \bibinfo {author} {\bibfnamefont {T.}~\bibnamefont
  {{Nishimichi}}}, \ and\ \bibinfo {author} {\bibfnamefont {F.}~\bibnamefont
  {{Bernardeau}}},\ }\href {\doibase 10.1103/PhysRevD.87.083509} {\bibfield
  {journal} {\bibinfo  {journal} {\prd}\ }\textbf {\bibinfo {volume} {87}},\
  \bibinfo {eid} {083509} (\bibinfo {year} {2013})},\ \Eprint
  {http://arxiv.org/abs/1301.3624} {arXiv:1301.3624 [astro-ph.CO]} \BibitemShut
  {NoStop}%
\bibitem [{\citenamefont {{Gil-Mar{\'{\i}}n}}\ \emph
  {et~al.}(2012)\citenamefont {{Gil-Mar{\'{\i}}n}}, \citenamefont {{Wagner}},
  \citenamefont {{Fragkoudi}}, \citenamefont {{Jimenez}},\ and\ \citenamefont
  {{Verde}}}]{2012JCAP...02..047G}%
  \BibitemOpen
  \bibfield  {author} {\bibinfo {author} {\bibfnamefont {H.}~\bibnamefont
  {{Gil-Mar{\'{\i}}n}}}, \bibinfo {author} {\bibfnamefont {C.}~\bibnamefont
  {{Wagner}}}, \bibinfo {author} {\bibfnamefont {F.}~\bibnamefont
  {{Fragkoudi}}}, \bibinfo {author} {\bibfnamefont {R.}~\bibnamefont
  {{Jimenez}}}, \ and\ \bibinfo {author} {\bibfnamefont {L.}~\bibnamefont
  {{Verde}}},\ }\href {\doibase 10.1088/1475-7516/2012/02/047} {\bibfield
  {journal} {\bibinfo  {journal} {JCAP}\ }\textbf {\bibinfo {volume} {2}},\
  \bibinfo {eid} {047} (\bibinfo {year} {2012})},\ \Eprint
  {http://arxiv.org/abs/1111.4477} {arXiv:1111.4477 [astro-ph.CO]} \BibitemShut
  {NoStop}%
\bibitem [{\citenamefont {{Scoccimarro}}\ and\ \citenamefont
  {{Couchman}}(2001)}]{2001MNRAS.325.1312S}%
  \BibitemOpen
  \bibfield  {author} {\bibinfo {author} {\bibfnamefont {R.}~\bibnamefont
  {{Scoccimarro}}}\ and\ \bibinfo {author} {\bibfnamefont {H.~M.~P.}\
  \bibnamefont {{Couchman}}},\ }\href {\doibase
  10.1046/j.1365-8711.2001.04281.x} {\bibfield  {journal} {\bibinfo  {journal}
  {Mon. Not. Roy. Astron. Soc.}\ }\textbf {\bibinfo {volume} {325}},\ \bibinfo
  {pages} {1312} (\bibinfo {year} {2001})},\ \Eprint
  {http://arxiv.org/abs/astro-ph/0009427} {astro-ph/0009427} \BibitemShut
  {NoStop}%
\bibitem [{\citenamefont {{Smith}}\ \emph {et~al.}(2008)\citenamefont
  {{Smith}}, \citenamefont {{Sheth}},\ and\ \citenamefont
  {{Scoccimarro}}}]{2008PhRvD..78b3523S}%
  \BibitemOpen
  \bibfield  {author} {\bibinfo {author} {\bibfnamefont {R.~E.}\ \bibnamefont
  {{Smith}}}, \bibinfo {author} {\bibfnamefont {R.~K.}\ \bibnamefont
  {{Sheth}}}, \ and\ \bibinfo {author} {\bibfnamefont {R.}~\bibnamefont
  {{Scoccimarro}}},\ }\href {\doibase 10.1103/PhysRevD.78.023523} {\bibfield
  {journal} {\bibinfo  {journal} {Phys. Rev.}\ }\textbf {\bibinfo {volume}
  {78}},\ \bibinfo {eid} {023523} (\bibinfo {year} {2008})},\ \Eprint
  {http://arxiv.org/abs/0712.0017} {arXiv:0712.0017} \BibitemShut {NoStop}%
\bibitem [{\citenamefont {Yamamoto}\ \emph {et~al.}(2016)\citenamefont
  {Yamamoto}, \citenamefont {Nan},\ and\ \citenamefont
  {Hikage}}]{Yamamoto:2016anp}%
  \BibitemOpen
  \bibfield  {author} {\bibinfo {author} {\bibfnamefont {K.}~\bibnamefont
  {Yamamoto}}, \bibinfo {author} {\bibfnamefont {Y.}~\bibnamefont {Nan}}, \
  and\ \bibinfo {author} {\bibfnamefont {C.}~\bibnamefont {Hikage}},\
  }\href@noop {} {\  (\bibinfo {year} {2016})},\ \Eprint
  {http://arxiv.org/abs/1610.03665} {arXiv:1610.03665 [astro-ph.CO]}
  \BibitemShut {NoStop}%
\bibitem [{\citenamefont {{Rampf}}\ and\ \citenamefont
  {{Wong}}(2012)}]{2012JCAP...06..018R}%
  \BibitemOpen
  \bibfield  {author} {\bibinfo {author} {\bibfnamefont {C.}~\bibnamefont
  {{Rampf}}}\ and\ \bibinfo {author} {\bibfnamefont {Y.~Y.~Y.}\ \bibnamefont
  {{Wong}}},\ }\href {\doibase 10.1088/1475-7516/2012/06/018} {\bibfield
  {journal} {\bibinfo  {journal} {JCAP}\ }\textbf {\bibinfo {volume} {6}},\
  \bibinfo {eid} {018} (\bibinfo {year} {2012})},\ \Eprint
  {http://arxiv.org/abs/1203.4261} {arXiv:1203.4261 [astro-ph.CO]} \BibitemShut
  {NoStop}%
\bibitem [{\citenamefont {{Komatsu}}\ \emph {et~al.}(2011)\citenamefont
  {{Komatsu}}, \citenamefont {{Smith}}, \citenamefont {{Dunkley}},
  \citenamefont {{Bennett}}, \citenamefont {{Gold}}, \citenamefont {{Hinshaw}},
  \citenamefont {{Jarosik}}, \citenamefont {{Larson}}, \citenamefont {{Nolta}},
  \citenamefont {{Page}}, \citenamefont {{Spergel}}, \citenamefont {{Halpern}},
  \citenamefont {{Hill}}, \citenamefont {{Kogut}}, \citenamefont {{Limon}},
  \citenamefont {{Meyer}}, \citenamefont {{Odegard}}, \citenamefont {{Tucker}},
  \citenamefont {{Weiland}}, \citenamefont {{Wollack}},\ and\ \citenamefont
  {{Wright}}}]{2011ApJS..192...18K}%
  \BibitemOpen
  \bibfield  {author} {\bibinfo {author} {\bibfnamefont {E.}~\bibnamefont
  {{Komatsu}}}, \bibinfo {author} {\bibfnamefont {K.~M.}\ \bibnamefont
  {{Smith}}}, \bibinfo {author} {\bibfnamefont {J.}~\bibnamefont {{Dunkley}}},
  \bibinfo {author} {\bibfnamefont {C.~L.}\ \bibnamefont {{Bennett}}}, \bibinfo
  {author} {\bibfnamefont {B.}~\bibnamefont {{Gold}}}, \bibinfo {author}
  {\bibfnamefont {G.}~\bibnamefont {{Hinshaw}}}, \bibinfo {author}
  {\bibfnamefont {N.}~\bibnamefont {{Jarosik}}}, \bibinfo {author}
  {\bibfnamefont {D.}~\bibnamefont {{Larson}}}, \bibinfo {author}
  {\bibfnamefont {M.~R.}\ \bibnamefont {{Nolta}}}, \bibinfo {author}
  {\bibfnamefont {L.}~\bibnamefont {{Page}}}, \bibinfo {author} {\bibfnamefont
  {D.~N.}\ \bibnamefont {{Spergel}}}, \bibinfo {author} {\bibfnamefont
  {M.}~\bibnamefont {{Halpern}}}, \bibinfo {author} {\bibfnamefont {R.~S.}\
  \bibnamefont {{Hill}}}, \bibinfo {author} {\bibfnamefont {A.}~\bibnamefont
  {{Kogut}}}, \bibinfo {author} {\bibfnamefont {M.}~\bibnamefont {{Limon}}},
  \bibinfo {author} {\bibfnamefont {S.~S.}\ \bibnamefont {{Meyer}}}, \bibinfo
  {author} {\bibfnamefont {N.}~\bibnamefont {{Odegard}}}, \bibinfo {author}
  {\bibfnamefont {G.~S.}\ \bibnamefont {{Tucker}}}, \bibinfo {author}
  {\bibfnamefont {J.~L.}\ \bibnamefont {{Weiland}}}, \bibinfo {author}
  {\bibfnamefont {E.}~\bibnamefont {{Wollack}}}, \ and\ \bibinfo {author}
  {\bibfnamefont {E.~L.}\ \bibnamefont {{Wright}}},\ }\href {\doibase
  10.1088/0067-0049/192/2/18} {\bibfield  {journal} {\bibinfo  {journal}
  {Astrophys. J. Suppl.}\ }\textbf {\bibinfo {volume} {192}},\ \bibinfo {eid}
  {18} (\bibinfo {year} {2011})},\ \Eprint {http://arxiv.org/abs/1001.4538}
  {arXiv:1001.4538 [astro-ph.CO]} \BibitemShut {NoStop}%
\bibitem [{\citenamefont {{Bernardeau}}\ \emph {et~al.}(2002)\citenamefont
  {{Bernardeau}}, \citenamefont {{Colombi}}, \citenamefont {{Gazta{\~n}aga}},\
  and\ \citenamefont {{Scoccimarro}}}]{Bernardeau:2002}%
  \BibitemOpen
  \bibfield  {author} {\bibinfo {author} {\bibfnamefont {F.}~\bibnamefont
  {{Bernardeau}}}, \bibinfo {author} {\bibfnamefont {S.}~\bibnamefont
  {{Colombi}}}, \bibinfo {author} {\bibfnamefont {E.}~\bibnamefont
  {{Gazta{\~n}aga}}}, \ and\ \bibinfo {author} {\bibfnamefont {R.}~\bibnamefont
  {{Scoccimarro}}},\ }\href {\doibase 10.1016/S0370-1573(02)00135-7} {\bibfield
   {journal} {\bibinfo  {journal} {Physics Reports}\ }\textbf {\bibinfo
  {volume} {367}},\ \bibinfo {pages} {1} (\bibinfo {year} {2002})},\ \Eprint
  {http://arxiv.org/abs/astro-ph/0112551} {astro-ph/0112551} \BibitemShut
  {NoStop}%
\bibitem [{\citenamefont {{Scoccimarro}}(2015)}]{2015PhRvD..92h3532S}%
  \BibitemOpen
  \bibfield  {author} {\bibinfo {author} {\bibfnamefont {R.}~\bibnamefont
  {{Scoccimarro}}},\ }\href {\doibase 10.1103/PhysRevD.92.083532} {\bibfield
  {journal} {\bibinfo  {journal} {Phys. Rev. D}\ }\textbf {\bibinfo {volume}
  {92}},\ \bibinfo {eid} {083532} (\bibinfo {year} {2015})},\ \Eprint
  {http://arxiv.org/abs/1506.02729} {arXiv:1506.02729} \BibitemShut {NoStop}%
\bibitem [{\citenamefont {{Blot}}\ \emph {et~al.}(2015)\citenamefont {{Blot}},
  \citenamefont {{Corasaniti}}, \citenamefont {{Alimi}}, \citenamefont
  {{Reverdy}},\ and\ \citenamefont {{Rasera}}}]{Blot:2015}%
  \BibitemOpen
  \bibfield  {author} {\bibinfo {author} {\bibfnamefont {L.}~\bibnamefont
  {{Blot}}}, \bibinfo {author} {\bibfnamefont {P.~S.}\ \bibnamefont
  {{Corasaniti}}}, \bibinfo {author} {\bibfnamefont {J.-M.}\ \bibnamefont
  {{Alimi}}}, \bibinfo {author} {\bibfnamefont {V.}~\bibnamefont {{Reverdy}}},
  \ and\ \bibinfo {author} {\bibfnamefont {Y.}~\bibnamefont {{Rasera}}},\
  }\href {\doibase 10.1093/mnras/stu2190} {\bibfield  {journal} {\bibinfo
  {journal} {Mon. Not. Roy. Astron. Soc.}\ }\textbf {\bibinfo {volume} {446}},\
  \bibinfo {pages} {1756} (\bibinfo {year} {2015})},\ \Eprint
  {http://arxiv.org/abs/1406.2713} {arXiv:1406.2713} \BibitemShut {NoStop}%
\bibitem [{\citenamefont {{Prunet}}\ \emph {et~al.}(2008)\citenamefont
  {{Prunet}}, \citenamefont {{Pichon}}, \citenamefont {{Aubert}}, \citenamefont
  {{Pogosyan}}, \citenamefont {{Teyssier}},\ and\ \citenamefont
  {{Gottloeber}}}]{Prunet:2008}%
  \BibitemOpen
  \bibfield  {author} {\bibinfo {author} {\bibfnamefont {S.}~\bibnamefont
  {{Prunet}}}, \bibinfo {author} {\bibfnamefont {C.}~\bibnamefont {{Pichon}}},
  \bibinfo {author} {\bibfnamefont {D.}~\bibnamefont {{Aubert}}}, \bibinfo
  {author} {\bibfnamefont {D.}~\bibnamefont {{Pogosyan}}}, \bibinfo {author}
  {\bibfnamefont {R.}~\bibnamefont {{Teyssier}}}, \ and\ \bibinfo {author}
  {\bibfnamefont {S.}~\bibnamefont {{Gottloeber}}},\ }\href {\doibase
  10.1086/590370} {\bibfield  {journal} {\bibinfo  {journal} {{The
  Astrophysical Journal Supplement Series}}\ }\textbf {\bibinfo {volume}
  {178}},\ \bibinfo {eid} {179-188} (\bibinfo {year} {2008})},\ \Eprint
  {http://arxiv.org/abs/0804.3536} {arXiv:0804.3536} \BibitemShut {NoStop}%
\bibitem [{\citenamefont {{Teyssier}}(2002)}]{Teyssier:2002}%
  \BibitemOpen
  \bibfield  {author} {\bibinfo {author} {\bibfnamefont {R.}~\bibnamefont
  {{Teyssier}}},\ }\href {\doibase 10.1051/0004-6361:20011817} {\bibfield
  {journal} {\bibinfo  {journal} {Astronomy \& Astrophysics}\ }\textbf
  {\bibinfo {volume} {385}},\ \bibinfo {pages} {337} (\bibinfo {year}
  {2002})},\ \Eprint {http://arxiv.org/abs/astro-ph/0111367} {astro-ph/0111367}
  \BibitemShut {NoStop}%
\bibitem [{\citenamefont {{Rasera}}\ \emph {et~al.}(2014)\citenamefont
  {{Rasera}}, \citenamefont {{Corasaniti}}, \citenamefont {{Alimi}},
  \citenamefont {{Bouillot}}, \citenamefont {{Reverdy}},\ and\ \citenamefont
  {{Balm{\`e}s}}}]{Rasera:2014}%
  \BibitemOpen
  \bibfield  {author} {\bibinfo {author} {\bibfnamefont {Y.}~\bibnamefont
  {{Rasera}}}, \bibinfo {author} {\bibfnamefont {P.-S.}\ \bibnamefont
  {{Corasaniti}}}, \bibinfo {author} {\bibfnamefont {J.-M.}\ \bibnamefont
  {{Alimi}}}, \bibinfo {author} {\bibfnamefont {V.}~\bibnamefont {{Bouillot}}},
  \bibinfo {author} {\bibfnamefont {V.}~\bibnamefont {{Reverdy}}}, \ and\
  \bibinfo {author} {\bibfnamefont {I.}~\bibnamefont {{Balm{\`e}s}}},\ }\href
  {\doibase 10.1093/mnras/stu295} {\bibfield  {journal} {\bibinfo  {journal}
  {"Mon. Not. Roy. Astron. Soc."}\ }\textbf {\bibinfo {volume} {440}},\
  \bibinfo {pages} {1420} (\bibinfo {year} {2014})},\ \Eprint
  {http://arxiv.org/abs/1311.5662} {arXiv:1311.5662} \BibitemShut {NoStop}%
\bibitem [{\citenamefont {Scoccimarro}\ \emph {et~al.}(1998)\citenamefont
  {Scoccimarro}, \citenamefont {Colombi}, \citenamefont {Fry}, \citenamefont
  {Frieman}, \citenamefont {Hivon},\ and\ \citenamefont
  {Melott}}]{Scoccimarro:1997st}%
  \BibitemOpen
  \bibfield  {author} {\bibinfo {author} {\bibfnamefont {R.}~\bibnamefont
  {Scoccimarro}}, \bibinfo {author} {\bibfnamefont {S.}~\bibnamefont
  {Colombi}}, \bibinfo {author} {\bibfnamefont {J.~N.}\ \bibnamefont {Fry}},
  \bibinfo {author} {\bibfnamefont {J.~A.}\ \bibnamefont {Frieman}}, \bibinfo
  {author} {\bibfnamefont {E.}~\bibnamefont {Hivon}}, \ and\ \bibinfo {author}
  {\bibfnamefont {A.}~\bibnamefont {Melott}},\ }\href {\doibase 10.1086/305399}
  {\bibfield  {journal} {\bibinfo  {journal} {Astrophys. J.}\ }\textbf
  {\bibinfo {volume} {496}},\ \bibinfo {pages} {586} (\bibinfo {year}
  {1998})},\ \Eprint {http://arxiv.org/abs/astro-ph/9704075}
  {arXiv:astro-ph/9704075 [astro-ph]} \BibitemShut {NoStop}%
\bibitem [{\citenamefont {Takahashi}\ \emph {et~al.}(2008)\citenamefont
  {Takahashi} \emph {et~al.}}]{Takahashi:2008wn}%
  \BibitemOpen
  \bibfield  {author} {\bibinfo {author} {\bibfnamefont {R.}~\bibnamefont
  {Takahashi}} \emph {et~al.},\ }\href {\doibase
  10.1111/j.1365-2966.2008.13731.x} {\bibfield  {journal} {\bibinfo  {journal}
  {Mon. Not. Roy. Astron. Soc.}\ }\textbf {\bibinfo {volume} {389}},\ \bibinfo
  {pages} {1675} (\bibinfo {year} {2008})},\ \Eprint
  {http://arxiv.org/abs/0802.1808} {arXiv:0802.1808 [astro-ph]} \BibitemShut
  {NoStop}%
\bibitem [{\citenamefont {Heitmann}\ \emph {et~al.}(2010)\citenamefont
  {Heitmann}, \citenamefont {White}, \citenamefont {Wagner}, \citenamefont
  {Habib},\ and\ \citenamefont {Higdon}}]{Heitmann:2008eq}%
  \BibitemOpen
  \bibfield  {author} {\bibinfo {author} {\bibfnamefont {K.}~\bibnamefont
  {Heitmann}}, \bibinfo {author} {\bibfnamefont {M.}~\bibnamefont {White}},
  \bibinfo {author} {\bibfnamefont {C.}~\bibnamefont {Wagner}}, \bibinfo
  {author} {\bibfnamefont {S.}~\bibnamefont {Habib}}, \ and\ \bibinfo {author}
  {\bibfnamefont {D.}~\bibnamefont {Higdon}},\ }\href {\doibase
  10.1088/0004-637X/715/1/104} {\bibfield  {journal} {\bibinfo  {journal}
  {Astrophys. J.}\ }\textbf {\bibinfo {volume} {715}},\ \bibinfo {pages} {104}
  (\bibinfo {year} {2010})},\ \Eprint {http://arxiv.org/abs/0812.1052}
  {arXiv:0812.1052 [astro-ph]} \BibitemShut {NoStop}%
\bibitem [{\citenamefont {Heitmann}\ \emph {et~al.}(2009)\citenamefont
  {Heitmann}, \citenamefont {Higdon}, \citenamefont {White}, \citenamefont
  {Habib}, \citenamefont {Williams},\ and\ \citenamefont
  {Wagner}}]{Heitmann:2009cu}%
  \BibitemOpen
  \bibfield  {author} {\bibinfo {author} {\bibfnamefont {K.}~\bibnamefont
  {Heitmann}}, \bibinfo {author} {\bibfnamefont {D.}~\bibnamefont {Higdon}},
  \bibinfo {author} {\bibfnamefont {M.}~\bibnamefont {White}}, \bibinfo
  {author} {\bibfnamefont {S.}~\bibnamefont {Habib}}, \bibinfo {author}
  {\bibfnamefont {B.~J.}\ \bibnamefont {Williams}}, \ and\ \bibinfo {author}
  {\bibfnamefont {C.}~\bibnamefont {Wagner}},\ }\href {\doibase
  10.1088/0004-637X/705/1/156} {\bibfield  {journal} {\bibinfo  {journal}
  {Astrophys. J.}\ }\textbf {\bibinfo {volume} {705}},\ \bibinfo {pages} {156}
  (\bibinfo {year} {2009})},\ \Eprint {http://arxiv.org/abs/0902.0429}
  {arXiv:0902.0429 [astro-ph.CO]} \BibitemShut {NoStop}%
\bibitem [{\citenamefont {{Lawrence}}\ \emph {et~al.}(2010)\citenamefont
  {{Lawrence}}, \citenamefont {{Heitmann}}, \citenamefont {{White}},
  \citenamefont {{Higdon}}, \citenamefont {{Wagner}}, \citenamefont {{Habib}},\
  and\ \citenamefont {{Williams}}}]{2010ApJ...713.1322L}%
  \BibitemOpen
  \bibfield  {author} {\bibinfo {author} {\bibfnamefont {E.}~\bibnamefont
  {{Lawrence}}}, \bibinfo {author} {\bibfnamefont {K.}~\bibnamefont
  {{Heitmann}}}, \bibinfo {author} {\bibfnamefont {M.}~\bibnamefont {{White}}},
  \bibinfo {author} {\bibfnamefont {D.}~\bibnamefont {{Higdon}}}, \bibinfo
  {author} {\bibfnamefont {C.}~\bibnamefont {{Wagner}}}, \bibinfo {author}
  {\bibfnamefont {S.}~\bibnamefont {{Habib}}}, \ and\ \bibinfo {author}
  {\bibfnamefont {B.}~\bibnamefont {{Williams}}},\ }\href {\doibase
  10.1088/0004-637X/713/2/1322} {\bibfield  {journal} {\bibinfo  {journal}
  {\apj}\ }\textbf {\bibinfo {volume} {713}},\ \bibinfo {pages} {1322}
  (\bibinfo {year} {2010})},\ \Eprint {http://arxiv.org/abs/0912.4490}
  {arXiv:0912.4490 [astro-ph.CO]} \BibitemShut {NoStop}%
\bibitem [{\citenamefont {Heitmann}\ \emph {et~al.}(2014)\citenamefont
  {Heitmann}, \citenamefont {Lawrence}, \citenamefont {Kwan}, \citenamefont
  {Habib},\ and\ \citenamefont {Higdon}}]{Heitmann:2013bra}%
  \BibitemOpen
  \bibfield  {author} {\bibinfo {author} {\bibfnamefont {K.}~\bibnamefont
  {Heitmann}}, \bibinfo {author} {\bibfnamefont {E.}~\bibnamefont {Lawrence}},
  \bibinfo {author} {\bibfnamefont {J.}~\bibnamefont {Kwan}}, \bibinfo {author}
  {\bibfnamefont {S.}~\bibnamefont {Habib}}, \ and\ \bibinfo {author}
  {\bibfnamefont {D.}~\bibnamefont {Higdon}},\ }\href {\doibase
  10.1088/0004-637X/780/1/111} {\bibfield  {journal} {\bibinfo  {journal}
  {Astrophys. J.}\ }\textbf {\bibinfo {volume} {780}},\ \bibinfo {pages} {111}
  (\bibinfo {year} {2014})},\ \Eprint {http://arxiv.org/abs/1304.7849}
  {arXiv:1304.7849 [astro-ph.CO]} \BibitemShut {NoStop}%
\bibitem [{\citenamefont {Nishimichi}\ \emph {et~al.}(2009)\citenamefont
  {Nishimichi} \emph {et~al.}}]{Nishimichi:2008ry}%
  \BibitemOpen
  \bibfield  {author} {\bibinfo {author} {\bibfnamefont {T.}~\bibnamefont
  {Nishimichi}} \emph {et~al.},\ }\href {\doibase 10.1093/pasj/61.2.321}
  {\bibfield  {journal} {\bibinfo  {journal} {Publ. Astron. Soc. Jap.}\
  }\textbf {\bibinfo {volume} {61}},\ \bibinfo {pages} {321} (\bibinfo {year}
  {2009})},\ \Eprint {http://arxiv.org/abs/0810.0813} {arXiv:0810.0813
  [astro-ph]} \BibitemShut {NoStop}%
\bibitem [{\citenamefont {{Baumann}}\ \emph {et~al.}(2012)\citenamefont
  {{Baumann}}, \citenamefont {{Nicolis}}, \citenamefont {{Senatore}},\ and\
  \citenamefont {{Zaldarriaga}}}]{2012JCAP...07..051B}%
  \BibitemOpen
  \bibfield  {author} {\bibinfo {author} {\bibfnamefont {D.}~\bibnamefont
  {{Baumann}}}, \bibinfo {author} {\bibfnamefont {A.}~\bibnamefont
  {{Nicolis}}}, \bibinfo {author} {\bibfnamefont {L.}~\bibnamefont
  {{Senatore}}}, \ and\ \bibinfo {author} {\bibfnamefont {M.}~\bibnamefont
  {{Zaldarriaga}}},\ }\href {\doibase 10.1088/1475-7516/2012/07/051} {\bibfield
   {journal} {\bibinfo  {journal} {JCAP}\ }\textbf {\bibinfo {volume} {7}},\
  \bibinfo {eid} {051} (\bibinfo {year} {2012})},\ \Eprint
  {http://arxiv.org/abs/1004.2488} {arXiv:1004.2488 [astro-ph.CO]} \BibitemShut
  {NoStop}%
\bibitem [{\citenamefont {{Carrasco}}\ \emph {et~al.}(2012)\citenamefont
  {{Carrasco}}, \citenamefont {{Hertzberg}},\ and\ \citenamefont
  {{Senatore}}}]{2012JHEP...09..082C}%
  \BibitemOpen
  \bibfield  {author} {\bibinfo {author} {\bibfnamefont {J.~J.~M.}\
  \bibnamefont {{Carrasco}}}, \bibinfo {author} {\bibfnamefont {M.~P.}\
  \bibnamefont {{Hertzberg}}}, \ and\ \bibinfo {author} {\bibfnamefont
  {L.}~\bibnamefont {{Senatore}}},\ }\href {\doibase 10.1007/JHEP09(2012)082}
  {\bibfield  {journal} {\bibinfo  {journal} {Journal of High Energy Physics}\
  }\textbf {\bibinfo {volume} {9}},\ \bibinfo {eid} {82} (\bibinfo {year}
  {2012})},\ \Eprint {http://arxiv.org/abs/1206.2926} {arXiv:1206.2926
  [astro-ph.CO]} \BibitemShut {NoStop}%
\bibitem [{\citenamefont {{Hertzberg}}(2014)}]{2014PhRvD..89d3521H}%
  \BibitemOpen
  \bibfield  {author} {\bibinfo {author} {\bibfnamefont {M.~P.}\ \bibnamefont
  {{Hertzberg}}},\ }\href {\doibase 10.1103/PhysRevD.89.043521} {\bibfield
  {journal} {\bibinfo  {journal} {Physical Review D}\ }\textbf {\bibinfo
  {volume} {89}},\ \bibinfo {eid} {043521} (\bibinfo {year} {2014})},\ \Eprint
  {http://arxiv.org/abs/1208.0839} {arXiv:1208.0839} \BibitemShut {NoStop}%
\bibitem [{\citenamefont {Baldauf}\ \emph {et~al.}(2015)\citenamefont
  {Baldauf}, \citenamefont {Mercolli},\ and\ \citenamefont
  {Zaldarriaga}}]{Baldauf:2015aha}%
  \BibitemOpen
  \bibfield  {author} {\bibinfo {author} {\bibfnamefont {T.}~\bibnamefont
  {Baldauf}}, \bibinfo {author} {\bibfnamefont {L.}~\bibnamefont {Mercolli}}, \
  and\ \bibinfo {author} {\bibfnamefont {M.}~\bibnamefont {Zaldarriaga}},\
  }\href {\doibase 10.1103/PhysRevD.92.123007} {\bibfield  {journal} {\bibinfo
  {journal} {Physical Review D}\ }\textbf {\bibinfo {volume} {92}},\ \bibinfo
  {pages} {123007} (\bibinfo {year} {2015})},\ \Eprint
  {http://arxiv.org/abs/1507.02256} {arXiv:1507.02256 [astro-ph.CO]}
  \BibitemShut {NoStop}%
\bibitem [{\citenamefont {{Sato}}\ and\ \citenamefont
  {{Nishimichi}}(2013)}]{2013PhRvD..87l3538S}%
  \BibitemOpen
  \bibfield  {author} {\bibinfo {author} {\bibfnamefont {M.}~\bibnamefont
  {{Sato}}}\ and\ \bibinfo {author} {\bibfnamefont {T.}~\bibnamefont
  {{Nishimichi}}},\ }\href {\doibase 10.1103/PhysRevD.87.123538} {\bibfield
  {journal} {\bibinfo  {journal} {Phys. Rev. D}\ }\textbf {\bibinfo {volume}
  {87}},\ \bibinfo {eid} {123538} (\bibinfo {year} {2013})},\ \Eprint
  {http://arxiv.org/abs/1301.3588} {arXiv:1301.3588} \BibitemShut {NoStop}%
\end{thebibliography}%

\end{document}